\documentclass[twoside,12pt]{article}
\usepackage{epsfig}
\usepackage{color}
\usepackage{amssymb}
\usepackage{graphicx}
\usepackage{latexsym}
\usepackage{amsmath}
\usepackage{bbm}
\usepackage{mathrsfs}
\usepackage{hyperref, cite, collref}

\newcommand{\SU}{\mathrm{SU}}
\newcommand{\U}{\mathrm{U}}
\newcommand{\SO}{\mathrm{SO}}

\newcommand{\Sp}{\mathrm{Sp}}

\newcommand{\Ssugra}{S_{\mbox{\tiny{sugra}}}}
\newcommand{\LambdaQCD}{\Lambda_{\tiny{\mbox{QCD}}}}

\newcommand{\Z}{\mathbb{Z}}

\newcommand{\ide}{\mathbbm{1}}
\newcommand{\real}{{\rm{Re}}}

\newcommand{\dd}{{\rm{d}}}
\newcommand{\Tr}{{\rm Tr}}
\newcommand{\tr}{{\rm Tr}}
\newcommand{\beq}{\begin{eqnarray}}
\newcommand{\eeq}{\end{eqnarray}}
\newcommand{\topsusc}{\chi_{\mbox{\tiny{topol}}}}

\begin{document}
\begin{flushright} 
HIP-2013-18/TH\\
IFT-UAM/CSIC-13-096
\end{flushright}
\title{Introductory lectures to large-$N$ QCD phenomenology and lattice results}
\author{Biagio~Lucini$^{1}$ and Marco~Panero$^{2,3}$\\
\\
$^1$ College of Science, Swansea University\\
Singleton Park, Swansea SA2 8PP, UK\\
\vspace{-2mm}\\
$^2$ Department of Physics and Helsinki Institute of Physics, University of Helsinki\\
FIN-00014 Helsinki, Finland\\
\vspace{-2mm}\\
$^3$ Instituto de F\'{\i}sica T\'eorica UAM/CSIC, Universidad Aut\'onoma de Madrid\\
E-28049 Madrid, Spain
\vspace{-2mm}\\}

\begingroup
\let\newpage\relax
\maketitle
\endgroup

\begin{abstract} An elementary, pedagogical introduction to the large-$N$ limit of QCD and to its phenomenological implications is presented, and a survey of lattice results in the 't~Hooft limit is briefly discussed.
\end{abstract}
\eject
\tableofcontents

\section{Introduction}

The current theoretical understanding of the physics of elementary particles is based on the so-called Standard Model: a framework incorporating quantum physics and special relativity that describes the strong, electromagnetic and weak interactions, in terms of renormalizable gauge theories.

The Standard Model predicts phenomena over a huge range of energy scales, and its validity is confirmed by experiments (in some cases, to a striking level of numerical precision). The experimental discovery of a particle with properties compatible with those of an elementary Higgs boson at the LHC is a further, impressive piece of evidence supporting the validity of the Standard Model---at least in the energy range accessible to present particle accelerators.

There exist well-grounded theoretical arguments to expect that the Standard Model is \emph{not} the ultimate theory of nature, but, rather, arises as an effective low-energy description of a more fundamental theory. In particular, the Standard Model does not include a description of \emph{general} relativity in a way consistent with quantum field theory. This, however, does not lead to observable effects in present laboratory experiments, because the energy scale at which quantum effects become relevant for the gravitational interaction is the Planck scale $1/\sqrt{G} \sim 10^{19}$~GeV, which is way beyond the current reach of accelerator experiments. The Standard Model is also characterized by the so-called ``hierarchy problem'': if the Higgs field is the only elementary scalar field, then there appears to be no convincing explanation why the typical scale (of the order of $10^2$~GeV) of its mass does not get driven to much higher values by quantum fluctuations, given the strong (quadratic) dependence on the energy scale at which new physics may set in. In the absence of some fundamental (super)symmetry reason, it is unlikely that huge contributions to the electroweak scale from quantum fluctuations of different fields cancel almost completely against each other, unless the parameters of the model take delicately fine-tuned values. Such a situation, however, appears to be extremely ``unnatural''. The Standard Model is also somewhat inelegant, in that it includes about twenty (or more, if neutrino masses have to be included), a priori unspecified, numerical constants (mostly coefficients of Yukawa interactions) of arbitrary values. In addition, it does not really ``unify'' the strong and electroweak interactions in a non-trivial way (nor does it predict that the couplings match with each other at some high energy).

For many problems related to Cosmology, the Standard Model of elementary particles seems to be inadequate: it does not feature a ``natural'' candidate for the experimentally observed large amount of cold dark matter, and grossly fails to describe the scale of dark energy. Issues related to baryogenesis, as well as the fact that, on large distance scales, the Universe appears to be homogeneous and isotropic, also give indications that new physics may exist. 

The question why there is no evidence for physics beyond the Standard Model to date (but this situation might change, with the experimental searches scheduled to continue with proton-proton collisions at an increased total center-of-mass energy of 13 TeV in 2015) despite the apparently solid theoretical arguments for its existence is currently central to our understanding of particle physics. However, it is worth remarking that, even within the Standard Model, there still exist various open theoretical problems. In particular, although most popular media reported the experimental discovery of a Higgs-like boson with scientifically inaccurate headlines about ``the particle that gives you mass'', it is worth pointing out that \emph{most} of the ordinary mass of atomi is actually due to the finite mass of nucleons, which is much larger than that of their constituent quarks, and arises \emph{dynamically}, as a result of the highly non-linear and non-perturbative nature of the strong nuclear interaction described 
by quantum chromodynamics (QCD).

While most processes involving the electroweak interaction can be accurately described in terms of weak-coupling perturbative expansions, these tools cannot be applied to the study of physical states at low energy in QCD. In this respect, a physical QCD bound state, such as a nucleon or a meson, is qualitatively radically different from, say, a hydrogen atom or a positronium state in quantum electrodynamics (QED). At the core of this difference is the fact that, in contrast to electrons, isolated quarks do not exist as asymptotic states, due to the \emph{confining} nature the strong nuclear interaction. While \emph{asymptotic freedom} implies that perturbative QCD computations can be meaningfully carried out (and successfully compared with experimental results) for processes involving large momentum transfers between the interacting partons, the properties of strongly interacting matter at low energies are determined by dynamics not captured by the perturbative treatment. In particular, the spectrum of the 
lightest physical states in QCD is characterized by confinement into color-singlet states (hadrons) and spontaneous breakdown of chiral symmetry, both of which are intrinsically non-perturbative phenomena.

The regularization of QCD on a spacetime lattice, first proposed by Wilson in 1974\footnote{For a historical account on the development of lattice gauge theories, see e.g. ref.~\cite{Creutz:2003qy}.}, is among the very few approaches\footnote{An alternative approach is based on the light-cone quantization of the theory~\cite{Brodsky:1997de}.} that enable to define the theory in a gauge-invariant, mathematically rigorous way (avoiding the mathematical subtleties associated with the measure definition in quantum field theory in the continuum), from first principles and without relying on a perturbative approach. The lattice definition of QCD allows one to derive analytically certain interesting properties of the theory (at least at a qualitative level) using strong-coupling expansion methods: well-known examples include the existence of confinement and the finiteness of the mass gap in the pure Yang-Mills theory. While these strong-coupling results are interesting \emph{per se}, it should be remarked that they are obtained in an unphysical limit, in which the lattice theory \emph{does not} correspond to the 
continuum theory. However, the regime of the lattice theory, which is analytically connected to the continuum theory (i.e., the one corresponding to weak couplings) can be investigated by numerical methods, via Monte~Carlo simulations.

During the past decades, this approach has led to a number of important, increasingly accurate physical results, which confirm the validity of QCD as the fundamental theory describing the strong nuclear interaction. However, the lattice Monte~Carlo approach has the disadvantage that, besides numerical results, it does not provide a deeper understanding of the dynamics of the theory. In addition, there exist various classes of problems, for which the approach based on the regularization on a Euclidean lattice is not very suitable (these include, in particular, phenomena involving real-time dynamics) or faces tough fundamental challenges.

These reasons provide motivation to consider alternative non-perturbative approaches to the physics of strong interactions. In this review article, we aim at discussing one of them, originally proposed by 't~Hooft in 1974, which bore a number of very fruitful developments. The basic idea consists in considering a generalization of QCD, in which the number $N$ of color charges (three, in the real world) is considered as an arbitrary parameter, and taking the limit in which it becomes arbitrarily large. Somewhat surprisingly, it turns out that, with an appropriate definition of such ``large-$N$ limit'', the theory reveals many interesting simplifications, and one can obtain a simple, intuitive explanation for many properties characterizing the phenomenology of real-world QCD.

The structure of this article is the following: in section~\ref{sect:definitions} we introduce the definition of the 't~Hooft limit of QCD and discuss the associated ``large-$N$ counting rules''. We show that the latter account for a natural classification of Feynman diagrams with different topologies, which are associated with different powers of $1/N$. The existence of an analogous topological expansion in string theory suggested that string theory may provide a reformulation of the large-$N$ limit of a gauge theory. We discuss how this idea has been made more quantitative in the holographic duality, namely in the (conjectured) correspondence between gauge theories and string theories defined in a higher-dimensional, curved spacetime, and briefly highlight the meaning of some implicit assumptions that analytical computations based on the holographic correspondence rely on. In section~\ref{sect:phenomenology} we show how, under some general assumptions, these large-$N$ counting rules allow one to derive a 
number of interesting phenomenological properties for different hadronic physical states of the theory (glueballs, mesons and baryons). In section~\ref{sect:factorization}, we briefly review an interesting mathematical feature of the large-$N$ limit of QCD, namely the property of \emph{factorization} of expectation values of products of operators associated with physical observables, and some of the manifold deep implications stemming from it. These include, in particular, the emergence of a class of non-trivial equivalences between different physical theories in the large-$N$ limit (which, for historical reasons, are called \emph{orbifold equivalences}). In section~\ref{sect:lattice}, we present an overview of lattice studies of gauge theories with a different number of colors, assessing the question, whether the large-$N$ limit is \emph{quantitatively} relevant for real-world QCD, or, in other words, whether ``three can already 
be considered as a large number''. The last section~\ref{sect:conclusions} is devoted to a summary and to some concluding remarks.

We would like to remark that the purpose of this review consists in introducing the basic ideas and the most important research directions in this field: here, we do not aim at providing a complete mathematical discussion of the large-$N$ approach, nor an exhaustive list of the many relevant works that have been published on this topic during almost four decades. On the contrary, we try to highlight those that, in our view, are the main aspects of the topic, presenting them in a way which should be easily accessible not only for researchers, but also for graduate and undergraduate students. For a more rigorous discussion and a more complete list of references, we recommend the interested readers to refer to our general review on this topic~\cite{Lucini:2012gg} and to an overview of recent lattice results in this field in ref.~\cite{Panero:2012qx}. An incomplete list of recommended earlier review articles on the topic (or on aspects thereof) by other authors includes refs.~\cite{Brezin_Wadia, Witten:1979pi, Coleman:1980nk, Yaffe:1981vf, Migdal:1984gj, Das:1984nb, Lebed:2002tj, Jenkins:1998wy, Lebed:1998st, Teper:1998kw, Manohar:1998xv, Makeenko:1999hq, Makeenko:2004bz, Narayanan:2007fb, Teper:2008yi, Vicari:2008jw, Teper:2009uf}. Finally, we would like to remark that, although the topic has an almost four-decade long history, works proposing novel approaches to large-$N$ QCD continue to appear~\cite{Bochicchio:2013tfa, Bochicchio:2013eda, Kaplan:2013dca}.

\section{Gauge theories at large $N$: from 't~Hooft to Maldacena}
\label{sect:definitions}

\subsection{Basics about the large-$N$ limit of QCD}
\label{subsect:t_Hooft_limit}

The large-$N$ limit of QCD was first discussed in a seminal article by 't~Hooft, published in 1974~\cite{'tHooft:1973jz}. The idea of considering physical models characterized by invariance under a certain group of transformations, with ``size'' related to an integer parameter $N$, in the limit in which $N$ tends to infinity had already been successfully applied in other contexts~\cite{Stanley:1968gx, Ma:1973zu, Brezin:1972se} (see also~\cite{Brezin:1976qa, Bardeen:1976zh, Okabe:1978nn}). 't~Hooft extended this approach to the case of gauge theories: he took the parameter $N$ to be the number of color charges, considered a generalization of the gauge group of QCD to $\SU(N)$, and studied the properties of the theory in the $N \to \infty$ limit.

In considering this limit, the first, trivial, observation is that---unless there are cancellations with some other quantities going to zero at the same time---the limit is singular: many quantities, which grow with $N$ or with some increasing function thereof, would obviously be divergent in this limit. However, it is possible to have sensible, finite limits when $N \to \infty$, provided at the same time one takes the coupling to zero, $g \to 0$. In particular, perturbatively it is easy to show that, in the \emph{double limit} $N \to \infty$, $g \to 0$ keeping the product $\lambda = g^2 N$ fixed, one can obtain finite results. $\lambda$, which is called the 't~Hooft coupling, is thus considered as the actual fundamental coupling of the theory---and many of the interesting simplifications of large-$N$ QCD arise from the (partial or complete) compensation between divergent powers of $N$, and vanishing powers of $g$.

In addition to $N$ and $g$, another dimensionless parameter of QCD is the number of quark flavors $n_f$. In nature, there exists $n_f=6$ quark flavors (\emph{up}, \emph{down}, \emph{strange}, \emph{charm}, \emph{bottom} and \emph{top}, in order of increasing mass). When generalizing QCD to the large-$N$ limit, it is possible to assume that $n_f$ is held fixed~\cite{'tHooft:1973jz}, or that it is scaled with $N$, keeping the $n_f/N$ ratio fixed~\cite{Veneziano:1976wm}: at the perturbative level, both limits make sense. In particular, the limit when $n_f$ is taken to infinity at fixed $x_f=n_f/N$ is called the ``Veneziano limit''; the leading-order perturbative expression of the QCD $\beta$-function:
\begin{equation}
\label{LO_QCD_beta_function}
\mu \frac{d \lambda}{d \mu} = - \frac{11- 2x_f}{24\pi^2} \lambda^2 + O(\lambda^3)
\end{equation}
(which describes the dependence of the physical running 't~Hooft coupling on the momentum scale $\mu$) immediately reveals that the theory remains asymptotically free for all values of $x_f < 11/2$. However, it turns out that the large-$N$ limit of QCD at fixed $n_f$ ('t~Hooft limit) is characterized by simpler properties, so it has received more attention in the literature. In the following, we concentrate on the 't~Hooft limit of QCD.

Perturbative inspection shows that, in the 't~Hooft limit of QCD, most of the interesting properties arise from competing effects due to terms growing like some power of $N$, and terms vanishing like some power of $g$, as we mentioned above. In order to keep track of the powers of $N$, it is particularly convenient to introduce a \emph{double-line notation} for Feynman diagrams: in this notation, every line corresponds to one power of $N$. Since quark fields are in the \emph{fundamental} representation of the gauge group, the number of their color components is $N$, so their propagators can be represented by a single line. By contrast, gluons are fields in the \emph{adjoint} representation of the gauge group, hence their color multiplicity is $N^2-1$, i.e. $O(N^2)$ for $N \to \infty$; as a consequence, gluon propagators are associated with a pair of oppositely oriented\footnote{Note that the adjoint representation arises in the decomposition of the tensor product of the fundamental and the antifundamental 
representation. In addition, the orientation of fundamental lines is well-defined, since, for all $\SU(N>2)$, the fundamental and antifundamental representations are not unitarily equivalent.} lines in double-line notation.

Representing quark and gluon propagators in Feynman diagrams with this double-line notation, counting the number of independent powers of $N$ (one for each line) as well as the powers of $g$ associated to the various interaction vertices, and finally expressing all factors of $g$ in terms of $\lambda$, it is straightforward to show that, at any given order in the coupling, different types of diagrams come with different $N$ multiplicities. In particular, it turns out that the contributions to a given amplitude for a physical process, that are proportional to the largest powers of $N$ (which is not larger than $N^2$), are those corresponding to \emph{planar diagrams without dynamical quark loops}. Here and in the following, a diagram is called \emph{planar}, if, in the double-line notation, it can be drawn on the surface of a plane (or of a sphere) without crossing lines.

\begin{figure}[tb!]
\begin{center}
\begin{minipage}[t]{8 cm}
\epsfig{file=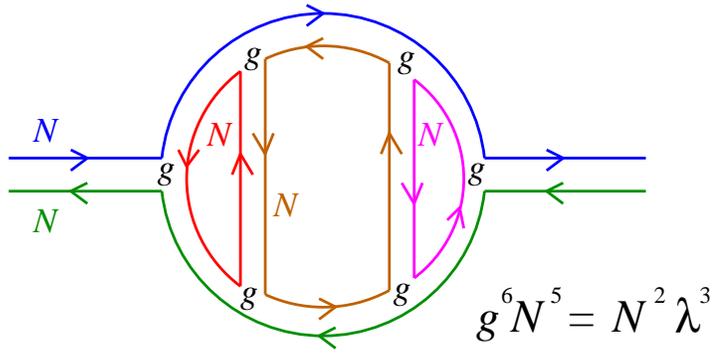,scale=0.55}
\vspace{1cm}\\
\epsfig{file=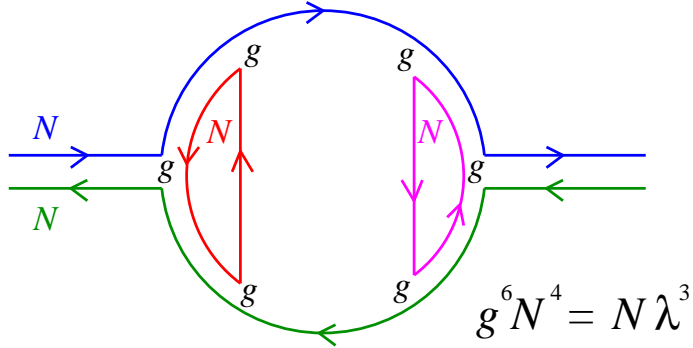,scale=0.55}
\vspace{1cm}\\
\epsfig{file=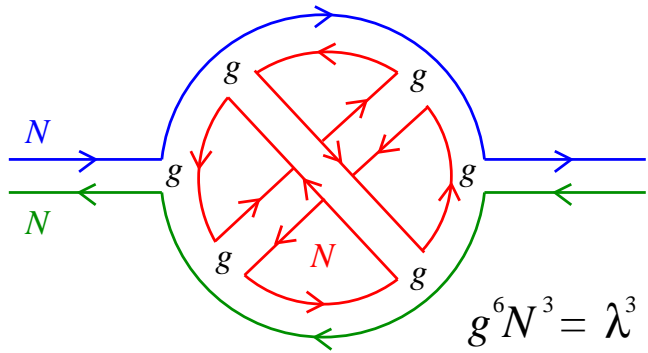,scale=0.55}
\vspace{5mm}\\
\end{minipage}
\begin{minipage}[t]{16.5 cm}
\caption{In the 't~Hooft limit of QCD, at any given order in an expansion in powers of the coupling, different diagrams yield contributions scaling with different powers of $N$, depending on the diagram topology. The dominant contributions come from planar diagrams without dynamical quark loops, like the one in the top panel. By contrast, diagrams including one (or more) internal quark loops, like the one in the central panel, are suppressed by one (or more) power of $1/N$. Similarly, diagrams of non-trivial topology (like the one in the bottom panel, which cannot be drawn on a planar surface without crossing lines) are also suppressed by two (or more) powers of $1/N$.~\label{fig:gluon_propagator}}
\end{minipage}
\end{center}
\end{figure}

As an example, it may be helpful to consider three different types of Feynman diagrams contributing to the gluon self-energy at three loops, as shown in fig.~\ref{fig:gluon_propagator}.

The diagram at the top of the figure is a planar one, and does not feature any internal quark loops: in the 't~Hooft limit, it is $O(N^2)$ (including the multiplicity associated with the number of degrees of freedom of the external gluon). The diagram at the center of the figure is also planar, but it includes an internal quark loop (at the center of the diagram). Since quark propagators are $O(N)$, while gluon propagators are $O(N^2)$ in the large-$N$ limit, the total multiplicity of the diagram is $O(N)$. Finally, the diagram at the bottom is, again, constructed out of gluons only, but its topology is non-trivial: the simplest Riemann surface on which the diagram can be drawn is a torus. The corresponding contribution to the gluon propagator is $O(N^0)$ in the large-$N$ limit. 

Although all the three diagrams shown in fig.~\ref{fig:gluon_propagator} correspond to the same power in the coupling, $O(\lambda^3)$, only one of them (the one in the top panel) yields a non-negligible contribution in the 't~Hooft limit. This feature (the dominance of planar diagrams without quark loops) is, in fact, a general property of the 't~Hooft limit of QCD. In particular, it implies that the amplitude $\mathcal{A}$ for a generic process can be expressed in \emph{double series}---not just in powers of the coupling, but also in powers of $1/N$, where the latter expansion has a \emph{topological nature}, i.e. the power of $1/N$ depends on the genus (or on the number $h$ of ``handles'') of the simplest Riemann surface on which the diagram can be drawn without crossing lines, and on the number $b$ of ``boundaries'' associated with quark loops:
\begin{equation}
\label{amplitude}
\mathcal{A} = \sum_{h,b=0}^{\infty} \left( \frac{1}{N} \right)^{2h+b-2}\sum_{n=0}^{\infty} c_{(h,b),n} \lambda^{n} \, .
\end{equation}

The fact that only a (small) subclass of Feynman diagrams gives non-negligible contributions for $N \to \infty$ led to early expectations that all of these diagrams could perhaps be summed \emph{exactly}, i.e. that QCD may be solved in the 't~Hooft limit. This expectation, however, turned out to be delusive: although the number of planar diagrams without quark loops grows only exponentially with the power of the coupling they correspond to (to be contrasted with the number of all diagrams at the same order of the coupling, which grows factorially), their resummation cannot be carried out explicitly.

An interesting observation is that a similar type of topological expansion is also found in string theory: the amplitude $\mathcal{A}_s$ associated with a generic string process can be written as:
\begin{equation}
\label{string_amplitude}
\mathcal{A}_s = \sum_{h,b=0}^{\infty} {g_s}^{2h+b-2} k_{h,b} \, ,
\end{equation}
with the string coupling $g_s$ playing a r\^ole analogous to $1/N$ in eq.~(\ref{amplitude}). Although the representation of Feynman diagrams like in fig.~\ref{fig:gluon_propagator} is in internal space (rather than in the physical space), this analogy suggested the idea that string theory could perhaps provide a reformulation of QCD in the 't~Hooft limit, so that the whole class of Feynman diagrams corresponding to the same power in $1/N$ (and for arbitrary powers of the coupling) may be resummed into the world sheet of a propagating string, with the same topology.

This intriguing idea, however, did not provide an obvious clue about what could be the parameter corresponding to the 't~Hooft gauge coupling $\lambda$, in the context of string theory. In fact, it is only with the holographic duality (proposed much later, during the second half of the 1990's) that this question found an answer.

\subsection{The large-$N$ limit and the gauge/gravity correspondence}
\label{subsect:holography}

The holographic duality was first discussed by Maldacena, by Gubser, Klebanov and Polyakov, and by Witten in a series of seminal papers~\cite{Maldacena:1997re, Gubser:1998bc, Witten:1998qj}; since then, it has been discussed in a very large number of works (reviews and introductory lecture notes on this topic include refs.~\cite{Aharony:1999ti, Petersen:1999zh, Klebanov:2000me, D'Hoker:2002aw, Maldacena:2003nj, Erdmenger:2007cm, Mateos:2007ay, Gubser:2009md, CasalderreySolana:2011us}). It is a conjectured correspondence relating gauge and string theories. One intriguing aspect is that, according to this conjecture, the gauge theory and the dual string theory are defined in spacetimes of different dimensions. This somewhat surprising feature of the correspondence implies, in particular, that, if a ``strong'' form of the duality holds (i.e. if a gauge theory and the corresponding string theory are actually \emph{completely equivalent}), then the information encoded in two theories defined in spaces of 
different dimension is the same. This is related to profound aspects of the relation between information and geometry in quantum physics, and it is the reason why the correspondence is called ``holographic''~\cite{'tHooft:1993gx, Stephens:1993an, Susskind:1994vu, Susskind:1998dq, Bousso:2002ju}.

In particular, the string theory is defined in a higher-dimensional spacetime and, as it will be discussed below, the extra-dimensions have a non-trivial counterpart in the gauge theory. We remark that this correspondence, in its full generality, is not rigorously proven yet. However, by now there exists very strong mathematical evidence supporting its validity (and no known counter-examples refuting it) at least in the most studied example, which relates the supersymmetric Yang-Mills (SYM) theory with four spinor supercharges\footnote{Note that, since in four dimensions each spinor has four degrees of freedom, this corresponds to sixteen supersymmetry generators.} and $\U(N)$ gauge group\footnote{In the literature, there has been some discussion whether the gauge group should be taken to be $\U(N)$ or, rather, $\SU(N)$~\cite{Witten:1998qj, Witten:1998zw}.} in four-dimensional Minkowski spacetime to type IIB superstring theory in a curved, ten-dimensional spacetime. A number of other examples are also known. Whether this indicates that a dual, predictive formulation in terms of a string theory necessarily exists for a generic gauge theory, is not known.

As we just mentioned, the most famous example of the gauge-string correspondence associates the $ \mathcal{N}=4 $ $ \U(N) $ Yang-Mills theory in four spacetime dimensions~\cite{Brink:1976bc, Gliozzi:1976qd} to type IIB superstring theory in ten spacetime dimensions~\cite{Green:1981yb, Schwarz:1983qr}. The $ \mathcal{N}=4 $ SYM theory is the supersymmetric non-Abelian gauge theory in four dimensions with the largest amount of supersymmetry. It includes a gauge field $A_\mu$, four Weyl fermions, and six real scalar fields. All of these fields transform under the adjoint representation of the $ \U(N) $ gauge group. The $\mathcal{R}$-symmetry of the theory is a global $\SU(4)$ symmetry: the gauge field is invariant under this symmetry, while the fermions transform according to the fundamental representation, and the scalar fields according to the two-index antisymmetric representation. An important property of this supersymmetric gauge theory is that it is invariant under scale transformations: the classical Lagrangian does not involve any dimensionful parameters, and thus ``looks the same at all energy scales'', and (in contrast to what happens in QCD) this property is not spoiled by quantum fluctuations either. In fact, this theory is not only invariant under scale transformations, but under the conformal group of transformations which includes Lorentz transformations, scale transformations, and special conformal transformations (the latter can be thought of as resulting from the product of an inversion, a translation, and a further inversion). It is possible to prove that the conformal invariance of $ \mathcal{N}=4 $ SYM theory holds perturbatively at all orders~\cite{Mandelstam:1982cb, Howe:1983sr} and that is also preserved at the non-perturbative level~\cite{Seiberg:1988ur}. A particular implication of this exact invariance of the theory is that its coupling does not get renormalized.

Type IIB superstring theory is a chiral supersymmetric string theory. Internal consistency requires it to be defined in ten spacetime dimensions. The theory has 32 supercharges, and admits both open and closed strings. Open strings with ends satisfying Dirichlet boundary conditions along $p$ spatial directions are constrained to start and end on hypersurfaces, which are called D$p$-branes~\cite{Polchinski:1995mt, Johnson:2000ch} (with the `D' standing for `Dirichlet'). In particular, type IIB superstring theory admits D$p$-branes with three spatial dimensions: these D3-branes prove crucial to the interpretation of the gauge/string duality, due to their twofold r\^ole in the theory. On the one hand, as we just said, they serve as loci on which the ends of \emph{open} strings can lie. On the other hand, they also have an interpretation as (heavy) topological solutions of the IIB theory in its supergravity limit, and this reveals their connection to a description in terms of \emph{closed} strings. When a set of $ N $ D3-branes are superimposed, open strings starting and ending on them can be thought of as describing \emph{gauge interactions} in the theory. At the same time, this setup also corresponds to a supergravity solution described by the following metric:
\begin{equation}
\label{brane_metric}
\dd s^2 = \sqrt{ 1 + \frac{R^4}{r^4} } \left( \dd r^2 + r^2 \dd \Omega_5^2 \right) + \frac{1}{\sqrt{ 1 + \frac{R^4}{r^4} } } \left( -\dd t^2 + \dd \mathbf{x}^2 \right),
\end{equation}
in which $r$ is the transverse distance from the branes (and the $r \to 0$ limit corresponds to a horizon), while $\Omega_5$ denotes the set of coordinates of a five-dimensional sphere, whereas $t$ is the time coordinate, and the $\mathbf{x}^i$'s are spatial coordinates on the brane. The parameter $R$ appearing  in eq.~(\ref{brane_metric}) can be interpreted as a ``curvature radius'' of the spacetime, in the presence of the $ N $ D3-branes. $R$ can be expressed in terms of the fundamental string theory parameters---i.e. the string length $l_s$ and the string coupling $g_s$---via
\begin{equation}
\label{R_versus_g_s_l_s}
R = l_s \sqrt[4]{4 \pi g_s N}.
\end{equation}

As eq.~(\ref{brane_metric}) shows, when the transverse distance from the branes is much larger than the curvature radius, the metric reduces to that of Minkowski spacetime with nine spatial dimensions plus time. On the contrary, when $r$ is much smaller than the curvature radius, the right-hand side of eq.~(\ref{brane_metric}) reduces to the metric of a spacetime which is the product of a five-dimensional anti-de~Sitter (AdS) spacetime times a five-dimensional sphere, $AdS_5 \times S^5$:
\begin{equation}
\label{AdS5_S5_metric}
\dd s^2 = \frac{R^2}{z^2} \left( -\dd t^2 + \dd \mathbf{x}^2 + \dd z^2 \right) + R^2 \dd \Omega_5^2, 
\end{equation}
where we introduced the new coordinate $z=R^2/r$. 

It is interesting to consider the global symmetries of the spacetime defined by eq.~(\ref{AdS5_S5_metric}). The anti-de~Sitter spacetime is the maximally symmetric Lorentz manifold characterized by constant negative scalar curvature: it is a (vacuum) solution to the Einstein equations, in the presence of a negative cosmological constant term. The symmetry of this spacetime is described by the $\SO(2,4)$ group. On the other hand, the symmetry of the five-dimensional sphere is described by the $\SO(6)$ group. 

The gauge/string duality can be interpreted as a duality between open and closed strings. On the one hand, the dynamics of the system of $N$ superimposed D3-branes can be described in terms of open strings starting and ending on the D3-branes. Their low-energy effective action takes the form of a Dirac-Born-Infeld action, and when the latter is expanded in derivatives, it reduces to the action of $\mathcal{N}=4$ super-Yang-Mills theory. So, in this case one ends up with an effective, low-energy description which is a supersymmetric gauge theory (with $\U(N)$ gauge group) in $(3+1)$ spacetime dimensions, with 4 supercharges. This effective description in terms of open strings is most convenient in the limit when $g_s N \ll 1$. Since $g_s N$ describes the strength of the coupling of $N$ D3-branes to gravity, this limit corresponds to the case in which the spacetime is almost flat.

On the other hand, the dynamics of the string theory with $N$ coincident D3-branes can also be described in terms of gravitational excitations (which is appropriate when $g_s N$ is large and the spacetime curvature radius is large). Gravitational excitations are associated with closed strings, which propagate in the bulk of the spacetime. In fact, when $g_s N$ is large, the string theory can be approximated by a low-energy effective theory, which is just supergravity in anti-de~Sitter spacetime.

The following aspects, related to the \emph{global} symmetries of the $\mathcal{N}=4$ supersymmetric theory, are particularly important.
\begin{enumerate}
\item Since the $\mathcal{N}=4$ SYM theory is conformal, it enjoys invariance under the conformal group. In $(D+1)$ spacetime dimensions, the conformal group is isomorphic to $\SO(1+1,D+1)$, hence for $D=3$ spatial dimensions one ends up with $\SO(2,4)$, which is the same symmetry group as the one describing the global symmetries of the $AdS_5$ spacetime.
\item The other global symmetry of the theory is the $\mathcal{R}$-symmetry, described by the $\SU(4)$ group. The algebra of the latter is the same as that of the $\SO(6)$ group, which, as we mentioned above, describes the global symmetry of the $S^5$ sphere.
\end{enumerate}
Hence, the global symmetries of the $\mathcal{N}=4$ SYM theory are equivalent to the global symmetries of the $AdS_5 \times S^5$ spacetime.

In particular, the isomorphism between the conformal symmetry group of the gauge theory and the symmetry group of the five-dimensional anti-de~Sitter spacetime is related to the interpretation of the radial coordinate $r$ (or, equivalently, of $z=R^2/r$) in eq.~(\ref{AdS5_S5_metric}), which parameterizes the energy scale of the dual gauge theory~\cite{Balasubramanian:1998sn, Balasubramanian:1998de, Danielsson:1998wt}.

The parameters of the $\mathcal{N}=4$ SYM theory---i.e the number of color charges $N$ and the 't~Hooft coupling $\lambda$---can also be related to the parameters of the string theory, via the relations
\begin{eqnarray}
\frac{\lambda}{N} &=& 4 \pi g_s, \label{lambda_over_N_vs_string} \\
\lambda &=& \frac{R^4}{l_s^4}. \label{lambda_vs_string}
\end{eqnarray}

Note that, according to eq.~(\ref{lambda_over_N_vs_string}), the large-$N$ limit of the gauge theory at fixed 't~Hooft coupling corresponds to the limit in which the string coupling $g_s$ tends to zero. This means that loop effects on the string side of the correspondence become irrelevant, and the string theory reduces to its \emph{classical} limit. 

In addition, eq.~(\ref{lambda_vs_string}) shows that in the limit of large 't~Hooft coupling for the gauge theory, the string length in the dual string theory becomes negligible: when $l_s$ is much smaller than the typical spacetime curvature radius, the stringy nature of gravitational interactions described by closed strings becomes irrelevant, so that the string theory reduces to its \emph{gravity} limit.

Hence, when both $N$ and $\lambda$ are large (i.e. in the 't~Hooft limit at strong coupling), the dual string theory reduces to a classical supergravity limit, which can be studied analytically.

At this point, a brief summary is in order.
\begin{enumerate}
\item The holographic duality is a conjectured correspondence relating gauge theories and string theories.
\item The theories are not defined in spaces of the same dimension, yet (at least to some extent) they encode equivalent physical information.
\item The correspondence is based on an open/closed string duality: the dynamics of the string theory can be described either in terms of open strings (representing gauge interactions) or of closed strings (associated with gravitational excitations.
\item The gauge theory and its dual string theory share the same global symmetries (although these symmetries have a different meaning in the two cases).
\item The parameters of the two theories are related in a non-trivial way.
\item The large-$N$ limit of the gauge theory at fixed 't~Hooft coupling corresponds to the classical limit of the string theory.
\item The strong-coupling limit of the gauge theory corresponds to the supergravity limit of the string theory.
\end{enumerate}

In order to carry out explicit calculations using the holographic correspondence (in the 't~Hooft and strong coupling limits of the gauge theory, so that the dual string theory becomes analytically tractable), one constructs an appropriate \emph{field-operator map}~\cite{Gubser:1998bc, Witten:1998qj}, which associates the generating functional of connected Green's functions in the gauge theory to the minimum of the supergravity action, with appropriate boundary conditions. Adding a source term which couples to a suitable operator $\int \dd^D x \mathcal{O}(x) J(x)$ to the Lagrangian of the gauge theory corresponds to including a bulk field $ \mathcal{J} $ (which reduces to $ J $ on the conformal boundary of the spacetime, up to inessential factors) in the dual string theory. So the mapping can be written as
\begin{equation}
\label{AdS_CFT_mapping}
\left\langle \mathcal{T}~\exp \int \dd ^D x \mathcal{O}(x) J(x) \right\rangle =
\exp \left\{ -\Ssugra \left[\mathcal{J}(x,r) \right] \right\} ,
\end{equation}
where $\Ssugra$ denotes the on-shell supergravity action, in the presence of the bulk field $ \mathcal{J} $. Starting from eq.~(\ref{AdS_CFT_mapping}), correlators of composite operators in the field theory can be computed, by taking appropriate functional derivatives with respect to the source terms, and carrying out the corresponding integrals in AdS space.

Although the gauge/string duality provides a tool to perform analytical computations for the non-perturbative regime of the $\mathcal{N}=4$ SYM theory, it should be noted that results for the latter are not directly relevant for QCD. This is due to the fact that, in vacuum, these two gauge theories have a number of \emph{qualitative} differences. In particular:
\begin{enumerate}
\item $\mathcal{N}=4$ SYM is maximally supersymmetric, QCD is not supersymmetric.
\item The field content is different: $\mathcal{N}=4$ SYM features fermions and scalars in the adjoint representation of the gauge group, while in QCD quarks are in the fundamental representation of the gauge group, and there exist no elementary scalar fields subject to the strong interaction.
\item $\mathcal{N}=4$ SYM is conformally invariant; by contrast, QCD (or even pure Yang-Mills theory) has a discrete spectrum of physical states, with a finite mass gap and is not conformally invariant at the quantum level.
\item In $\mathcal{N}=4$ SYM the bare coupling is a well-defined, physically meaningful parameter of the theory, which does not depend on the momentum scale. On the contrary, in QCD the bare coupling has no physical meaning; a physical, renormalized coupling can be defined, which runs with the momentum scale, and depends on the renormalization scheme.
\end{enumerate}

However, it should be noted that most of these qualitative differences disappear (or are at least mitigated), if one considers both theories at a finite temperature $T$. In particular, a finite temperature breaks explicitly the Lorentz symmetry (in a Euclidean setup, a finite temperature corresponds to a finite extent for the Euclidean time direction) and, as a consequence, also supersymmetry. Thermal boundary conditions along the Euclidean time direction are antiperiodic for fermions, while they are periodic for bosons. This implies that the lowest Matsubara frequency is zero for bosons, but it is of order $T$ for fermions, hence a thermal setup breaks the boson-fermion degeneracy. Thermal fluctuations also lift the scalars, which are not protected by supersymmetry anymore.  

The extension of the gauge/string duality discussed above to a finite-temperature setup was first discussed in ref.~\cite{Witten:1998zw}: it leads to an asymptotic boundary characterized by $S^3 \times S^1$ geometry and to a solution which is an AdS-Schwarzschild black hole with metric
\begin{equation}
\label{AdS_Schwarzschild}
\dd s^2 = \frac{r^2}{R^2} \left[ f(r) \dd \tau^2 + \dd \mathbf{x}^2 \right] + \frac{R^2}{r^2} \left[ \frac{1}{f(r)}\dd r^2 + r^2 \dd \Omega_5^2 \right], \qquad f(r) = 1 - \frac{r_H^4}{r^4},
\end{equation}
where $\tau$ denotes the Euclidean time coordinate. Note that $r=r_H$ corresponds to the black hole horizon: its Hawking temperature is $T= r_H/(\pi R^2)$, which is interpreted as the physical temperature of the dual gauge theory.

Following this construction, a number of results have been derived using the holographic correspondence, for the large-$N$ $\mathcal{N}=4$ super-Yang-Mills theory at finite temperature. In particular, we would like to mention at least two among the most celebrated ones: the computation of the entropy density in the strong-coupling limit~\cite{Gubser:1996de,Gubser:1998nz},
\begin{equation}
\label{s_over_s0}
s = \frac{\pi^2 N^2 T^3}{2} \left[ \frac{3}{4} +\frac{45}{32} \zeta(3) (2 \lambda)^{-3/2} + \dots \right]
\end{equation}
(where the $\pi^2 N^2 T^3/2$ prefactor appearing on the right-hand side is the value of the entropy density for the $\mathcal{N}=4$ plasma in the free limit) and the ratio between the shear viscosity $\eta$ and the entropy density~\cite{Policastro:2001yc}:
\begin{equation}
\label{one_over_fourpi}
\frac{\eta}{s}=\frac{1}{4\pi}.
\end{equation}
In addition, there exist also many applications of the gauge/string duality to finite-temperature phenomena involving real-time dynamics. These include, in particular, a growing sub-field of research combining analytical tools with numerical approaches to general relativity problems in anti-de~Sitter spacetime: for an overview, see ref.~\cite{novelnum12} and the links to the online talks therein.

There are a number of works extending the applicability of holographic techniques to gauge theories that are qualitatively more similar to QCD (see ref.~\cite{Erdmenger:2007cm} and references therein for an extensive discussion). For example, it is possible to modify the setup discussed above, by including a set of $n_f$ D7-branes~\cite{Karch:2000gx, Karch:2002sh} (``flavor branes''), with which one can mimic a set of quark fields (in the fundamental representation of the gauge group) of different flavors. This reduces the amount of supersymmetry of the theory, and enriches the resulting spectrum with new physical states. In particular, open strings stretching between D7- and D3-branes are then interpreted as massive ``quarks'', while open strings starting and ending on D7-branes are interpreted as the ``mesons'' of the theory. (We use the quotation marks to indicate, however, that these states are not really the \emph{actual} physical quarks and mesons, since the resulting theory is still different from QCD.) It is 
also possible to extend the gauge/string correspondence to non-supersymmetric theories, with a linearly rising confining potential~\cite{Polchinski:2000uf, Maldacena:2000yy, Klebanov:2000hb, Aharony:2002up, Kruczenski:2003be, Kruczenski:2003uq, Sakai:2004cn, Sakai:2005yt, Nunez:2010sf}.

An alternative (somewhat less rigorous, but more phenomenology-oriented) approach consists in constructing some \emph{ad~hoc} five-dimensional gravitational model that should reproduce the known features of QCD. A partial list of articles in which this approach has been followed includes refs.~\cite{Polchinski:2001tt, Polchinski:2002jw, Karch:2002xe, Son:2003et, Brodsky:2003px, deTeramond:2005su, DaRold:2005zs, Erlich:2005qh, Hirn:2005nr, Karch:2006pv, Csaki:2006ji, Csaki:2008dt, Gursoy:2010fj}.

A common feature of both the former (``top-down'') and the latter (``bottom-up'') approach---and of virtually all holographic computations---is that they rely on the validity of the large-$N$ limit, i.e. on the assumption that the features of the theory with a finite number of colors are approximated ``sufficiently well'' (up to trivial factors) by those of the large-$N$ theory. This, however, has some shortcomings. In particular, the classical supergravity approximation, which is valid in the large-$N$ and strongly interacting limits of the dual gauge theory, does not provide a completely satisfactory description of asymptotic freedom, and appears to miss certain details about the dynamics of the theory~\cite{Gubser:2009md}. In order to overcome these problems, it would probably be necessary to proceed to the inclusion of finite $\alpha^\prime$ string corrections, to take the finiteness of the gauge coupling into account~\cite{Reece:2011zz}. Related issues, and the connection between asymptotic freedom and the asymptotic behavior of correlators, have also been discussed in detail in ref.~\cite{Bochicchio:2013eda}.

We conclude this brief overview of the r\^ole of the large-$N$ limit in the gauge/string duality by mentioning that it is also important in studies of the integrability of $\mathcal{N}=4$ SYM theory. In this context, ``integrability'' means that, for this theory, in the large-$N$ limit it is possible to derive the scaling dimensions of local operators, as a function of the value of the coupling. This is done by mapping the integral equations obtained with a thermodynamic Bethe \emph{Ansatz} to a set of algebraic equations. Some of the original works discussing this topic include refs.~\cite{Minahan:2002ve, Bena:2003wd, Beisert:2003jj, Beisert:2003tq}, while a more thorough review can be found in ref.~\cite{Beisert:2010jr}.

\section{Phenomenological implications of the 't~Hooft limit of QCD}
\label{sect:phenomenology}

Besides the intriguing analogy with string theory alluded to in sect.~\ref{sect:definitions}, the 't~Hooft limit of QCD has a number of very interesting phenomenological implications, which can be easily obtained using the so-called \emph{large-$N$ counting rules}. As we already mentioned, these rules allow one to identify the dominant Feynman diagrams at fixed 't~Hooft coupling, as those associated with the largest power of $N$. \emph{Under the assumption that the 't~Hooft limit of QCD is a confining theory}, these rules can also be applied to physical hadronic states.

To be more quantitative, it is convenient to rewrite the QCD functional integral as:
\begin{equation}
\label{partition_function}
\mathcal{Z}[J] = \int \mathcal{D} A \mathcal{D} \overline{\psi} \mathcal{D} \psi \exp \left\{ i N \int \dd t~ \dd^3 x~\left[ \sum_{f=1}^{n_f} \overline{\psi}_f \left( i \gamma^\mu D_\mu - m_f \right) \psi_f-\frac{1}{4\lambda} \left( F^a_{\mu\nu} F^{a\,\mu\nu} \right) + J_a \mathcal{O}_a \right] \right\}\, ,
\end{equation}
where the gauge and fermion fields have been suitably rescaled (with respect to their conventional textbook normalization) to single out an overall $N$ factor in the exponent. Then the connected correlators of physical, local or composite operators $\mathcal{O}_a$, involving at most one trace over color indices, can be written as:
\begin{equation}
\label{vacuum_expectation_values}
\langle \mathcal{O}_1(x_1) \dots \mathcal{O}_n (x_n)  \rangle_{\mbox{\tiny{conn}}} = (iN)^{-n} \left\{ \frac{\delta}{\delta J_1(x_1)} \dots \frac{\delta}{\delta J_n(x_n)} \ln \mathcal{Z}[J] \right\}_{J=0}\, .
\end{equation}

As eq.~(\ref{amplitude}) shows, the sum of vacuum graphs in the 't~Hooft limit is $O(N^2)$ (the leading contribution comes from the $h=b=0$ term), while it is $O(N)$ in the presence of fermionic bilinears ($h=0$, $b=1$). Then it follows that the generic $n$-point connected correlator in eq.~(\ref{vacuum_expectation_values}), dominated by diagrams of planar gluon loops, is $O(N^{2-n})$, in the case of purely gluonic operators---or $O(N^{1-n})$, if quark bilinears are involved.

These laws imply that, when $\mathcal{O}_i$ is a Hermitian operator describing glueball states, the connected two-point correlation function of $\langle \mathcal{O}_i \mathcal{O}_i \rangle$ is $O(N^0)$ in the large-$N$ limit (so that, with these normalizations, $\mathcal{O}_i$ creates a glueball state with amplitude $O(N^0)$, when acting on the vacuum of the theory). By contrast, three-, four-, and higher-order $n$-point connected correlation functions, which can be associated with the decay of a glueball into two, three, or more glueballs (or with processes related to these by crossing symmetry) are suppressed in the 't~Hooft limit, as they scale at most like $O(N^{2-n})$ and hence tend to zero for $n \ge 3$.

The case of operators involving fermion bilinears (as appropriate for mesons) is analogous, but in this case $\langle \mathcal{O}_i \mathcal{O}_i \rangle$ is $O(1/N)$, thus the operator creating a meson state with amplitude of order $O(N^0)$ is $\sqrt{N}\mathcal{O}_i$ and connected correlators of three (or more) meson states are suppressed by one (or more) power(s) of $1/\sqrt{N}$ in the large-$N$ limit. Also suppressed are glueball-meson interactions, as well as more exotic objects, like molecules or tetraquarks (although the analysis of the latter involves some subtleties~\cite{Weinberg:2013cfa, Knecht:2013yqa, Lebed:2013aka}).

From these observations, it follows that, if QCD is confining in the 't~Hooft limit, then the lightest physical states in the spectrum are glueballs and mesons with masses $O(N^0)$. Their interactions are suppressed at least as $1/\sqrt{N}$ for $N \to \infty$, so that in the 't~Hooft limit QCD turns into \emph{a theory of stable, weakly interacting hadrons}. This drastic simplification of the theory implies, in particular, that quantities which, according to perturbation theory, are characterized by a \emph{logarithmic} dependence on the momentum scale involved, can be expressed in terms of sums of propagators of non-interacting hadrons. Since any finite sum of rational functions is a rational function, it follows that the number of light glueballs and mesons in large-$N$ QCD must be \emph{infinite}.

Another interesting phenomenological implication of the 't~Hooft limit is that it provides an intuitive explanation of the empirical observation, due to Okubo, Zweig and Iizuka~\cite{Okubo:1963fa, Zweig:1981pd, Iizuka:1966fk} and known as ``OZI rule'', that certain decays of mesons occur less frequently than others: experimental results indicate that QCD processes corresponding to Feynman diagrams, which can be made disconnected (``split in two'') by removing only internal gluon lines, are disfavored. One example can be found in the decay of the electrically neutral $\varphi$ meson to three pions, which---despite a much more limited phase space---turns out to be less frequent than the decay to a pair of charged kaons.

\begin{figure}[tb!]
\begin{center}
\begin{minipage}[t]{8 cm}
\epsfig{file=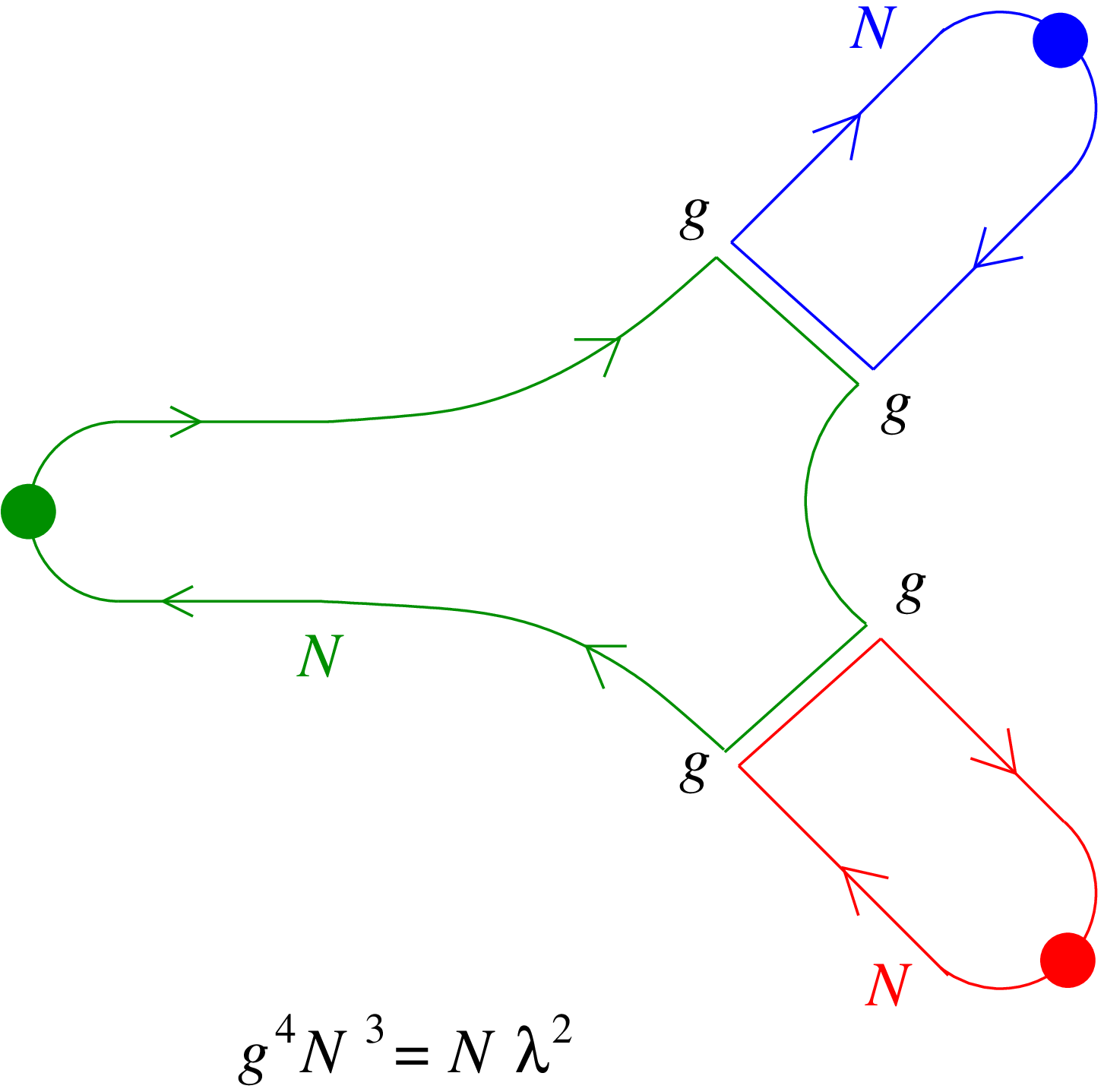,scale=0.5}
\vspace{1cm}\\
\epsfig{file=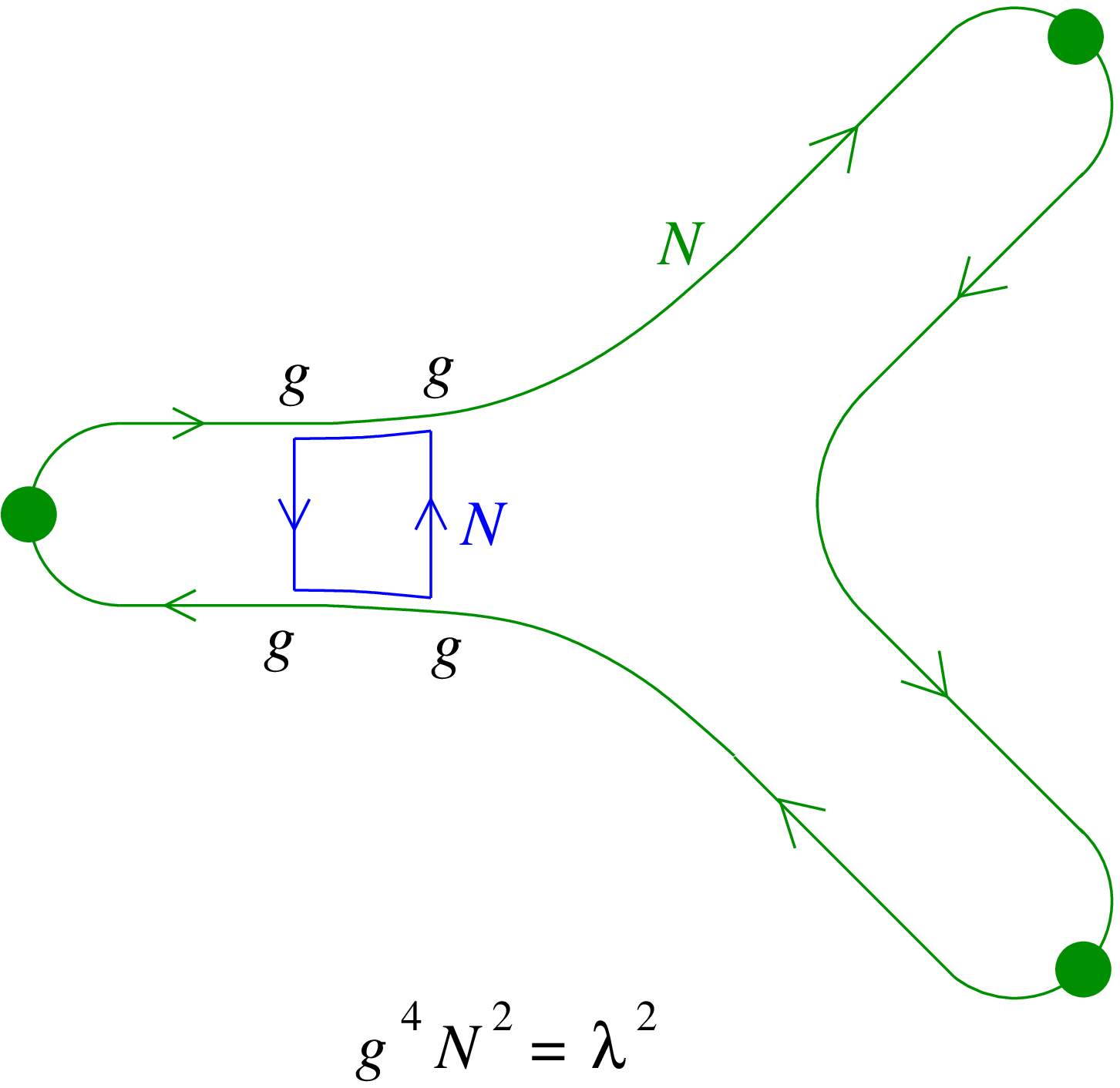,scale=0.5}
\vspace{5mm}\\
\end{minipage}
\begin{minipage}[t]{16.5 cm}
\caption{The 't~Hooft limit of QCD gives an intuitive explanation for the empirical rule first observed by Okubo, Zweig and Iizuka~\cite{Okubo:1963fa, Zweig:1981pd, Iizuka:1966fk}, stating that QCD processes described by Feynman diagrams involving an intermediate stage, which includes virtual gluons only, are suppressed. In the 't~Hooft limit, this rule can be interpreted as a suppression by a $1/N$ factor in these different types of diagrams. The figure shows two different $O(\alpha^2_{\mbox{\tiny{strong}}})$ Feynman diagrams relevant for the decay of an isospin-singlet meson (the blob on the left). The diagram in the bottom panel, in which the process goes through the complete annihilation of the valence quark/antiquark of the initial meson and the emission of a pair of gluons, is suppressed by one power of $1/N$ with respect to the diagram in the top panel, in which the ``valence'' quark and antiquark of the initial meson are still present in the final state (and, hence, also at all intermediate stages of the 
process).~\label{fig:ozi_rule}}
\end{minipage}
\end{center}
\end{figure}

In the 't~Hooft limit, the OZI rule can be interpreted as a suppression (by at least one power of $1/N$) of diagrams involving an intermediate state which contains only virtual gluons. This is best clarified by an explicit example. In fig.~\ref{fig:ozi_rule}, we show two different Feynman diagrams corresponding to the decay of a meson (the blob on the left-hand side) to two lighter mesons (on the right-hand side). Both diagrams represent processes involving the exchange of two gluons, i.e. proportional to $\alpha_{\mbox{\tiny{strong}}}^2$. However, in the diagram at the top of the figure the line corresponding to the propagation of the ``valence'' quark (and antiquark) of the initial meson survives throughout the whole process, and it still appears in the final states. This implies that, for the diagram at the top of the figure, there exists no intermediate stage of the decay including virtual gluons only. A different way in which the decay process can occur (at least for an isospin-singlet initial meson) is depicted in the bottom panel of the figure: it involves the annihilation of the valence quark and antiquark of the initial state, with the emission of two virtual gluons (which are then absorbed on the fermion line of the valence quarks of the final states). Counting the number of independent fundamental color indices running through the two different diagrams, one sees that the one at the bottom of the figure is suppressed by one power of $1/N$ with respect to the one at the top, in agreement with the OZI rule (at least qualitatively\footnote{For a more quantitative analysis, non-trivial dynamics must be taken into account~\cite{Geiger:1996re}.}). Note that the diagram in the top panel of fig.~\ref{fig:ozi_rule} can be drawn on a plane with one ``hole'', corresponding to $h=0$, $b=1$ in eq.~(\protect\ref{amplitude}), while the one in the bottom panel of the figure can only be drawn on a plane with at least two holes ($h=0$, $b=2$).

Other interesting phenomenological implications of the large-$N$ limit for mesons can be derived from the analysis of low-energy models described in terms of effective chiral Lagrangians. In particular, by writing the partition function of these models in a form in which the number of colors is explicitly factored out in the expression of the Lagrangian---like in eq.~(\ref{partition_function})---, it is easy to see that these models become exact at tree level in the large-$N$ limit. In other words, for these low-energy effective theories, the 't~Hooft limit is equivalent to the \emph{classical} limit, and a number of interesting phenomenological implications can be derived: for a detailed discussion, see refs.~\cite{Lebed:1998st, Pich:2002xy} and references therein.

In view of the fact that an overall $N$ factor also appears in the exponent in eq.~(\ref{partition_function}), one may wonder if the large-$N$ limit is equivalent to the classical limit \emph{also} for full QCD---as for its low-energy description in terms of an effective model for mesons. The answer is no: in the QCD partition function, the dependence on $N$ is not only in the factor appearing in the exponent, but also in the functional \emph{measure}, since the number of gluon degrees of freedom is $O(N^2)$---and the number of fermion degrees of freedom is $O(N)$---in the large-$N$ limit. By contrast, the degrees of freedom of the effective meson Lagrangian are color-singlet, \emph{hadronic} states, whose number is $O(N^0)$ in the large-$N$ limit.\footnote{Nevertheless, as it will be shown later, it is still possible to give an interpretation of the large-$N$ limit for the full theory, in terms of an analogy with a sort of classical limit (as long as one introduces appropriate definitions for a ``classical'' Hamiltonian and a ``classical'' configuration space).}

Finally, one important phenomenological implication for the meson sector in the 't~Hooft limit of QCD is related to the $\eta^\prime$ meson, which is the lightest isoscalar, pseudoscalar, electrically neutral meson. The fact that this particle has a mass of $957.78(6)$~MeV~\cite{Nakamura:2010zzi}, much heavier than the other pseudoscalar mesons not involving heavy valence quarks (i.e. pions, kaons and the $\eta$) posed a long-standing puzzle (the so-called ``$\U(1)$ problem''), related to the interpretation of the light pseudoscalar mesons as the (pseudo-)Nambu-Goldstone bosons associated with the spontaneous breakdown of part of the global chiral symmetry characterizing the classical QCD Lagrangian with $n_l$ (nearly) massless quark flavors. The issue can be briefly summarized as follows: QCD (in contrast to the electro-weak theory) is a \emph{vector} theory, i.e. strong nuclear interactions act in the same way on the left- and right-handed components of the quark fields. In addition, QCD is also ``blind'' 
to quark flavor: the only explicit flavor dependence in the QCD Lagrangian in eq.~(\ref{partition_function}) is in the different quark masses in the Dirac operator. As a consequence, if the theory features $n_l$ exactly massless quark flavors, then classically there exists a global $\U(n_l)_{\mbox{\tiny{L}}} \times \U(n_l)_{\mbox{\tiny{R}}}$ symmetry: in the absence of mass terms, the left- and right-handed complex components of the quark fields are independent of each other, and different flavors can be arbitrarily rotated into each other. This classical symmetry can be rewritten in an equivalent way, by considering ``vector'' (and ``axial'') transformations, which---roughly speaking---act on both the left- and right-handed quark field components, by rotating them in the same (respectively: in the opposite) way in flavor space. In addition, it is convenient to factor out a $\U(1)$ subgroup from each $\U(n_l)$ group, so that the classical chiral symmetry of the QCD Lagrangian with $n_l$ massless quarks can be 
written as:
\begin{equation}
\label{classical_chiral_symmetry}
\SU(n_l)_{\mbox{\tiny{V}}} \times \U(1)_{\mbox{\tiny{V}}} \times \SU(n_l)_{\mbox{\tiny{A}}} \times \U(1)_{\mbox{\tiny{A}}} \,.
\end{equation}

Upon quantization (and continuing to neglect effects due to the finiteness of the light quark masses, as well as effects due to electroweak interactions), the different factors appearing in eq.~(\ref{classical_chiral_symmetry}) have different fates. The $\SU(n_l)_{\mbox{\tiny{V}}}$ symmetry remains exact: in real-world QCD, it manifests itself (for example) in the approximate degeneracy of the proton and neutron masses. Also the $\U(1)_{\mbox{\tiny{V}}}$ symmetry is preserved at the quantum level, and corresponds to baryon number conservation in QCD. On the contrary, the $\SU(n_l)_{\mbox{\tiny{A}}}$ is spontaneously broken in the QCD vacuum: the existence of a non-vanishing chiral condensate $\langle \bar{\psi} \psi \rangle$ implies that the ground state of the quantum theory is not invariant under $\SU(n_l)_{\mbox{\tiny{A}}}$ transformations, and that the spectrum includes a multiplet of $(n_l^2-1)$ massless Nambu-Goldstone bosons: the pions (and the kaons and the $\eta$, if also the \emph{strange} quark is 
considered as approximately massless). Finally, the puzzle of the axial $\U(1)$ problem is the following: were this symmetry preserved at the quantum level, the spectrum of physical states would include mass-degenerate particles of opposite parity---but this is not seen experimentally. On the other hand, if the $\U(1)_{\mbox{\tiny{A}}}$ were spontaneously broken, then there would exist an associated massless Nambu-Goldstone boson---a pseudoscalar state in the iso-scalar sector: the $\eta^\prime$ meson. However, the experimental evidence shows that the $\eta^\prime$ meson is much heavier than the other light pseudoscalar mesons, hence it cannot be interpreted as the Nambu-Goldstone boson associated with a spontaneously broken $\U(1)_{\mbox{\tiny{A}}}$ symmetry. The resolution of the puzzle is that, at the quantum level, the $\U(1)_{\mbox{\tiny{A}}}$ symmetry is neither preserved, nor spontaneously broken: it is \emph{explicitly} broken by the quantum measure---the $\mathcal{D} \overline{\psi} \mathcal{D} \psi$ term in eq.~(\ref{partition_function}) is not invariant under $\U(1)_{\mbox{\tiny{A}}}$ transformations. An explicit calculation shows that the corresponding anomaly is related to the \emph{topological charge} of the QCD vacuum, and proportional to $g^2$. (In an ideal QCD world with exactly massless quarks,) it is this anomaly that it is responsible for the non-vanishing mass of the $\eta^\prime$ meson. Since the anomaly is proportional to $g^2$, however, it is vanishing in the 't~Hooft limit ($g^2=\lambda/N$, and the 't~Hooft limit is the $N \to \infty$ limit at fixed $\lambda$). As a consequence, in the 't~Hooft limit of QCD the $\U(1)_{\mbox{\tiny{A}}}$ symmetry is not anomalous---rather, it gets spontaneously broken by the QCD vacuum, so that in the spectrum there are $n_l^2$ pseudoscalar Nambu-Goldstone bosons, including the $\eta^\prime$ meson. In fact, the large-$N$ limit entails quite a large number of phenomenological implications for mesons, particularly from the analysis of effective Lagrangians. As an example, following ref.~\cite{Kaiser:2005eu}, consider the low-energy description of the lightest mesons (with \emph{up}, \emph{down} and \emph{strange} valence quarks): packaging the physical degrees of freedom in a matrix-valued field $\mathcal{U}$, the low-energy dynamics can be described in terms of an effective Lagrangian, whose terms can be classified in terms of powers of covariant derivatives and masses. Their coefficients can be interpreted as low-energy constants, whose size can be estimated on the basis of large-$N$ counting rules, and accounting for the effect of heavier resonance states. The results turn out to be in remarkably good agreement with estimates from phenomenological models based on experimental input, and with lattice QCD computations: see table 1 in ref.~\cite{Pich:2002xy} or table 2 in ref.~\cite{Kaiser:2005eu}. Another example of quantitative analysis of large-$N$ phenomenology for mesons can be found in ref.~\cite{Uehara:2003ax}. For a more detailed discussion, see the review~\cite{Pich:2002xy} and references therein, as well as refs.~\cite{Pelaez:2003dy, Pelaez:2004xp, Pelaez:2006nj, Geng:2008ag}.

Thus far, we have only discussed two types of hadrons: glueballs and mesons. Of course, the theory also allows the construction of baryons, i.e. color-singlet hadronic states built from $N$ valence quarks. Their very definition implies that, in contrast to mesons or glueballs, even the operator structure for baryons depends explicitly on $N$. Perturbatively, it is easy to show that all the leading contributions to baryon masses in the 't~Hooft limit are $O(N)$, using elementary combinatorics arguments: for example, (besides the contribution from possibly non-vanishing valence quark masses,) contributions from diagrams involving a one-gluon exchange between valence quarks involve two powers of the coupling $g$, and can occur in $O(N^2)$ different ways (the number of independent pairs of valence quarks among which a gluon can be exchanged is $(N^2-N)/2$), resulting, again, in an $O(N)$ contribution at fixed $\lambda$. Including a one-gluon-loop correction for the propagator of the exchanged gluon, the new diagram involves four powers of $g$ and, in addition to the $O(N^2)$ multiplicity related to the choice of the pair of valence quarks, one further, independent fundamental color index runs in the interior of the virtual gluon loop, so that the multiplicity of the diagram at fixed $\lambda$ is, again $O(N)$. If two gluons are subsequently exchanged between the same two valence quarks, the diagram involves four factors of $g$ and an internal color loop, so that the corresponding contribution scales like $N^2 \times N \times g^4 = N \lambda^2$, i.e. is $O(N)$ in the 't~Hooft limit. If the two exchanged gluons interact at a four-gluon vertex (proportional to $g^2$), then the corresponding contribution is proportional to a factor $O(N^2)$ from the quark pair choice, times $g^6$, and it involves two internal fundamental indices running in the two loops, so that the resulting contribution is proportional to $N^4 \times g^6 = N \lambda^3$. If the two gluons are exchanged between three different valence quarks, the combinatorial factor associated with the number of possibilities to choose the valence quarks is $O(N^3)$, and four powers of $g$ are involved: once again, the net result scales as $O(N)$ in the 't~Hooft limit. 

Note, however, that the case in which two gluons are exchanged between four different valence quarks---or generalizations thereof---seems to violate this scaling law, being $O(N^2)$ (four powers of the coupling and a combinatorial factor $O(N^4)$ from the choice of the two pairs of valence quarks). In fact, this apparent breakdown of the scaling with $N$ of the different contributions to the baryon mass (which might make the large-$N$ limit meaningless for baryons) is misleading: such terms appear only because the propagation of a baryon is described by the \emph{exponential} of its energy, and arise once the exponential is expanded in a Taylor series.

From the point of view of the interpretation of large-$N$ QCD as a theory of almost free glueballs and mesons (with interactions characterized by a coupling suppressed like some power of $1/\sqrt{N}$), baryons can be interpreted as the \emph{solitons} of the theory: in the limit when the coupling becomes perturbative, they become arbitrarily heavy---and the $N$-dependence of their masses can never be captured at any order in a $1/N$-expansion.

The fact that in the large-$N$ limit baryons become arbitrarily heavy objects also means that they become \emph{non-relativistic}. Indeed, in the 't~Hooft limit it is possible to see how the connection between QCD and certain non-relativistic and Skyrme models for strong interactions arises.

An even more interesting class of \emph{quantitative} implications for the baryon sector in large-$N$ QCD arises when one combines the expectations from large-$N$ counting rules with the requirement that the theory be \emph{unitary} (the latter condition is necessary to ensure that the evolution of all observable physical states is such, that the sum of probabilities of the possible different event outcomes is equal to unity)~\cite{Dashen:1993jt, Jenkins:1993zu, Gervais:1983wq, Carone:1993dz, Luty:1993fu, Jenkins:1996de, Manohar:1998xv, Jenkins:1998wy, Lutz:2001yb}. In particular, following this approach it is possible to show that baryon states can be described in terms of a contracted spin-flavor algebra, and a \emph{systematic} $1/N$ expansion can be derived, for various quantities, including axial couplings, form factors, masses and magnetic moments of different states, the nucleon-nucleon potential and scattering, and various quantities related to the baryon structure. The accuracy of these results for real-world QCD with $N=3$ colors appears to be good: for example, as discussed in ref.~\cite{Lebed:1998st}, the large-$N$ prediction for the relative mass difference between the nucleon and the $\Delta$ baryon is $1/3$, while the experimental value is about $0.27$---and even better agreement can be obtained for certain mass combinations. However, there exist also cases in which the experimental results tend to deviate from the corresponding large-$N$ predictions: this is probably due to non-trivial dynamics in $N=3$ QCD, which is missed by large-$N$ computations.

\section{From factorization to orbifold dualities}
\label{sect:factorization}

As we discussed in sect.~\ref{sect:phenomenology}, the large-$N$ counting rules associated with the 't~Hooft limit of QCD entail many phenomenological implications, and often provide intuitive (qualitative or quantitative) explanations for poorly understood features of the real-world theory with $N=3$ color charges.

In addition to these phenomenological aspects, the large-$N$ counting rules also imply a number of consequences at a more ``formal'' or ``fundamental'' level, and reveal surprising properties. As we will discuss in this section, these properties include, in particular, correspondences between theories defined in spaces of different volume, or with different field content---including correspondences between supersymmetric and non-supersymmetric theories!

To expose how the large-$N$ counting rules lead to these equivalences, consider the expectation values or correlation functions of gauge-invariant physical operators $\mathcal{O}_a$, such as:
\begin{itemize}
\item local, gauge-invariant operators constructed from purely gluonic fields,
\item closed loop operators (e.g. Wilson loops), or
\item color-singlet operators involving fermionic bilinears.
\end{itemize}
In the $N \to \infty$ limit at fixed 't~Hooft coupling, the large-$N$ counting rules immediately imply that the leading contributions to correlation functions of such operators are associated with disconnected diagrams, as they feature the maximum number of color traces, and, hence, the largest number of independent color indices. More precisely:
\begin{equation}
\label{factorization}
\langle \mathcal{O}_1 \mathcal{O}_2 \rangle = \langle \mathcal{O}_1 \rangle \langle \mathcal{O}_2 \rangle + O(1/N).
\end{equation}

Eq.~(\ref{factorization}) offers two different, interesting interpretations of the large-$N$ limit.

As first pointed out in ref.~\cite{Haan:1981ks}, upon interpreting $N$ as the physical volume $V$ of a system (note that both $N$ and $V$ are related to the ``number'' of degrees of freedom---a quantity, on which both the functional measure in the partition function of a large-$N$ field theory and the statistical measure in the partition function of a statistical system do depend) eq.~(\ref{factorization}) can be interpreted in analogy with the cluster decomposition in statistical mechanics or statistical field theory. In the presence of a finite correlation length, averages of products of physical operators over a sufficiently large physical volume factorize into products of the averages of each operator---up to corrections suppressed as a function of $1/V$.

A different interpretation of eq.~(\ref{factorization}) is based on its analogy with the equation describing the suppression of quantum fluctuations in the classical limit of a quantum system~\cite{Yaffe:1981vf}. A simple way to show how a quantum theory reduces to its classical limit when $\hbar \to 0$ is by considering a coherent state basis. Coherent states have the properties that
\begin{itemize}
\item they form an overcomplete basis, and encode the full information of the system operators in their \emph{diagonal elements}, and
\item they have vanishing overlaps in the $\hbar \to 0$ limit.
\end{itemize}
The combination of these features implies that in the classical limit expectation values of operator products factorize, and the quantum uncertainties associated with conjugate variables tend to zero. In particular, it is possible to define the classical phase space as the manifold of coordinates that  label coherent states.

As discussed in ref.~\cite{Yaffe:1981vf} (see also the references therein for further details), a suitable basis of coherent states can also be constructed (at least formally) for the large-$N$ limit of a quantum field theory, or a statistical system (e.g. an $N$-component spin model). An oversimplistic, but intuitive, argument suggesting that for $N \to \infty$ the theory reduces to a sort of classical limit simply comes from the observation that an overall $N$ can be factored out of the Lagrangian of the theory: see eq.~(\ref{partition_function}). This implies that the functional integral is dominated by a set of ``classical'' field configurations, where $N$ plays a r\^ole analogous to $1/\hbar$. Does this mean that the large-$N$ limit is fully equivalent to the classical limit? Actually, no: as we have already mentioned, the analogy between the $1/N \to 0$ and the $\hbar \to 0$ limits is not complete, in the sense that, for the large-$N$ theory, also the \emph{measure} in the functional integral depends on $N$. Nevertheless, the two limits share many interesting features. 

At a more formal level, the construction of a ``classical'' analogue of a large-$N$ theory goes as follows. Given a family of theories, labelled by the parameter $N$, ``large-$N$ coherent states'' can be defined, by introducing a suitable coherence group~\cite{Klauder_Skagerstam}: this group is defined in terms of appropriate ``coordinates'' and ``momenta'', and generalizes the Heisenberg group of quantum mechanics. Upon acting on the ``vacuum'' state of each theory, this coherence group generates coherent states. Given the set $\mathcal{S}$ of operators which have well-defined (properly normalized) matrix elements in the basis of coherent states for $N \to \infty$, it is then possible to introduce ``classically equivalent'' coherent states, defined as those for which the matrix elements of operators in $\mathcal{S}$ become equal in the large-$N$ limit. The equivalence classes introduced by this relation can then be identified with the classical states: for $N \to \infty$, the theory reduces to a classical theory defined on the coadjoint orbit of the coherence group, and the (properly normalized) Hamiltonian of the theory tends to a classical Hamiltonian in the corresponding phase space.

Although this construction defines unambiguously an algorithm to derive the solution of a theory in the large-$N$ limit (which can be obtained by solving the corresponding classical Hamiltonian problem), unfortunately it leads to explicit solutions only for certain types of models, including vector~\cite{Coleman:1974jh}, single-matrix~\cite{Brezin:1977sv, Marchesini:1979yq} and one-plaquette lattice models~\cite{Jevicki:1980zq, Wadia:1980cp}. On the contrary, it does not yield direct solutions for the case of gauge theories. 

An interesting implication of the factorization of expectation values of operator products given by eq.~(\ref{factorization}) is that in the large-$N$ limit the theory tends to become spacetime independent. As first observed in ref.~\cite{Witten:1979pi}, for $\mathcal{O}_1=\mathcal{O}_2$, eq.~(\ref{factorization}) implies the suppression of quantum fluctuations in the 't~Hooft limit. This suggests that the functional path integral should receive contributions essentially from just one configuration (up to gauge transformations). This configuration is usually called the ``master field''~\cite{Coleman:1980nk}, and it is expected to satisfy a quenched Langevin equation~\cite{Greensite:1982mf}. Since vacuum expectation values are Poincar\'e invariant, such should be the master field, too (at least in one gauge): the theory, then, would be completely spacetime independent!

It was later discovered that actually this intuitive picture is not correct, and the master field cannot be interpreted as a classical field~\cite{Haan:1981ks}. However, as shown in refs.~\cite{Douglas:1994kw, Douglas:1994zu, Gopakumar:1994iq}, the idea can be formulated more rigorously in the context of non-commutative probability theory~\cite{Singer:1994zz, Accardi:1994gd, www.uni-math.gwdg.de/mitch/free.pdf}. 

Building on ideas related to large-$N$ factorization and spacetime independence, it was also found that in the 't~Hooft limit one can write a closed set of Schwinger-Dyson equations, that have to be satisfied by the expectation values of physical operators~\cite{Makeenko:1979pb}. A complete solution for these ``loop equations'', however, has not been found.

An interesting consequence of factorization is that, when examining the Schwinger-Dyson equations satisfied by Wilson loops in the large-$N$ theory on the lattice, it turns out that they are independent of the physical hypervolume of the system \emph{provided center symmetry is unbroken}: this is the so-called Eguchi-Kawai (EK) volume reduction~\cite{Eguchi:1982nm}. In principle, this property would allow one to study the large-$N$ theory in arbitrarily small volumes, either by analytical techniques (reducing the original theory to a matrix model), or by numerical simulations on a single-site lattice. However, it is well-known that center symmetry \emph{does} get broken in a small volume in the continuum limit: this can already be seen at the perturbative level, for all $D>2$. In order to preserve center symmetry, various fixes have been proposed, since the 1980's: for example, in the quenched EK model~\cite{Bhanot:1982sh}, one studies the dynamics of the single-site model for a fixed set of eigenvalues of the link variables along the various directions, and then averages over a center-symmetric distribution for the eigenvalues. However, this method has recently been shown to fail~\cite{Bringoltz:2008av}, due to the fact that the quenching prescription fixes the eigenvalues of the link matrices in the four directions only up to cyclic permutations, and dynamical fluctuations lead to non-trivial correlations among the eigenvalues along different directions. Another approach to preserve center symmetry in the reduced EK model is based on imposing twisted boundary conditions~\cite{GonzalezArroyo:1982hz, GonzalezArroyo:1982ub}: interestingly, this approach can also be used for a non-perturbative definition of field theories defined on a non-commutative spacetime~\cite{GonzalezArroyo:1983ac, Aoki:1999vr, Ambjorn:1999ts, Ambjorn:2000nb, Ambjorn:2000cs} (for an alternative regularization of such theories, see, e.g., refs.~\cite{Panero:2006bx, Panero:2006cs} and references therein). Volume independence in the twisted EK model holds both at strong coupling and in the perturbative regime, but, at least for the simplest definition of the twist, it has been found to fail at intermediate couplings, in a range which appears to increase when $N$ grows~\cite{Teper:2006sp, Azeyanagi:2007su}. However, a couple of years ago, the authors who originally suggested the twisted EK model came up with a new formulation of the twist~\cite{GonzalezArroyo:2010ss}, which they are currently studying numerically~\cite{GonzalezArroyo:2012fx, Perez:2013dra}. The extrapolation of results for the string tension obtained from large volume simulations (at moderate values of $N$) compares well with the result from a single-site simulation in the new version of the twisted model, at a much larger value of $N$.

\begin{figure}[tb!]
\begin{center}
\begin{minipage}[t]{16.5 cm}
\begin{centering}
\epsfig{file=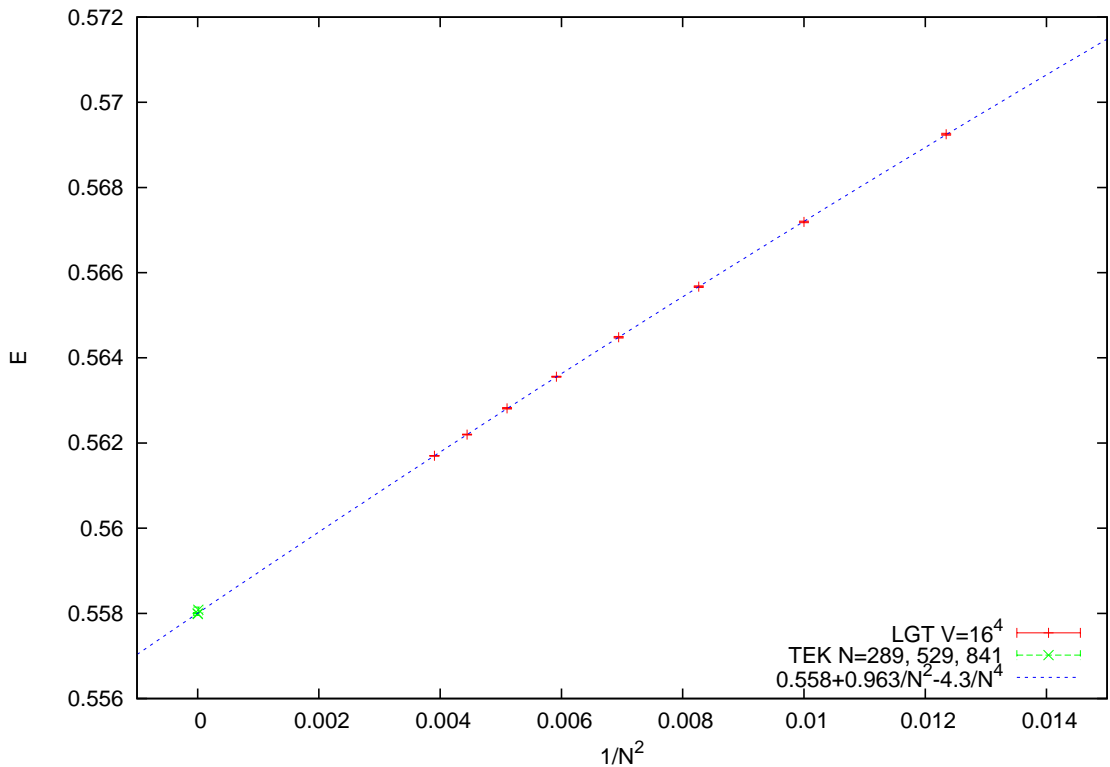,scale=1.2}
\end{centering}
\vspace{5mm}\\
\end{minipage}
\begin{minipage}[t]{16.5 cm}
\caption{The results for the average lattice plaquette (see sec.~\protect\ref{sect:lattice} for a precise definition), which can be interpreted as a lattice counterpart of the Euclidean action density, obtained in ref.~\cite{Gonzalez-Arroyo:2013bta} from simulations of the EK model with twisted boundary conditions and adjoint Dirac fermions, show evidence for the validity of volume reduction at large $N$. The extrapolation of results obtained from simulations at smaller values of $N$ on larger volumes (red symbols) is consistent with the data obtained at large $N$ in a small volume (green symbols).\label{fig:plaquette_13056253}}
\end{minipage}
\end{center}
\end{figure}

Another possibility to preserve center symmetry in EK models is based on the inclusion of dynamical adjoint fermions (obeying periodic boundary conditions in all directions): this idea was initially proposed in ref.~\cite{Kovtun:2007py}, and has since been studied both analytically and numerically by a number of authors~\cite{Hollowood:2009sy, Azeyanagi:2010ne, Catterall:2010gx, Bringoltz:2009kb, Bringoltz:2011by, Cossu:2009sq, Hietanen:2009ex, Hietanen:2010fx, Hanada:2013ota, Lee:2013hk, Okawa_parallel, Gonzalez-Arroyo:2013bta}. In particular, recent results seem to indicate that, indeed, EK volume reduction with adjoint Dirac fermions works as expected, both with $n_f=1$ and $n_f=2$ flavors~\cite{Bringoltz:2011by}. However, numerical investigations of this model are still in progress. As an example of recent numerical studies, fig.~\ref{fig:plaquette_13056253} shows the results for the lattice Euclidean action density obtained in ref.~\cite{Gonzalez-Arroyo:2013bta}, combining adjoint Wilson fermions and symmetric twisted boundary conditions with non-vanishing flux: the results from small-volume simulations at very large values of $N$ (green symbols) are in perfect agreement with the extrapolation of those from large lattices at smaller $N$ (red symbols).

\begin{figure}[tb!]
\begin{center}
\begin{minipage}[t]{16.5 cm}
\begin{centering}
\epsfig{file=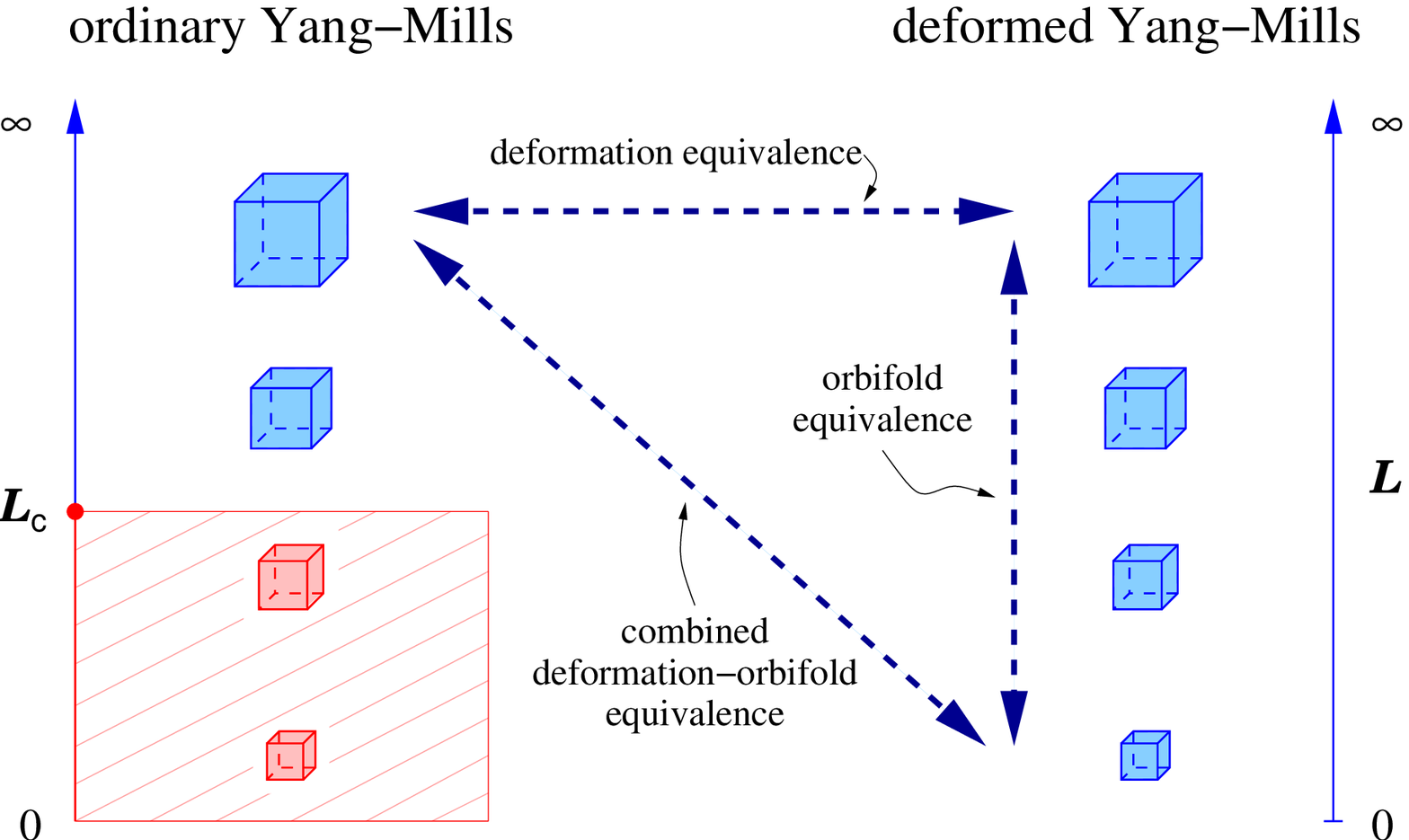, scale=0.8}
\end{centering}
\vspace{5mm}\\
\end{minipage}
\begin{minipage}[t]{16.5 cm}
\caption{As discussed in ref.~\cite{Unsal:2008ch} (from which this figure is taken), by combining the equivalence of undeformed and deformed large-$N$ Yang-Mills theories in a large volume and the volume independence of the deformed theory, it is possible to derive non-perturbative information on large-$N$ Yang-Mills theory in a large volume via studies of a model defined in an arbitrarily small spacetime. The spontaneous breakdown of center symmetry at a critical system size $L_c$ prevents a direct approach in the undeformed theory, for which complete volume reduction does not hold.\label{fig:mappings_08030344}}
\end{minipage}
\end{center}
\end{figure}

Another possible way to enforce center symmetry in EK models is based on double-trace deformations~\cite{Unsal:2008ch}: essentially, one modifies the usual Yang-Mills (YM) action, with the addition of (products of) traces of Polyakov loops, with positive coefficients, which explicitly suppress center-symmetry breaking configurations, at the cost of corrections to observables, that are suppressed in the large-$N$ limit:
\begin{equation}
S_{\mbox{\tiny{YM}}} \longrightarrow S_{\mbox{\tiny{YM}}} + \frac{1}{N_t^3} \sum_{\vec{x}} \sum_{n=1}^{\lfloor N/2 \rfloor} a_n |\mbox{tr}(L^n(\vec{x}))|^2.
\end{equation}
The strategy, then (as nicely summarized in fig.~\ref{fig:mappings_08030344}, taken from ref.~\cite{Unsal:2008ch}) consists in exploiting first the equivalence of ordinary YM theory with its deformed counterpart in large volume, and then the equivalence of the latter with deformed YM in an arbitrarily small volume. This allows one to extract non-perturbative information on large-$N$ YM in a large volume, from the study of the volume-reduced deformed model.

Related ideas have also been discussed in the context of $\SU(N)$ YM theory at finite temperature~\cite{Myers:2007vc}. Dedicated numerical algorithms to study the EK model with double-trace deformation have been devised~\cite{Vairinhos:2010ha}. A nice feature of this approach is that one can reduce only one (or a few) direction(s), while keeping the others large.

Neuberger and collaborators proposed the partial reduction approach to EK~\cite{Narayanan:2003fc, Kiskis:2003rd}: the idea is to simulate the large-$N$ theory in lattices which are small, but still larger than the critical size at which center symmetry gets spontaneously broken, corresponding to the inverse of the deconfinement temperature $T_c$. As an example of results obtained in this approach, in ref.~\cite{Kiskis:2009rf} the confining potential was computed up to distances equal to $9$ lattice spacings from simulations on a lattice of linear size $L=6a$.

Volume reduction and volume independence in large-$N$ gauge theories can be interpreted as a form of ``orbifold'' equivalence~\cite{Bershadsky:1998cb, Strassler:2001fs}, namely as a correspondence based on projections under some discrete subgroup of the global symmetries of two different theories~\cite{Kovtun:2007py, Kovtun:2003hr, Kovtun:2004bz, Kovtun:2005kh, Unsal:2006pj}: under the assumption that the discrete symmetry used in this projection is not spontaneously broken, the vev's and correlation functions of invariant (or ``neutral'') sectors of observables of the original (``parent'') and projected (``daughter'') theories are equal---up to a trivial rescaling of coupling constants and volume factors. Such orbifold equivalences do not relate only theories defined in different volumes, but also theories with different field content: for example, the orientifold planar equivalence~\cite{Armoni:2003gp, Armoni:2003fb} (investigated on the lattice in ref.~\cite{Lucini:2010kj}) can be interpreted as a  correspondence between two different daughter theories obtained by orbifold projections from a common parent theory~\cite{Unsal:2006pj}. 

Finally, orbifold projections are also relevant for lattice formulations of supersymmetry~\cite{Catterall:2009it}---see also refs.~\cite{Unsal:2006qp, Hanada:2007ti, Anagnostopoulos:2007fw, Ishii:2008ib, Ishiki:2008te}.

\section{Large-$N$ gauge theories on the lattice}
\label{sect:lattice}

In this section, we first introduce the basics about the formulation of non-Abelian gauge theories on a lattice in subsection~\ref{subsec:lattice_definitions}, then we present an overview of lattice results for $\SU(N)$ \ gauge theories at large $N$. In particular, we mostly discuss results in $(3+1)$ spacetime dimensions (subsection~\ref{subsec:results_4D}), but we also review the results that have been obtained in lower-dimensional spacetimes (subsection~\ref{subsec:results_less_than_4D}).

\subsection{The lattice formulation of non-Abelian gauge theories}
\label{subsec:lattice_definitions}

In the Standard Model of elementary particle physics, strong nuclear interactions are described by QCD: a gauge theory based on the unbroken non-Abelian $\SU(3)$ ``color'' gauge group. The Lagrangian of the theory reads
\begin{equation}
\label{QCD_Lagrangian}
\mathcal{L} = -\frac{1}{2} \Tr \left( F_{\alpha\beta}F^{\alpha\beta} \right) + \sum_{f=1}^{n_f} \overline{\psi}_f \left( i \gamma^\alpha D_\alpha - m_f \right) \psi_f.
\end{equation}
Denoting the bare gauge coupling as $g$, the covariant derivative is defined as $D_\mu=\partial_\mu-igA_\mu^a(x) T^a$, where the $T^a$'s are the eight generators of the Lie algebra of $\SU(3)$, in their representation as Hermitian matrices of size $3 \times 3$ and vanishing trace. They are conventionally normalized as $\Tr (T^a T^b) = \delta^{ab}/2$. Thus, gluon fields are massless fields in the adjoint representation of the $\SU(3)$ algebra, and the pure-glue part of the Lagrangian is defined in terms of the non-Abelian field strength tensor $F_{\alpha\beta}=(i/g)[D_\alpha,D_\beta]$. For simplicity, we neglect a possible $\theta$-term.

The fermionic contribution to the Lagrangian is bilinear in the quark ($\psi(x)$) and antiquark ($\overline{\psi}(x)=\psi^\dagger(x) \gamma^0$) fields, where the $\gamma^\alpha$'s are Dirac matrices. Quark fields are in the fundamental representation of the gauge group, and occur in $n_f$ different ``flavors'', which are labelled by the $f$ subscript in the equation above. Their masses are generically different, and are denoted by $m_f$. As we already mentioned in section~\ref{sect:phenomenology}, QCD is a ``vector'' gauge theory, in the sense that the gauge interaction couples equally to the left- and right-handed components of the quark fields.

A quantum description of QCD can be obtained by functional integration over the gluon and fermion degrees of freedom, with a measure proportional to the exponential of ($i$ times) the action, in natural units:
\begin{equation}
\label{QCD_partition_function}
Z = \int \mathcal{D}A \mathcal{D} \overline{\psi} \mathcal{D} \psi \exp \left( i \int \dd^4 x \mathcal{L} \right).
\end{equation}

Although the elementary degrees of freedom of QCD are quark and gluons, which carry non-vanishing color charge, color is not observed \emph{directly} in experiments, but only \emph{indirectly}.\footnote{For example, evidence for the existence of three color charges can be obtained, by comparing the cross sections of processes involving decays to leptonic states \emph{versus} those involving decays to hadrons.} The physical states observed in nature are \emph{color-singlet} hadronic states: baryons and mesons (and glueballs). In fact, the low-energy spectrum of QCD is characterized by two striking properties:
\begin{enumerate}
\item confinement of color degrees of freedom into color-singlet states, and
\item spontaneous breakdown of the (approximate) chiral symmetry.
\end{enumerate}
On the other hand, QCD processes involving energies larger than, say, $2$~GeV, can be adequately described in terms of perturbative QCD computations. This is due to the fact that the \emph{physical} coupling of QCD is a (scheme-dependent) quantity which runs with the energy: it becomes small in the high-energy limit (\emph{asymptotic freedom}~\cite{Gross:1973id, Politzer:1973fx}), while it becomes large at low energies, at a typical scale $\LambdaQCD$, of the order of a few hundred MeV's. Thus, QCD (and, in general, most non-Abelian gauge theories) features a dynamically generated mass scale which characterizes the hadron spectrum.

Due to the running of the coupling, the study of physical, strongly interacting states at low energies necessarily requires an approach which does not rely on the smallness of the coupling: an intrinsically non-perturbative approach.

With the exception of the light-cone formalism~\cite{Brodsky:1997de}, the lattice regularization of QCD~\cite{Wilson:1974sk} is the unique non-perturbative, gauge-invariant formulation of QCD from its firsts principles: in fact, it provides the very non-perturbative \emph{definition} of QCD (while being fully consistent with the perturbative definition, too).

The basic idea underlying the lattice formulation of QCD consists in defining the theory in a gauge-invariant way, on a discrete spacetime grid---rather than in the continuum spacetime. This allows one to trade the continuous, infinite number of degrees of freedom appearing in eq.~(\ref{QCD_partition_function}) for a countable (and, if the theory is defined in a spacetime of finite extent, even finite) number of degrees of freedom.

This discretization makes the theory rigorously well-defined at a mathematical level: it replaces the functional integral of the continuum formulation with a product of ordinary integrals, and it provides a natural cutoff, inversely proportional to the lattice spacing $a$. 

The continuum theory is then recovered in the limit for $a \to 0$; more precisely, in this limit the continuum theory arises as an ``effective low-energy description'' of the lattice theory, valid at distance scales much longer than the lattice spacing. In order to define the continuum limit $a \to 0$, one has to provide a sensible definition for the lattice spacing $a$ in physical units. A way to do this consists in identifying a suitable dimensionful observable (which, on the lattice, can be expressed in the appropriate units of the lattice spacing) and fixing it to its physical value.

For example, if the observable is a certain correlation length $\xi$, the ratio $\xi/a$ is dimensionless. Then, taking the continuum limit of the lattice theory is possible, if the parameters of the theory can be tuned in such a way, that the $\xi/a$ ratio tends to infinity. This means that the continuum limit can be taken, when the lattice theory has a continuous transition, characterized by a diverging correlation length.

This is possible for non-Abelian gauge theories in four spacetime dimensions, which are known to be asymptotically free and possess an ultraviolet fixed point when the coupling tends to zero. For these theories, the possibility of defining a continuum limit is thus related to the fact that the coupling becomes weak at short distances. For the lattice theory, the bare coupling appearing in the lattice action has the meaning of a physical coupling at the distance of the lattice spacing; thus, asymptotic freedom implies $g \to 0$ for $a \to 0$.

An important issue is the restoration of the full continuum symmetries. For simplicity, let us consider the case of pure Yang-Mills theory. Clearly, the lattice regularization explicitly breaks translational and (Euclidean) rotational symmetries: on a hypercubic lattice, the group of continuum translations is broken down to its subgroup of translations by integer multiples of the lattice spacing (in each direction), while the group of rotations is broken down to rotations by angles which are integer multiples of $\pi/2$. However, gauge symmetry is kept exact at all values of the lattice spacings: the gauge degrees of freedom of the lattice theory are not the continuum gauge fields (taking values in the algebra of generators of the gauge group), but rather parallel transporters defined on the links between nearby lattice sites, taking values in the gauge group itself.

The lattice action and other lattice operators differ from their continuum counterparts by operators of higher dimension, which, being suppressed by some power of $a$, become irrelevant in the continuum limit.\footnote{The fact that the action of the lattice Yang-Mills theory has exact gauge invariance and invariance under discrete translations and rotations at any value of the lattice spacing implies that no undesired operators, not present in the original theory, are generated upon renormalization.}

The lattice formulation of QCD was initially proposed by Wilson in 1974~\cite{Wilson:1974sk}, by discretizing the four Euclidean dimensions; a related formulation, in which the time dimension was kept continuous, was discussed by Kogut and Susskind~\cite{Kogut:1974ag}.

The lattice formulation of QCD is based on the Feynman path integral approach, which, for quantum mechanics of one non-relativistic particle, consists in expressing the transition amplitude from an initial state to a final state as a weighted sum over all possible trajectories. For a particle with Hamiltonian $\hat{H}=\hat{p}^2/(2m)+\hat V (x)$ which propagates from $x$ to $y$ in a time interval $t$, the transition amplitude can be readily evaluated, by dividing the time interval into $n$ intervals, and repeatedly inserting the ``resolution of the identity'' $\ide = \int \dd x_i |x_i \rangle \langle x_i|$:
\begin{eqnarray}
\langle y | e^{-it \hat H} | x \rangle =  \left( \frac{m}{2\pi i \epsilon} \right)^{n/2} \int \dd x_1 \!\! \int \dd x_2 \dots \int \dd x_{n-1} \exp \left\{ i \frac{m}{2t} \left[ (y-x_1)^2 + (x_1-x_2)^2 + \dots \right. \right.\nonumber \\
\left. \left. + (x_{n-1}-x)^2 \right] -i \epsilon \left[ \frac{1}{2} V(y) + V(x_1) + V(x_2) + \dots + V(x_{n-1}) + \frac{1}{2} V(x) \right] \right\}. \nonumber
\end{eqnarray}
Finally, taking the $n \to \infty$ limit, one can express the amplitude as
\begin{equation}
\langle y | e^{-it \hat H} | x \rangle = \int \mathcal{D} x \; e^{i\int \dd t \mathcal{L}}  = \int \mathcal{D} x \; e^{i S}, \nonumber
\end{equation}
so that the quantum propagation of the particle is written as a \emph{weighted} sum over paths. 

The weight is a \emph{complex} phase factor, given by the exponential of $i$ times the action $S$ over $\hbar$. The complex nature of the weight implies that there are large cancellations, in particular for paths with action $S \gg \hbar$. The \emph{classical} limit, on the other hand, is recovered for paths making the action stationary: for $\hbar \to 0$, only the trajectory satisfying the classical equation of motion yields a non-negligible contribution to the path integral.

To avoid the large cancellations associated with a complex weight, it is convenient to perform a Wick rotation to Euclidean time, defining $\tau=it$. Then, the propagation amplitude can be re-expressed in terms of the Euclidean action $S_E$:
\begin{equation}
\int \mathcal{D} x \; \exp \left\{ - \int \dd \tau \left[ \frac{1}{2} m \left( \frac{dx}{d \tau} \right)^2 + V(x) \right] \right\} =  \int \mathcal{D} x \exp \left( -S_E \right) \nonumber = \mathcal{Z}, \nonumber
\end{equation}
where $\mathcal{D} x$ denotes the multiple integration over $x_i$ points. In this form, the weight of each path is a \emph{real positive} quantity, formally analogous to a ``Boltzmann factor''. This enables one to draw a connection with the partition function of a classical statistical mechanics system, and to use the corresponding computational techniques---both analytical (for example the counter-part of high-temperature expansions) and numerical (e.g. integration by Monte Carlo methods). 

This approach can be readily extended to ``second-quantized'' theories, i.e. generalized to quantum field theory (QFT). In Minkowski spacetime, the physically relevant information for a QFT is encoded in Wightman functions $\langle 0 | \hat\phi(x_1) \dots \hat\phi(x_n) | 0 \rangle$. Under certain well-defined mathematical conditions~\cite{Osterwalder:1973dx}, it is possible to carry out an analytical continuation to Euclidean spacetime, where the physical information is contained in a set of symmetric Schwinger functions. The symmetry of Schwinger functions is related to the fact that, in a certain sense, Euclidean quantum fields ``can be treated as'' classical variables. In particular, in the Euclidean formalism of QFT, bosonic fields are associated with classical (commuting) numbers, while fermionic fields are represented as Grassmann (anticommuting) variables. 
In fact, the Euclidean quantization is not carried out by mapping classical observables to Hermitian \emph{operators}, but, rather, by treating the fields as stochastic variables.

While the number of degrees of freedom in the continuum is infinite, the lattice regularization makes it discrete, and \emph{finite} for a system defined in a finite hypervolume. In particular, this opens up the possibility of carrying out the computation of the QFT partition functional (or, more precisely, of the functionals associated with expectation values of physical operators) by numerical means, via Monte Carlo simulations.

Let us now discuss in detail the Wilson lattice regularization for gauge theories. For simplicity, we will restrict our attention to the regularization on a (hyper)cubic lattice: this is by far the most common geometry that is used, since this type of lattice is defined in any (integer) dimension, it is uni-partite, and the number of the elementary vectors defining the lattice is equal to the number of spacetime dimensions $D$. Regularizations on  lattices of different geometry (including ones characterized by higher symmetry), however, are also possible and lead to the same continuum limit. We assume that the lattice spacing $a$ is the same in the four directions (although anisotropic lattices are used in some problems). The glue sector of the theory is defined in terms of parallel transporters along the oriented links joining nearest-neighbor lattice sites:
\begin{equation}
\label{link_definition}
U_\mu(x) = \exp \left[ i g a A_\mu (x+a \hat{\mu}/2) \right].
\end{equation}
$U_\mu(x)$ denotes the parallel transporter from the site $x$ to the site $(x+a\hat{\mu})$, where $\hat{\mu}$ is the versor in the Euclidean direction $\mu$, with $\mu = 1,\ \dots, D$. Note that, being parallel gauge transporters along paths of finite length $a$, the link variables $U_\mu(x)$ take values in the gauge group---rather than in its algebra. The coupling $g$ appearing in the expression on the r.h.s. of eq.~(\ref{link_definition}) is the bare lattice coupling, and it describes the strength of the gauge interaction at a distance $a$ \emph{in the lattice theory}.

Under a gauge transformation $\chi$, the link variable $U_\mu(x)$ transforms as:
\begin{equation}
U_\mu(x) \rightarrow \chi(x) U_\mu(x) \chi^\dagger(x+a\hat{\mu}).
\end{equation}
Gauge-invariant, purely gluonic lattice operators are given by traces of path-ordered products of link variables around closed contours. The simplest of them is the trace of the \emph{plaquette} $U_\Box$, which is obtained as the path-ordered product of links around an elementary $a \times a$ square on the lattice. 

\begin{figure}[tb!]
\begin{center}
\begin{minipage}[t]{16.5 cm}
\begin{centering}
\hspace{1cm} \epsfig{file=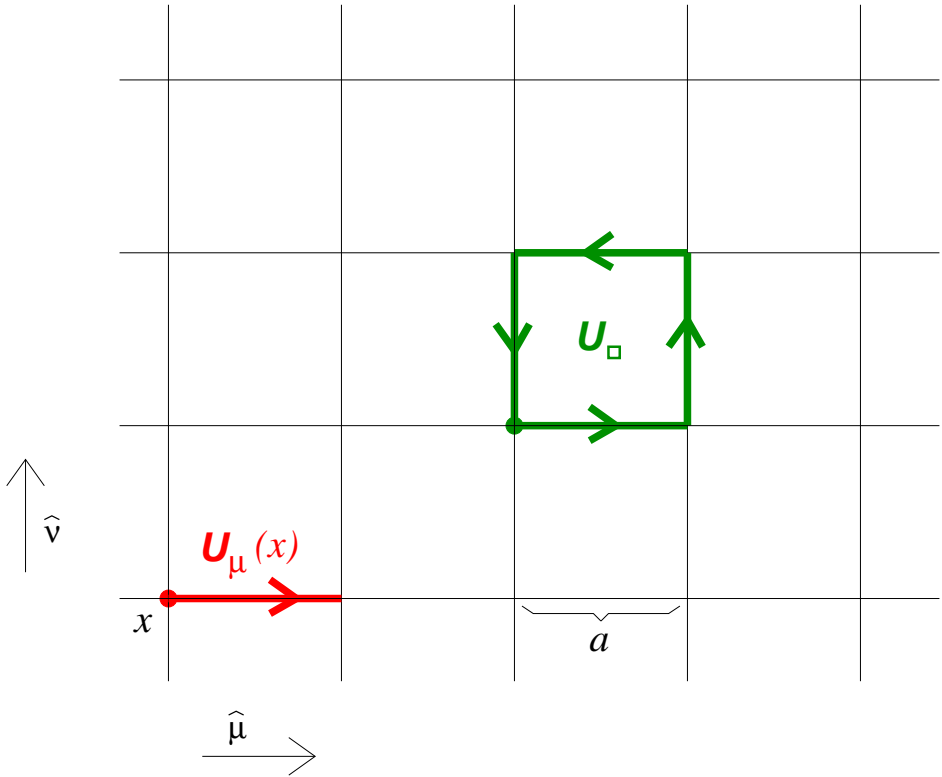,scale=0.5}
\end{centering}
\vspace{5mm}\\
\end{minipage}
\begin{minipage}[t]{16.5 cm}
\caption{The fundamental objects in the lattice formulation of Yang-Mills theory are a set of $U_\mu(x)$ matrices, defined in eq.~(\protect\ref{link_definition}), on the oriented links (of length $a$) of the lattice: they can be interpreted as parallel transporters between neighboring lattice sites. Gauge-invariant quantities are obtained from traces of path-oriented products of links around closed paths: the simplest example is the plaquette $U_\Box$, appearing in the Wilson action definition, eq.~(\protect\ref{Wilson_gauge_action}).\label{fig:lattice_action}}
\end{minipage}
\end{center}
\end{figure}

The simplest lattice action for the purely gluonic theory is the Wilson action, given by the sum over all lattice plaquettes
\begin{equation}
\label{Wilson_gauge_action}
S_W = \beta \sum_{\Box} \frac{1}{N}\real \Tr \left( \ide - U_\Box \right),
\end{equation}
with $\beta=2N/g^2$, see fig.~\ref{fig:lattice_action}.

\begin{figure}[tb!]
\begin{center}
\begin{minipage}[t]{8 cm}
\epsfig{file=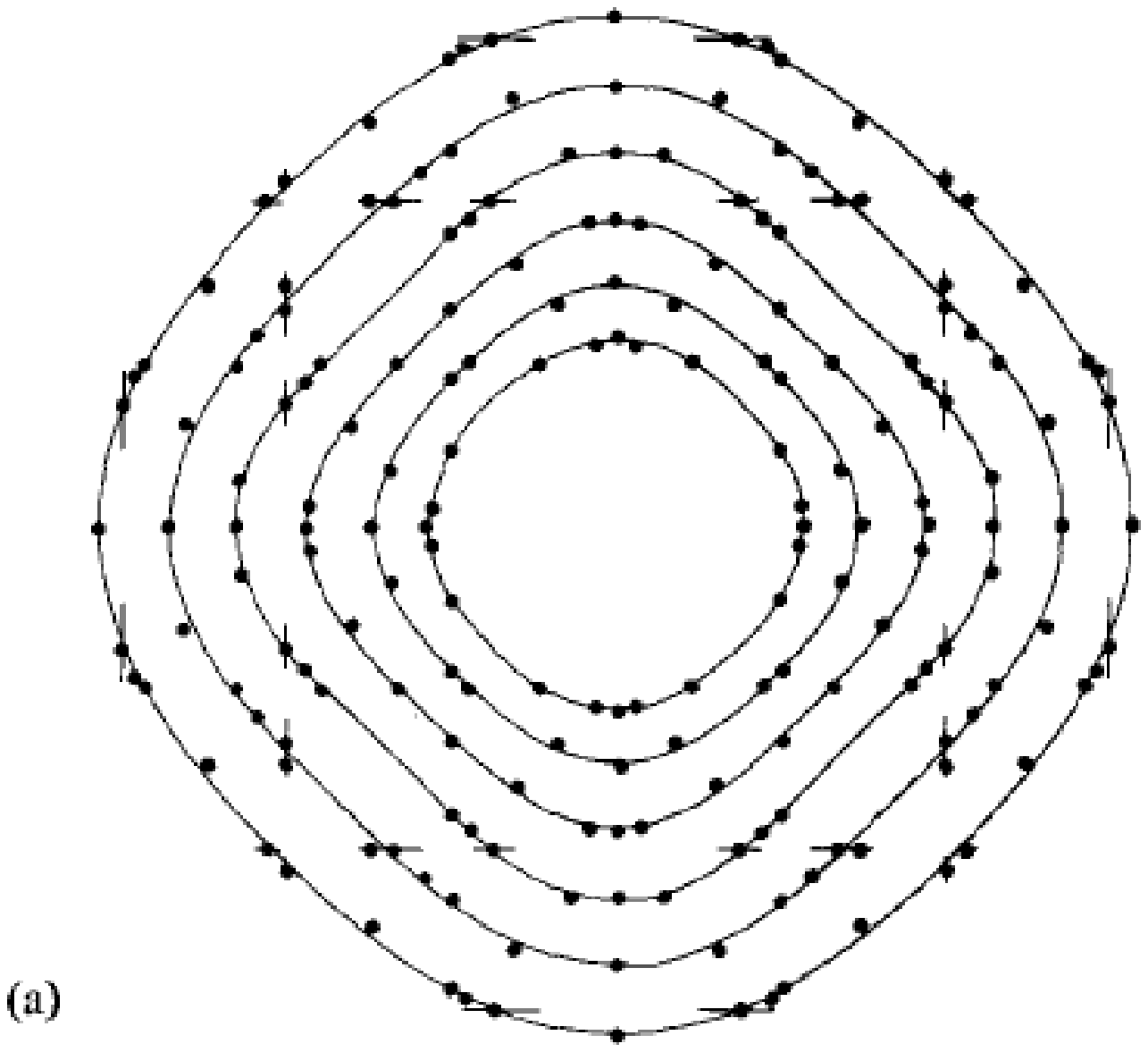,scale=0.55}
\vspace{1cm}\\
\epsfig{file=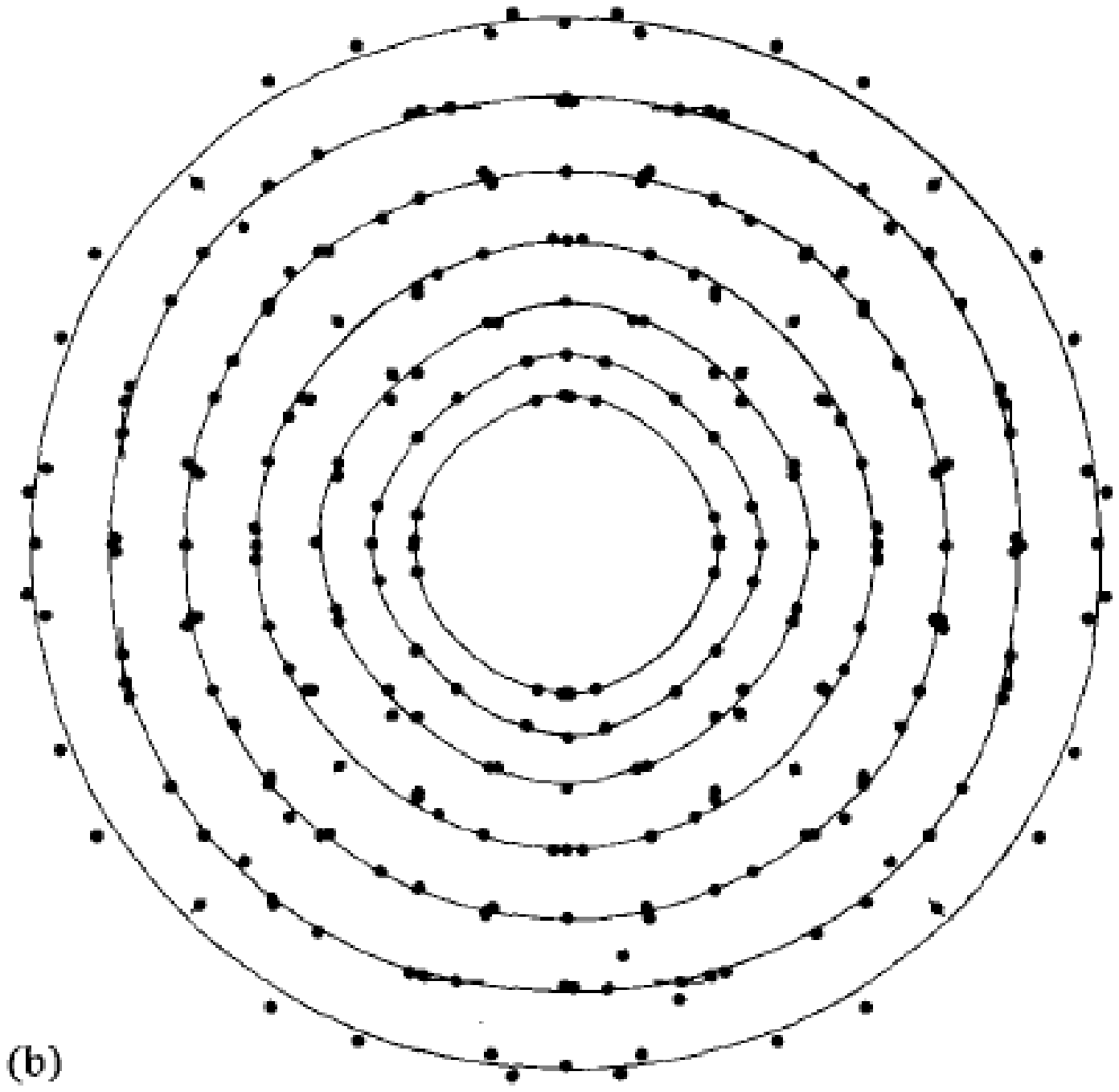,scale=0.55}
\vspace{5mm}\\
\end{minipage}
\begin{minipage}[t]{16.5 cm}
\caption{Restoration of continuum rotational symmetry in the confining potential in $\SU(2)$ Yang-Mills theory, as observed in ref.~\cite{Lang:1982tj}. The solid lines represent the loci of equal potential, in the presence of a static color source at the center of the figure. The top panel (a) displays results obtained on a coarser lattice ($\beta=2$), while the data in the bottom panel (b) were obtained on a finer lattice ($\beta=2.25$). The restoration of the continuum rotational symmetry when $a$ tends to zero is manifest.\label{fig:rotational_symmetry_restoration}}
\end{minipage}
\end{center}
\end{figure}

It is a trivial exercise to prove that, in the limit $a \to 0$, the Wilson action defined in eq.~(\ref{Wilson_gauge_action}) tends to the continuum Euclidean Yang-Mills action
\begin{equation}
\label{Wilson_action_to_YM_action}
\lim_{a \to 0} S_W = \frac{1}{4} \int \dd^4 x (F_{\mu\nu}^a)^2.
\end{equation}
At finite values of $a$, $S_W$ differs from the continuum Yang-Mills action by relative corrections $O(a^2)$. These corrections, which are responsible for the explicit breakdown of continuum symmetries through lattice artifacts, are suppressed for $a \to 0$. Fig.~\ref{fig:rotational_symmetry_restoration}, taken from ref.~\cite{Lang:1982tj}, shows an explicit numerical example of this, as observed in the study of the confining potential in $\SU(2)$ Yang-Mills theory. The solid lines in the figure represent ``isopotential curves'', namely lines along which the potential associated with the strong interaction (in the presence of a static fundamental color source at the center) takes constant values. The top panel (a) shows results obtained from simulations at $\beta=2$, while the data displayed in the bottom panel (b) were obtained from simulations at $\beta=2.25$: since $\beta$ is inversely proportional to the square lattice coupling $g^2$, and due to asymptotic freedom and to the (logarithmic) running of the coupling, the lattice spacing at $\beta=2.25$ is finer than at $\beta=2$. Correspondingly, one expects that lattice discretization effects, generically proportional to some power of the lattice spacing, get reduced when $\beta$ is increased. This is indeed observed in the numerical results shown in the figure: the isopotential curves in panel (b) are nearly perfectly symmetric under continuous rotations, whereas those in panel (a), obtained on a coarser lattice, are only invariant under discrete rotations by angles which are multiple of $\pi/2$.

The convergence to the continuum limit can be improved by adding counter-terms which compensate for the leading discretization effects~\cite{Curci:1983an, Weisz:1983bn, Luscher:1984xn, Luscher:1985zq}.

The partition function of the lattice Yang-Mills theory is given by
\begin{equation}
\mathcal{Z} = \int \prod_{x, \mu} \dd U_\mu(x) \exp \left( - S_W \right).
\end{equation}
At each value of the lattice spacing $a$, it is invariant under gauge transformations, discrete rotations and translations, as well as under parity, Euclidean time reversal, and charge conjugation. Note that the discreteness of the lattice implies that the ultraviolet divergences plaguing the continuum theory are regularized (in a gauge-invariant way) by an intrinsic momentum cutoff scale $\pi/a$.

As we mentioned above, physical observables in the Yang-Mills lattice theory are built from traces of ordered products of link variables along closed lattice paths. Their vacuum expectation values, defined by
\begin{equation}
\langle \mathcal{O} \rangle = \frac{1}{\mathcal{Z}} \int \prod_{x, \mu} \dd U_\mu(x) \mathcal{O} \exp \left( - S_W \right),
\end{equation}
can be evaluated using different methods, either analytical (lattice strong coupling expansions or perturbative expansions on the lattice) or numerical (Monte Carlo simulations based on importance sampling) ones.

Examples of interesting physical observables in the pure Yang-Mills theory include:
\begin{enumerate}
\item Wilson loops:
\begin{equation}
\mathcal{W}(r,L) = \real \tr \prod U_\mu(x).
\end{equation}
They are the defined as the holonomies of the gauge connection around a given closed path. On the lattice, they are expressed as path-ordered products of link variables around a closed loop. In particular, a rectangular Wilson loop of sizes $r$ and $L$, lying in a plane parallel to the Euclidean time direction, can be interpreted in terms of the process associated with the creation, propagation over a Euclidean time interval $L$ and annihilation of an infinitely heavy (static) quark-antiquark pair at a relative distance $r$. The potential $V(r)$ associated with the heavy quark-antiquark pair at a distance $r$ can be extracted from the expectation value of $\mathcal{W}(r,L)$ according to the formula
\begin{equation}
V(r) = - \lim_{L \to \infty } \frac{1}{L} \ln \langle \mathcal{W}(r,L) \rangle .
\end{equation}
\item Polyakov loops $\mathcal{P}$: they are defined analogously to Wilson loops, except that they correspond to non-contractible loops winding around a periodic direction (of extent $L$) in the system. When the periodic direction is regarded as the Euclidean time, a Polyakov loop corresponds to the world line of a static color source at finite temperature $T=1/L$. Then, the free energy $F$ of a bare static source at temperature $T$ can be obtained from the relation
\begin{equation}
F = - T \ln \langle \mathcal{P} \rangle .
\end{equation}
In the confining phase of Yang-Mills theory (at zero or low temperatures) the expectation value of $\mathcal{P}$ is vanishing, so that the free energy associated with an isolated color source is infinite: quarks cannot exist as asymptotic states. By contrast, the expectation value of $\mathcal{P}$ is finite (and, as a consequence, such is its free energy) in the deconfined phase of Yang-Mills theory at sufficiently high temperatures.
\item Linear combinations of traces of holonomies along loops of different shapes (transforming according to well-defined irreducible representations of the group of discrete spatial rotations of the lattice, and with well-defined parity and charge conjugation quantum numbers): they correspond to glueball operators, i.e. operators creating gauge-invariant, color-singlet states with well-defined $J^{PC}$ quantum numbers. In particular, these operators can be projected onto their zero-momentum Fourier component by averaging over the spatial coordinates in a fixed-time lattice slice: then, the decay of the logarithm of two-point correlation function at large Euclidean time separations $\Delta L$ becomes linear in $\Delta L$, with a slope given by minus the mass of the lightest physical state in the spectrum with the quantum numbers considered. This is the basis of spectroscopy computations on the lattice.
\end{enumerate}

In order to obtain physical information from lattice simulations of gauge theories, however, it is important to understand that the simulation results at finite lattice spacing depend on:
\begin{itemize}
\item The number of spacetime dimensions (typically: $D=4$ or $3$).
\item The lattice geometry: in particular, the type of lattice (hypercubic, $F_4$, hyperdiamond, \dots) and its sizes. As we already mentioned, the high-temperature regime of a QFT can be described, in a Euclidean setup, by a system with a compact direction of finite size $L$ (with periodic boundary conditions for bosonic fields, and antiperiodic boundary conditions for fermionic fields). The physical temperature is equal (in natural units) to the inverse of the shortest size of the system.
\item The gauge group ($\SU(N)$, $\U(1)$, $\Z_N$, $\Sp(N)$, $\SO(N)$, \dots) and the field representation.
\item The lattice action and the value of its coupling(s). For the Wilson action introduced above, large values of $\beta$ correspond to small values of the bare coupling $g$, and, hence, small values of the spacing $a$.
\end{itemize}

To extract results relevant for the continuum theory, an extrapolation to the continuum limit $a \to 0$ has to be carried out. This is possible when the lattice theory has a continuous transition. As we mentioned, for non-Abelian gauge theories, this is possible due to asymptotic freedom: $g=0$ is an ultraviolet fixed point.

Assuming the lattice size to be arbitrarily large (which corresponds, in particular, to zero temperature, and which allows one to neglect undesired finite-volume effects), the phase structure of some of the historically most studied pure gauge theories on the lattice can be summarized as follows:
\begin{itemize}
\item For both $D=4$ and $D=3$ Euclidean spacetime dimensions, all $\SU(N)$ Yang-Mills theories are confining, at all values of the bare coupling.
\item In $D=4$ dimensions, compact $\U(1)$ lattice gauge theory has a confining phase at strong coupling (i.e., at small $\beta=1/e^2$), and a Coulomb phase at weak coupling. The confining phase at strong coupling is a lattice artifact, and has no direct connection with the continuum theory. Rather, it can serve as a lattice toy model for other confining theories. Since the deconfining transition at finite $\beta$ is a weakly first-order one, strictly speaking it is not possible to define a \emph{bona fide} continuum limit for the confining regime of this theory.
\item In $D=3$ dimensions, the $\Z_2$ lattice gauge theory has a confining phase at strong  coupling, and a deconfined phase at weak coupling. In this case, the deconfining transition is of second order.
\end{itemize}

We emphasize that, although non-Abelian lattice gauge theories at strong bare coupling (i.e. at small $\beta$) can be rigorously proved to be confining and to have a finite mass-gap (i.e. the mass of the lightest physical state in the spectrum is finite), this \emph{does not} provide a solution to the confinement problem. The reason is that the strong coupling regime of lattice gauge theories corresponds to \emph{large} values of the lattice spacing $a$ and is dominated by severe lattice artifacts---so, strictly speaking, it has no direct connection to continuum physics. In particular, the validity of the lattice strong-coupling expansions, on which these analytical results are based, is limited by a finite radius of convergence. For $\SU(N)$ Yang-Mills theories with more than $4$ color charges, it turns out that the strong-coupling phase of the lattice theory is separated from the weak-coupling phase (which is analytically connected to the continuum theory) by an unphysical, first-order bulk transition. Nevertheless, we remark that strong-coupling techniques in lattice gauge theory provide many interesting qualitative insights, and continue to be actively studied even in recent years~\cite{deForcrand:2009dh, Langelage:2009jb, Langelage:2010yn, Langelage:2010yr, Fromm:2011qi, Fromm:2012eb}.

As we mentioned above, in a lattice theory at finite spacing $a$, the symmetries and the properties of the continuum theory are approximately recovered, for physical quantities at momentum scales much lower than $\pi/a$. Since in all numerical simulations the results are expressed as \emph{dimensionless} quantities, in units of the lattice spacing, one has to set the physical scale (namely: to define the value of $a$ in physical units, at a given finite value of the lattice coupling) non-perturbatively. A common way to do this consists in using some low-energy quantity as a reference.

One possibility is based on the large-distance behavior of the confining static potential in a pure Yang-Mills theory. First, one carries out simulations at a given value of $\beta$ on a lattice of given sizes, computing the expectation values of large Wilson loops, $\langle \mathcal{W}(r, L \rangle $, with $r/a$ and $L/a \gg 1$. Then, these values can be fitted to the expected area-law behavior (which characterizes confining theories),
\begin{equation}
\langle \mathcal{W}(r, L) \rangle \propto \exp \left( -\sigma a^2 \cdot \frac{r}{a} \cdot \frac{L}{a}\right).
\end{equation}
This allows one to extract a numerical value for $\sigma a^2$. Then, one can obtain $a$ in fm by defining the string tension $\sigma$ to have its ``phenomenological'' value $\sigma=(440 \mbox{MeV})^2$ (with 197 MeV $\simeq 1$~fm$^{-1}$ in natural units). Finally, the continuum expectation value for a given observable $\mathcal{O}$ is obtained by evaluating $\langle \mathcal{O} \rangle $ on finer and finer lattices, and extrapolating for $a \to 0$.

The convergence to the continuum limit is determined by how fast the discretization effects (in the lattice action and in the operators associated with the observables) are suppressed for $a \to 0$. In particular, a tree-level analysis shows that, for the Wilson discretization on a hypercubic lattice of finite spacing $a$, the dominating discretization effects in the action are $O(a^2)$ (as we are ``approximating derivatives by finite differences''). In addition, quantum effects also introduce $g$-dependent corrections. In order to improve the convergence to the continuum limit, one can modify the definition of the lattice action and observables, by inclusion of extra terms that remove the leading discretization artifacts. The obvious advantage is a faster approach to the continuum limit, which enables one to perform more reliable continuum extrapolations from simulation results obtained on lattices with a smaller number of sites. Generally, this advantage largely overcompensates the drawback that more 
complicated definitions of the lattice operators tend to slow down numerical simulations, hence the use of improved lattice actions is quite common---particularly for the most resource-demanding simulations involving dynamical fermions.

The quark contribution to the continuum QCD action is of the form
\begin{equation}
\int \dd^4 x \sum_{f} \bar \psi_f (x) \left(m_f + \sum_\mu \gamma^\mu D_\mu \right) \psi(x),
\end{equation} 
with $D_\mu = \partial_\mu -ig_0 T^a A_\mu^a$. The lattice discretization of fermion fields, however, involves various subtleties. To begin with, in the path integral formulation, the traces over fermionic variables are rewritten as formal integrals over classical anticommuting numbers, i.e. Grassmann variables. The latter do not admit a direct computer implementation in terms of local variables, but the bilinearity of the fermionic action allows one to treat the fermionic contribution to the action exactly, by writing it as a fermionic determinant, which can be evaluated numerically. The fermionic determinant, however, is a non-local function of the gauge fields, and this leads to a major computational overhead with respect to the simulation of pure Yang-Mills theory. For this reason, most of the lattice QCD computations carried out until the late 1990's were performed in the \emph{quenched approximation}, which consists in neglecting the effect of dynamical quarks altogether, and in evaluating 
operators involving valence quarks on configurations generated according to a quantum weight depending only on the pure Yang-Mills action.

Yet, it is clear that in general the fermionic determinant \emph{must} be included, in order to get the correct description of actual physical phenomena. Luckily, the computer-power and algorithmic progress during the last fifteen years is making the quenched approximation obsolete.

Fermion simulations, however, also involve other, much more fundamental, subtleties. In particular, it is well-known that a na\"{\i}ve lattice discretization of the continuum Dirac operator leads to the doubling problem, i.e. to the existence of $(2^D-1)$ unphysical lattice modes. This problem is related to the fact that the Dirac operator involves a first-order derivative: its lattice discretization
\begin{equation}
\sum_{x} a^4 \sum_{f} \left\{ m_f \bar \psi_f (x) \psi_f (x) + \frac{1}{2a} \bar \psi_f (x)  \sum_\mu \gamma_\mu [ U_\mu(x) \psi(x+a\hat{\mu}) - U^\dagger_\mu(x-a\hat{\mu}) \psi(x-a\hat{\mu})] \right\}
\end{equation}
leads to a periodic dispersion relation, which exhibits unphysical zeros for momenta with components $\pi/a$.

One possibility to solve this problem was proposed by Wilson~\cite{Wilson:1975id}: the unphysical doublers can be removed, by adding an extra term to the quark lattice action, which is proportional to (the lattice discretization of) a Laplacian:
\begin{equation}
-\frac{ra^3}{2} \sum_{x,f,\mu} \bar \psi_f (x) [ U_\mu(x) \psi_f(x+a\hat{\mu}) - 2\psi(x) + U^\dagger_\mu(x-a\hat{\mu}) \psi_f(x-a\hat{\mu})].
\end{equation}
This term has energy dimension five, and hence becomes irrelevant in the continuum limit. However, it has the effect of removing the doublers, by giving masses $O(a^{-1})$ to the modes with at least one $p_\mu=\pi/a$ component. One important feature of the Wilson Dirac operator is that, at finite values of the lattice spacing, it explicitly breaks the chiral symmetry that one expects for $m_f=0$. In addition, it leads to additive mass renormalization in the interacting theory: this implies that the chiral limit has to be achieved by fine tuning of the bare lattice parameters.

An alternative (partial) solution to the doubling problem was proposed by Kogut and Susskind, and goes under the name of ``staggered fermions''. The idea is to perform a \emph{local} redefinition of the lattice fermion fields, which leads to a spin diagonalization,
\begin{equation}
\psi(x) = \gamma_1^{x_1} \gamma_2^{x_2} \gamma_3^{x_3} \gamma_4^{x_4} \chi(x),
\end{equation}
followed by a projection leaving only one of the spinor components. This reduces the number of doublers down to $2^{\lfloor D/2 \rfloor}$ (where $\lfloor x \rfloor$ denotes the largest integer not larger than $x$) i.e. to four, in four spacetime dimensions, and leads to a formulation in which different components of the original spinor are ``staggered'' over nearby sites (within ``blocks'' of $2^D$ sites). This formulation has a close connection to Dirac-K\"ahler fermions in the continuum. The staggered action reads
\begin{equation}
\sum_{x,f} \left\{ \bar m_f \chi_f(x) \chi_f(x) + \frac{\bar\chi_f(x)}{2a} \sum_{\mu} (-1)^{\sum_{\nu<\mu} x_\nu} [ U_\mu(x) \chi_f(x+a\hat{\mu}) - U^\dagger_\mu(x-a\hat{\mu}) \chi_f(x-a\hat{\mu})] \right\}.
\end{equation}
The staggered lattice Dirac operator preserves a remnant of chiral symmetry, and does not lead to additive mass renormalization. In addition, the reduced number of components makes it computationally efficient (and, hence, very popular). The remaining degeneracy of the free staggered operator is referred to in terms of ``quark tastes'' (to distinguish them from the physical quark flavors); however, taste degeneracy is broken by interactions. A commonly used method in staggered simulations where one wants to simulate two light quark flavors is the so-called ``rooting trick'', i.e. taking the square root of the determinant of the staggered operator, in order to reduce the number of physical flavors down to two. This procedure is valid at the perturbative level, although during the past few years there has been some debate whether it is valid also non-perturbatively.

There exist also formulations of lattice fermions that respect chiral symmetry. As it is well-known, in the continuum the latter is an important global symmetry for massless quarks, and in nature it is approximately realized for the light \emph{up}, \emph{down} (and, to a certain extent, also \emph{strange}) quark flavors. As we mentioned, the spontaneous breakdown of chiral symmetry plays an important r\^ole in the hadronic spectrum, being associated to the existence of light pseudo-Nambu-Goldstone modes: the pions (and the kaons and $\eta$, if the strange quark is also considered as ``light''). On the lattice, however, a well-known no-go theorem due to Nielsen and Ninomiya states that either chiral symmetry is explicitly broken, or there exist unphysical doublers~\cite{Nielsen:1981xu}. The solution consists then in formulating lattice fermions satisfying a modified form of chiral symmetry, known as the Ginsparg-Wilson relation~\cite{Ginsparg:1981bj, Luscher:1998pqa}
\begin{equation}
\left\{ D, \gamma_5 \right\} = a D \gamma_5 D
\end{equation}
and a modified chiral rotation
\begin{equation}
\psi \to \psi + \delta \psi, \;\; \delta \psi = i\epsilon\gamma_5 (1 - a D/2) \psi .
\end{equation}
An explicit construction of a lattice Dirac operator satisfying these requirements was proposed in ref.~\cite{Neuberger:1997fp}, and goes under the name of ``overlap fermions''
\begin{equation}
D = D_{ov} = \frac{1}{a} \left[ 1 + \gamma_5 \mbox{sign}(\gamma_5 D_W) \right].
\end{equation}
This formulation realizes the (modified) chiral symmetry exactly at every value of the lattice spacing, satisfies the Atiya-Singer theorem~\cite{Atiyah:1968mp}, and leads to exactly one massless physical flavor in the continuum limit, with no need for parameter fine-tuning. 

A different formulation of lattice fermions satisfying the Ginsparg-Wilson relation is based on the domain-wall construction~\cite{Kaplan:1992bt, Shamir:1993zy, Furman:1994ky}, whereby a chiral fermion is obtained by introducing an unphysical fifth dimension (on which the gauge fields do not depend), along which the bare fermion mass changes sign. Despite the superficial differences, one can prove that this construction is essentially equivalent to the overlap operator. In both cases, the main numerical drawbacks of Ginsparg-Wilson lattice fermions are related to the fact that they are computationally much more expensive than Wilson or staggered fermions.

Having discussed the basic concepts underlying the lattice formulation of gauge and fermion fields, we now turn to a brief review of the main numerical results in the large-$N$ limit.

\subsection{Lattice results for large-$N$ gauge theories in $(3+1)$ spacetime dimensions}
\label{subsec:results_4D}

One of the main non-perturbative issues to be studied via lattice simulations at large $N$ was the confining nature of non-Abelian gauge theory in the 't~Hooft limit. 

Early works addressing this question were presented in refs.~\cite{Teper:1998kw, Teper:1997tq}, which considered Yang-Mills theories with $\SU(2)$, $\SU(3)$ and $\SU(4)$ gauge groups. These works studied correlation functions of zero-transverse-momentum, gluonic, string-like operators winding around a spatial direction of the lattice (``torelons'') of length $L$, and found numerical evidence that they decay exponentially with the torelon-torelon separation $\tau$, $\exp [ - m(L) \tau ]$. The torelon energy per unit length is approximately constant ($\sigma$) for long torelons, indicating confinement.

\begin{figure}[tb!]
\begin{center}
\begin{minipage}[t]{16.5 cm}
\begin{centering}
\hspace{1cm} \epsfig{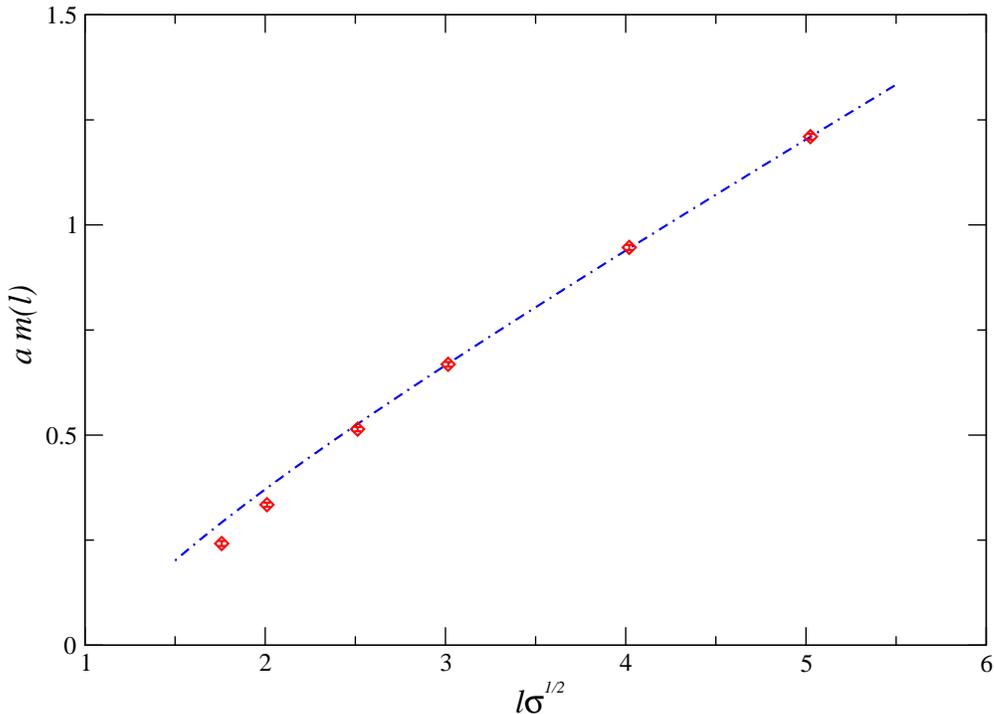}
\end{centering}
\vspace{5mm}\\
\end{minipage}
\begin{minipage}[t]{16.5 cm}
\caption{The mass $m$ of a long torelon, as a function of its length $l$ (in the appropriate units of the string tension $\sigma$) can be described very accurately in terms of a confining string model, including the L\"uscher term associated with string fluctuations. The dashed line in this plot is not a fit to the data, but a prediction of this bosonic string model for the low-energy description of long flux tubes. The figure is adapted from ref.~\cite{Teper:2009uf}.\label{fig:loop_mass}}
\end{minipage}
\end{center}
\end{figure}

In fact, on the lattice it is also possible to accurately study the corrections to the linear dependence of the torelon energy, which become non-negligible at intermediate values of the torelon length $L$. This issue has been addressed in many studies~\cite{Teper:1997tq, Teper:1998kw, Lucini:2000qp, Lucini:2001nv, Lucini:2001ej, DelDebbio:2001sj, DelDebbio:2001kz, DelDebbio:2002yp, DelDebbio:2003tk, Meyer:2004hv, Lucini:2004my, Teper:2009uf, Athenodorou:2010cs, Lohmayer:2012ue, Mykkanen:2012dv}, including also for flux tubes in higher representations, or for excited string states. The results indicate that the leading correction to the linear dependence of $m$ on $L$ can be expressed in terms of a ``L\"uscher term'', due to the quantum fluctuations of the torelon~\cite{Luscher:1980fr, Luscher:1980ac},
\begin{equation}
\label{torelon_mass}
m(L)= \sigma L - \frac{\pi}{3L} + \dots ,
\end{equation}
and that subleading corrections are captured rather well by a simple bosonic Nambu-Got{\={o}} string model~\cite{Nambu:1974zg, Goto:1971ce}. The latter model assumes that the flux tube can be described as an infinitesimally thin, fluctuating string, with an action proportional to the surface of the world-sheet it spans during its time evolution. The plot in fig.~\ref{fig:loop_mass}, from ref.~\cite{Teper:2009uf}, shows the results of a lattice calculation of torelon masses in $\SU(6)$ gauge theory. The formula in eq.~(\ref{torelon_mass}) describes the numerical results very well, for all torelon lengths larger than approximately $3/\sqrt{\sigma}$.

While the fact that at large $N$ confining flux tubes can be modelled very well in terms of an effective, low-energy bosonic string model~\cite{Polchinski:1992vg, Luscher:2004ib, Drummond:2004yp, Billo:2006zg, Meyer:2006qx, Aharony:2009gg, Aharony:2010cx, Aharony:2011ga, Aharony:2010db, Aharony:2011gb, Billo:2012da, Gomis:2012ki, Dubovsky:2012sh, Gliozzi:2012cx}, the existence of such string-like behavior is by no means a feature that characterizes only the large-$N$ limit. On the contrary, it appears to be quite a generic phenomenon in confining gauge theories, having been observed also in $\SU(N)$ gauge theories for $N=2$ or $3$~\cite{Bali:1994de, Luscher:2002qv, Juge:2002br, Bonati:2011nt}, as well as in the confining, strong-coupling phase of compact $\U(1)$ gauge theory~\cite{Koma:2003gi, Panero:2004zq, Panero:2005iu, Amado:2013rja} and in gauge theories based on exceptional gauge groups~\cite{Greensite:2006sm}. However, there are intriguing theoretical arguments suggesting that a very simple effective string 
model could become exact in the 't~Hooft limit~\cite{Polchinski:1992vg}.

\begin{figure}[tb!]
\begin{center}
\begin{minipage}[t]{16.5 cm}
\begin{centering}
\epsfig{file=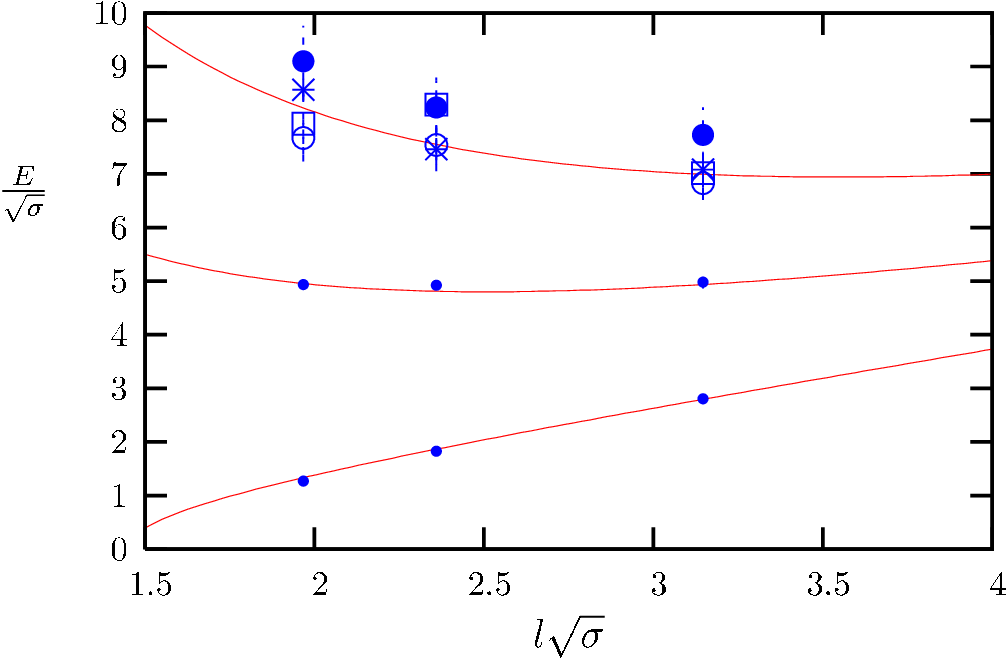,scale=1.5}
\end{centering}
\vspace{5mm}\\
\end{minipage}
\begin{minipage}[t]{16.5 cm}
\caption{The closed string spectrum in $\SU(5)$ Yang-Mills theory, as determined in ref.~\cite{Athenodorou:2010cs}, can be accurately compared with the predictions from an effective Nambu-Got{\={o}} string model.\label{fig:4D_strings}}
\end{minipage}
\end{center}
\end{figure}

The validity of a string-like picture as an effective model for confining flux tubes in large-$N$ gauge theories is also confirmed by some recent, high-precision studies of excited string states, like those reported in ref.~\cite{Athenodorou:2010cs} (see fig.~\ref{fig:4D_strings}).

Having confirmed the confining nature of non-Abelian gauge theories in the large-$N$ limit, another interesting question to be addressed is whether the 't~Hooft coupling $\lambda = g^2 N$ is really the ``natural'' coupling characterizing the theory in such a limit. As we mentioned, the introduction of the 't~Hooft coupling is very intuitive from a perturbative point of view, but, a priori, non-perturbative effects could make the issue more complicated.

\begin{figure}[tb!]
\begin{center}
\begin{minipage}[t]{16.5 cm}
\begin{centering}
\epsfig{file=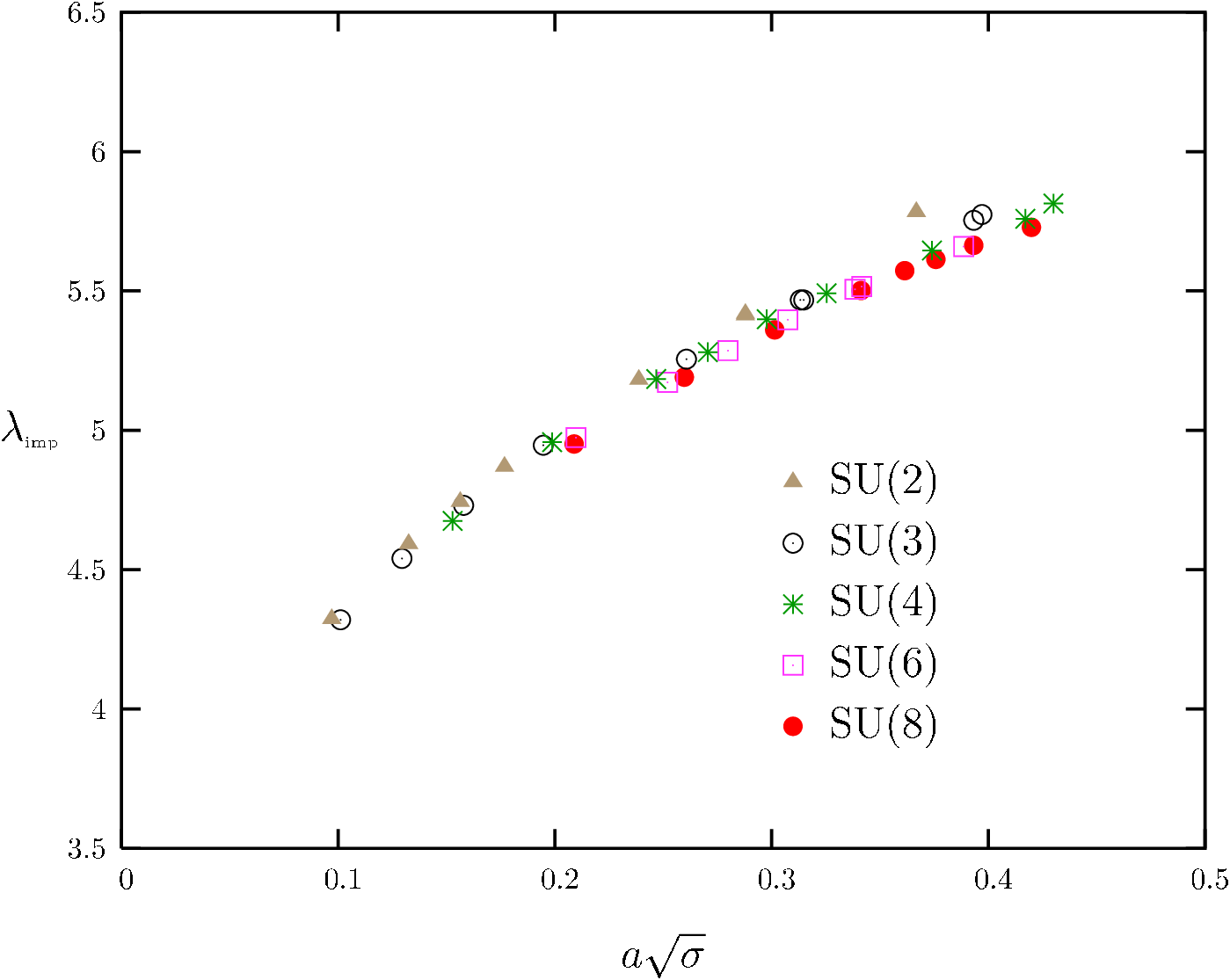,scale=1.1}
\end{centering}
\vspace{5mm}\\
\end{minipage}
\begin{minipage}[t]{16.5 cm}
\caption{Comparing $\SU(N)$ Yang-Mills theories with a different number of colors, the lattice spacing $a$ is seen to depend only on the mean-field improved lattice 't~Hooft coupling (assuming that the scale is set by the value of the string tension $\sigma$). The figure shows numerical results from ref.~\cite{Allton:2008ty}.\label{fig:Trivini}}
\end{minipage}
\end{center}
\end{figure}

On the lattice, there is quite clear evidence that the 't~Hooft coupling $\lambda$ is, indeed, the appropriate one to describe the large-$N$ limit. This can already be seen at the level of the ``bare'' lattice coupling appearing in the definition of $\beta=2N/g^2$, which can be interpreted as a sort of ``physical'' coupling (for the lattice theory) at distances of the order of the lattice spacing $a$. If the large-$N$ limit at fixed 't~Hooft coupling is a physically sensible definition of the large-$N$ limit also at the non-perturbative level, then one would expect that different $\SU(N)$ gauge theories should be characterized by the same dynamically generated $ \LambdaQCD $ scale, provided they are compared at the same value of $\lambda$. Equivalently, the running of the coupling in different $\SU(N)$ theories should be such, that the lattice spacing $a$ should only depend on the coupling $\lambda$, but not on $N$ and $g$ separately. This is indeed observed in numerical results of lattice simulations~\cite{Lucini:2001ej, DelDebbio:2001sj, Allton:2008ty}, in particular if one uses an appropriate ``mean-field improved'' definition of the  lattice coupling~\cite{Parisi:1980pe, Lepage:1992xa}, which reduces the impact of lattice artifacts --- see fig.~\ref{fig:Trivini}, taken from ref.~\cite{Allton:2008ty}.

\begin{figure}[tb!]
\begin{center}
\begin{minipage}[t]{16.5 cm}
\begin{centering}
\epsfig{file=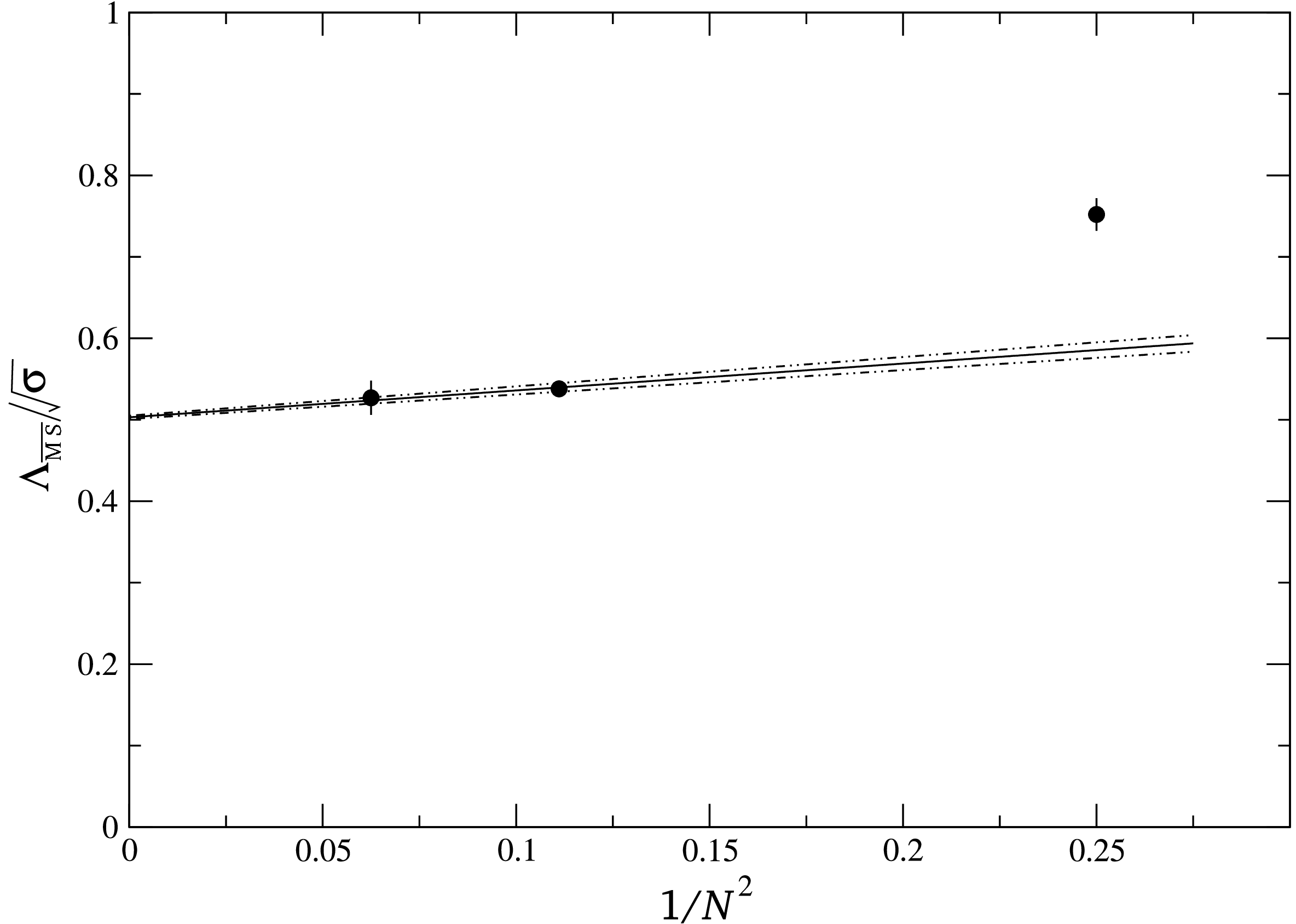,scale=0.6}
\end{centering}
\vspace{5mm}\\
\end{minipage}
\begin{minipage}[t]{16.5 cm}
\caption{The dynamically generated $\LambdaQCD$ scale (in the modified minimal subtraction scheme, and in units of $\sqrt{\sigma}$), shows a mild dependence on the number of colors~\cite{Lucini:2008vi}.\label{fig:Moraitis}}
\end{minipage}
\end{center}
\end{figure}

The definition of a coupling at energy scales lower than the lattice cutoff can be formulated in different schemes. One possibility is given by the Schr\"odinger functional (SF) scheme~\cite{Luscher:1991wu, Luscher:1992an, Sint:1993un}, which is defined in terms of the effective action of a system with fixed boundary conditions in the temporal direction. If the separation between the fixed temporal boundaries of the system is $L$ (and the other directions are taken to be sufficiently large), the SF running coupling at the length scale $L$ can be obtained, by studying the effective action corresponding to different boundary conditions, depending on a certain parameter $\eta$. With this method, the authors of ref.~\cite{Lucini:2008vi} studied the running coupling in $\SU(4)$ Yang-Mills theory, and discussed a large-$N$ extrapolation, comparing their results with those obtained in the $\SU(2)$~\cite{Luscher:1992zx} and $\SU(3)$ theories~\cite{Luscher:1993gh}). The numerical data reveal that the running 
coupling agrees nicely with the two-loop perturbative $\beta$-function for all energies larger than a few hundreds MeV. In addition, the $\LambdaQCD$ scale in the modified minimal subtraction scheme, when expressed in the appropriate units of $\sigma$, has a mild dependence on $N$: the leading corrections are proportional to $1/N^2$, and the value for the theory with three colors is close to the extrapolated large-$N$ limit (see fig.~\ref{fig:Moraitis}).

\begin{figure}[tb!]
\begin{center}
\begin{minipage}[t]{16.5 cm}
\begin{centering}
\hspace{2cm} \epsfig{file=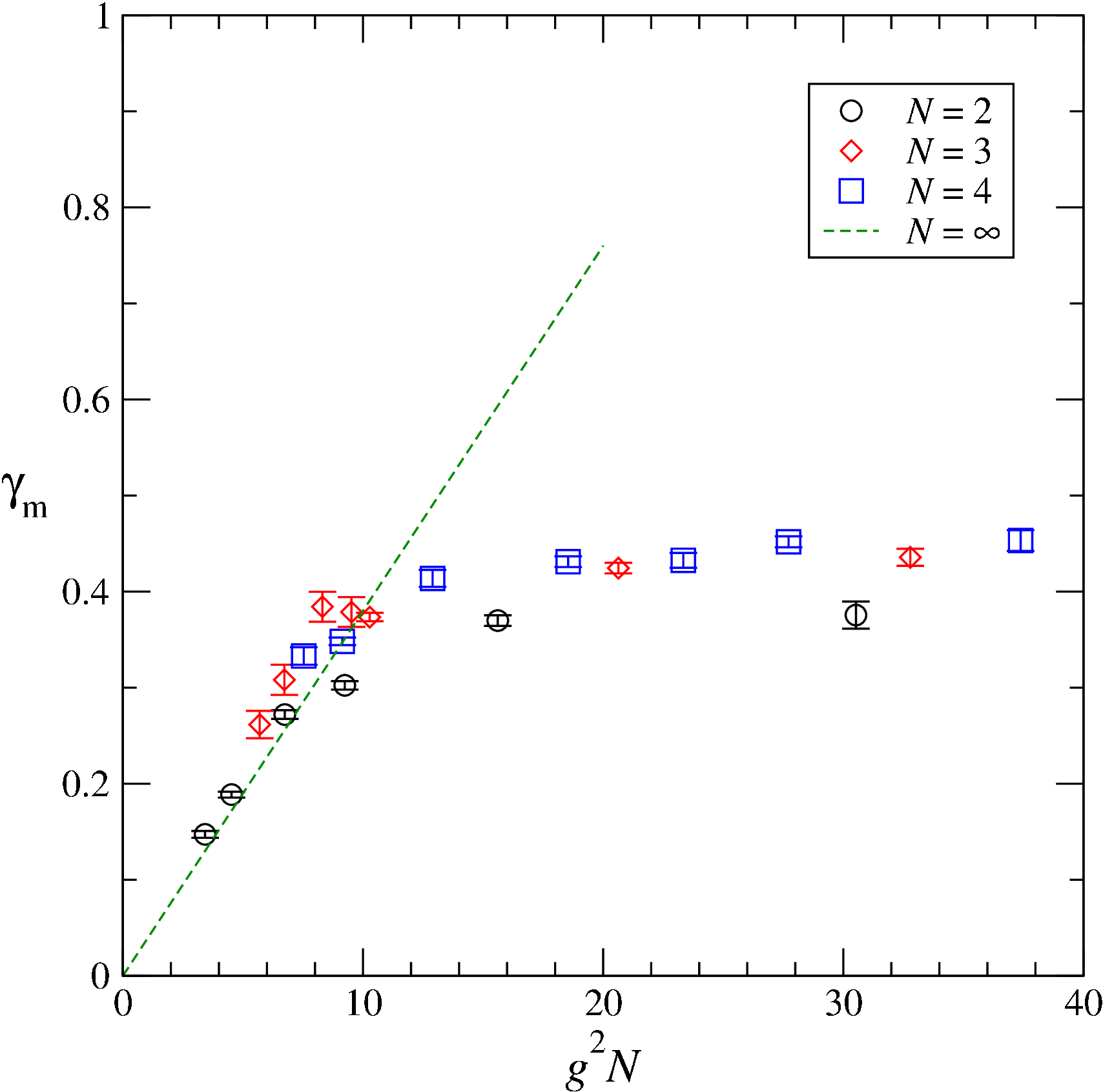,scale=0.65}
\end{centering}
\vspace{5mm}\\
\end{minipage}
\begin{minipage}[t]{16.5 cm}
\caption{The mass anomalous dimension $\gamma_m$ (which describes the dependence of the pseudoscalar renormalization constant $Z_{\mbox{\tiny{PS}}}$ on the inverse energy scale $L$ via $Z_{\mbox{\tiny{PS}}}(L) \propto L^{-\gamma_m}$) in theories with $N=2$, $3$ and $4$ colors and two flavors of dynamical fermions in the two-index symmetric representation, as a function of the 't~Hooft coupling, as determined in ref.~\cite{DeGrand:2012qa}. The dashed line is the leading-order perturbative prediction for $N \to \infty$.\label{fig:Svetitsky}}
\end{minipage}
\end{center}
\end{figure}

In fact, a similarly mild dependence on the number of color charges has also been observed in simulations with dynamical fermions (in the two-index symmetric representation of the gauge group)~\cite{DeGrand:2012qa, DeGrand:2013uha}: as shown in fig.~\ref{fig:Svetitsky}, the dependence of the anomalous dimension $\gamma_m$ on the 't~Hooft coupling is strikingly similar for theories with two, three and four color charges.

Let us now turn to the lattice results that have been obtained for the spectrum of physical states in large-$N$ non-Abelian gauge theory. Computations of the spectrum of glueballs as well as mesons and baryons have been reported in refs.~\cite{Teper:1998kw, Lucini:2001ej, Lucini:2004my, Meyer:2004gx, Meyer:2004jc, DelDebbio:2007wk, Bali:2008an, Hietanen:2009tu, Lucini:2010nv, DeGrand:2012hd, Bali:2013kia, DeGrand:2013nna}: all of them have been carried out in pure Yang-Mills theory, or in the quenched approximation.

\begin{figure}[tb!]
\begin{center}
\begin{minipage}[t]{16.5 cm}
\begin{centering}
\epsfig{file=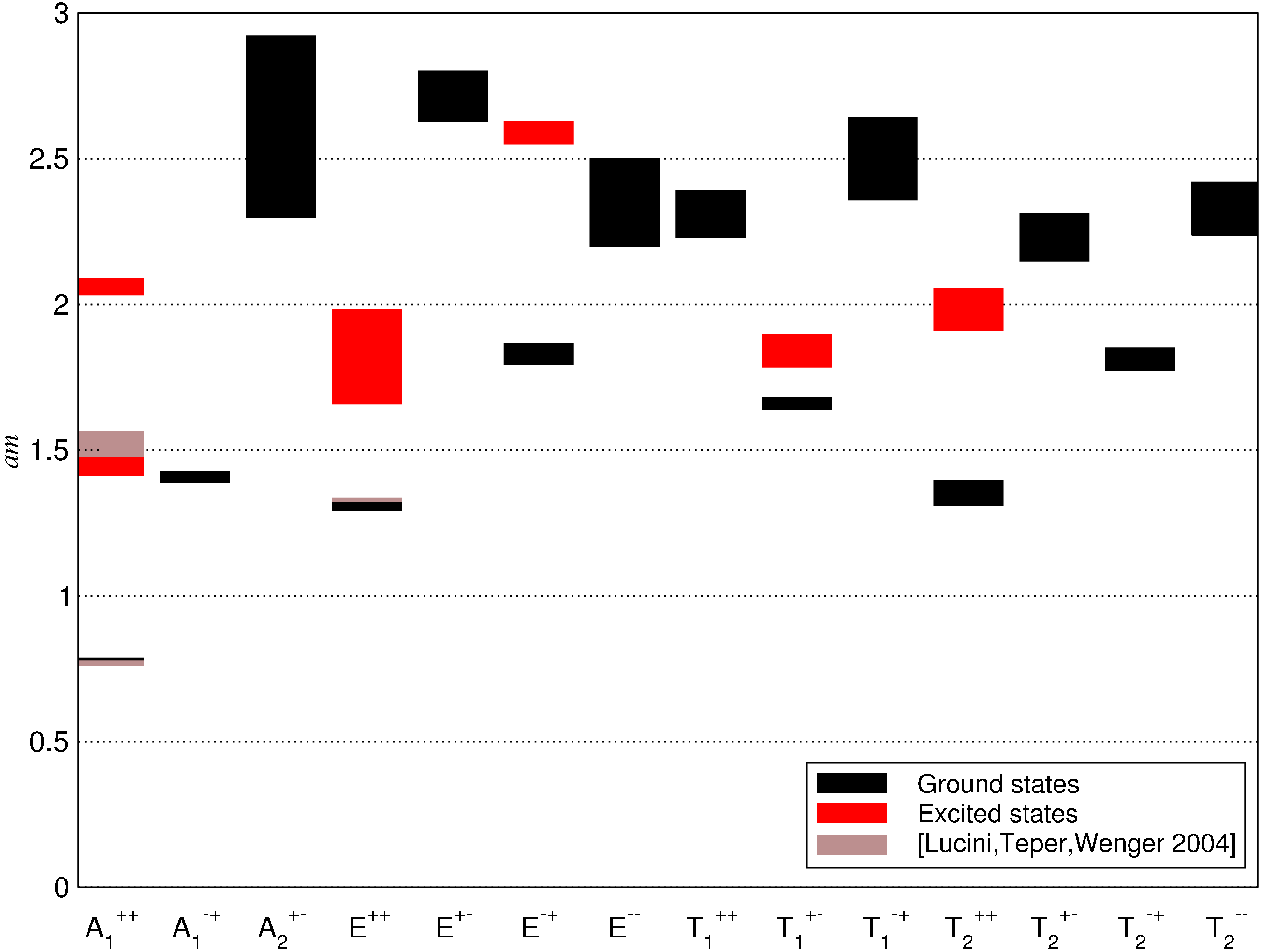,scale=0.55}
\end{centering}
\vspace{5mm}\\
\end{minipage}
\begin{minipage}[t]{16.5 cm}
\caption{Glueball spectrum in the large-$N$ limit of $\SU(N)$ Yang-Mills theory, at a fixed lattice spacing, from ref.~\cite{Lucini:2010nv}. A comparison with the continuum-extrapolated results from ref.~\cite{Lucini:2004my} for the ground state and for the first excited $J^{PC}=0^{++}$ glueball, and for the $J^{PC}=2^{++}$ ground state, is also shown.\label{fig:4D_glueballs}}
\end{minipage}
\end{center}
\end{figure}

Assuming that the different $\SU(N)$ theories are characterized by the same string tension, all glueball masses exhibit a very mild dependence on the number of colors $N$. For all values of $N \ge 3$ (or even for $N=2$), the lattice results for the lightest state in a channel with given quantum numbers can be fitted well by a constant plus a term linear in $1/N^2$, in agreement with the expectation that the leading finite-$N$ corrections are quadratic in $1/N$ for purely gluonic states in Yang-Mills theory. For the lightest states in the spectrum, conclusive results\footnote{These results are compatible with those reported in ref.~\cite{Meyer:2004jc}, within the uncertainties of the calculation (including statistical errors and systematic uncertainties related to technical aspects of the numerical study). Note, however, that in the $\SU(8)$ theory at the smallest simulated lattice spacing Meyer and Teper found a lower mass for the  $0^{++\star}$ state. A clarification of this result would require further investigation.} in the continuum limit have been reported in refs.~\cite{Lucini:2001ej,Lucini:2004my}:
\begin{eqnarray}
 & & \frac{m_{0^{++}}}{\sqrt{\sigma}} = 3.28(8) + \frac{2.1(1.1)}{N^2}, \\
 & & \frac{m_{0^{++\star}}}  {\sqrt{\sigma}} = 5.93(17) - \frac{2.7(2.0)}{N^2} , \\
 & & \frac{m_{2^{++}}}{\sqrt{\sigma}} = 4.78(14) + \frac{0.3(1.7)}{N^2} .
\end{eqnarray}
More recently, heavier states (including some excitations) were studied in ref.~\cite{Lucini:2010nv} (see fig.~\ref{fig:4D_glueballs}). Since this is a computation at just one finite lattice spacing, and no continuum extrapolation was performed, the states are not classified according to the irreducible representations of the group of rotations in continuum tridimensional space, but rather according to the five irreducible representations of the cubic group, as appropriate for a study carried out on a hypercubic lattice. However, the comparison with continuum-extrapolated results from an earlier work~\cite{Lucini:2004my} indicates that these results are already quite close to the continuum limit.

\begin{figure}[tb!]
\begin{center}
\begin{minipage}[t]{16.5 cm}
\begin{centering}
\hspace{1cm} \epsfig{file=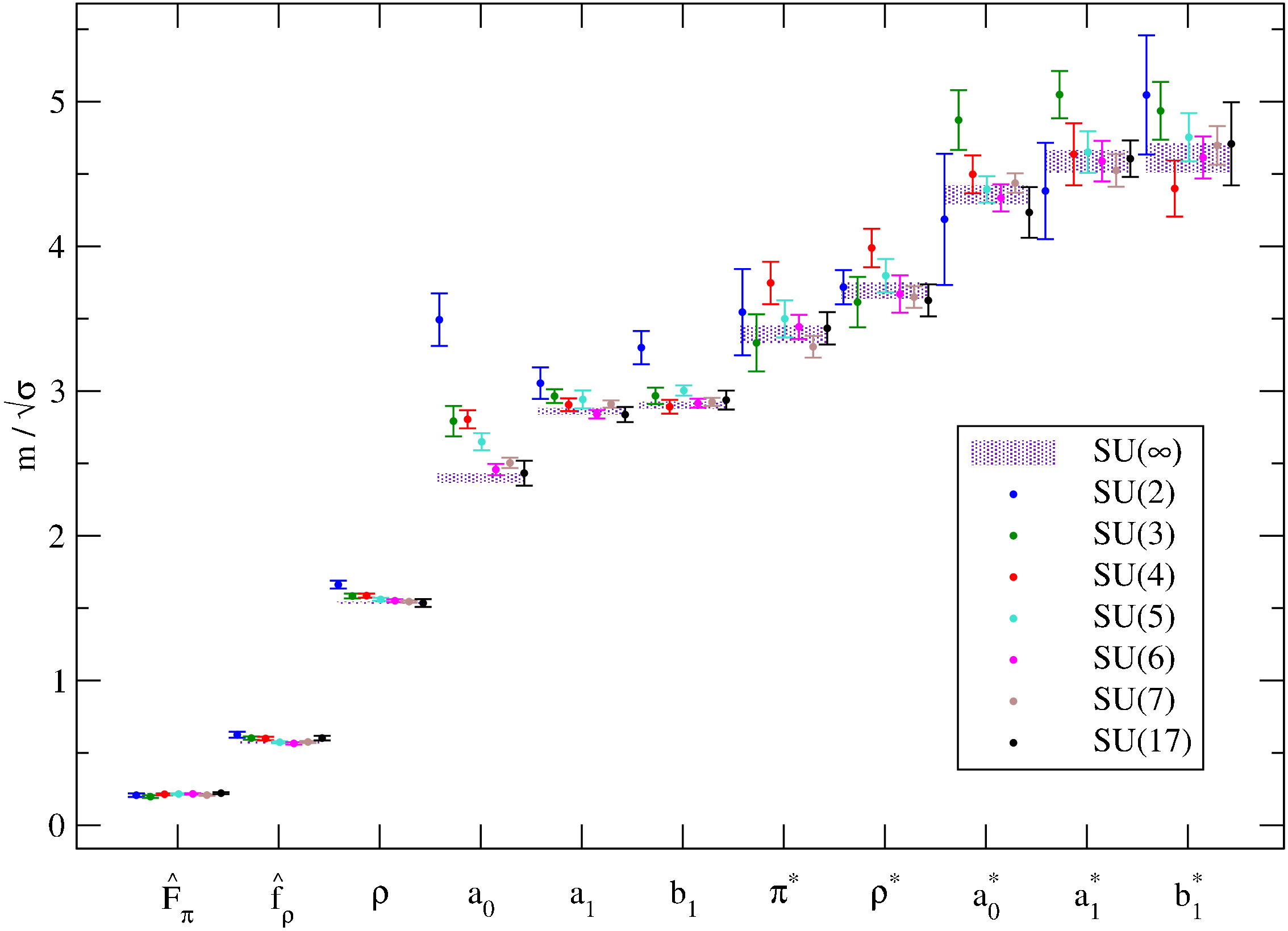,scale=0.555}
\end{centering}
\vspace{5mm}\\
\end{minipage}
\begin{minipage}[t]{16.5 cm}
\caption{Masses (and some decay constants) of mesons in large-$N$ QCD, in the chiral limit, from ref.~\cite{Bali:2013kia}. The horizontal bands denote the extrapolations to the 't~Hooft limit.\label{fig:meson_spectrum}}
\end{minipage}
\end{center}
\end{figure}

The meson spectrum at large $N$ has been studied in refs.~\cite{DelDebbio:2007wk, Bali:2008an, Hietanen:2009tu, DeGrand:2012hd, Bali:2013kia}. The results from four of these studies~\cite{DelDebbio:2007wk, Bali:2008an, DeGrand:2012hd, Bali:2013kia} consistently indicate a smooth approach to the 't~Hooft limit for the masses of different states (including some excitations) and the decay constants. This is clearly shown in fig.~\ref{fig:meson_spectrum}, taken from the most recent work~\cite{Bali:2013kia}: symbols of different colors correspond to different values of $N$, while the band denotes the extrapolation to the 't~Hooft limit. These results confirm that, in the large-$N$ limit, the pion and $\rho$ masses are close to those in the real world:
\begin{equation}
\lim_{N \to \infty} \frac{m_\rho}{\sqrt{\sigma}} = 1.79(5),
\end{equation}
and in reasonable agreement with the holographic models reviewed in ref.~\cite{Erdmenger:2007cm}. In ref.~\cite{Bochicchio:2013sra}, large-$N$ lattice results for mesons~\cite{Bali:2013kia} and glueballs~\cite{Meyer:2004jc, Meyer:2004gx} have been shown to be in agreement with the predictions of a topological field theory underlying the large-$N$ limit of pure Yang-Mills~\cite{Bochicchio:2013eda}.

The other lattice study of the large-$N$ meson spectrum~\cite{Hietanen:2009tu} (see also ref.~\cite{Narayanan:2005gh}), however, found incompatible results, and came to the conclusion that the mass of the $\rho$ meson in the large-$N$ limit would be much larger than in the theory with $N=3$ colors. It is possible that this discrepancy with the other studies may be due to uncontrolled systematic errors, related, in particular, to contamination from excited states in the momentum-space evaluation of quark propagators carried out in ref.~\cite{Hietanen:2009tu}.

\begin{figure}[tb!]
\begin{center}
\begin{minipage}[t]{16.5 cm}
\begin{centering}
\epsfig{file=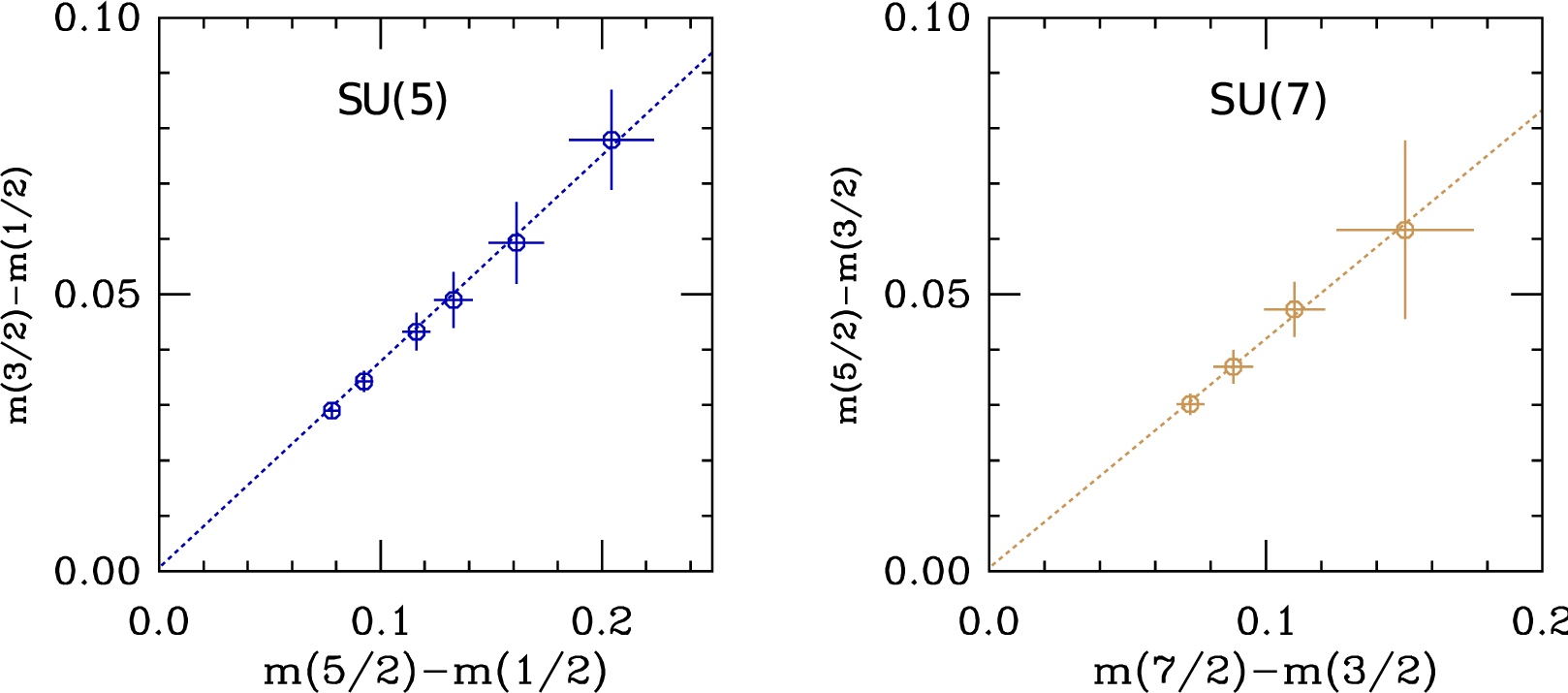, scale=0.9}
\end{centering}
\vspace{5mm}\\
\end{minipage}
\begin{minipage}[t]{16.5 cm}
\caption{Evidence for a rotor-type spectrum~\cite{Adkins:1983ya, Jenkins:1993zu} in the lattice results for baryons, in theories with $N=5$ (left-hand side panel) and $7$ (right-hand side panel) color charges, from ref.~\cite{DeGrand:2012hd}.\label{fig:baryon_spectrum}}
\end{minipage}
\end{center}
\end{figure}

The large-$N$ baryonic spectrum (for odd values of $N$) was studied in refs.~\cite{DeGrand:2012hd, DeGrand:2013nna}. In particular, it was shown that baryon masses are approximately linear in $N$, and that the masses of states of different spin are compatible with a rotor spectrum, as first predicted thirty years ago in ref.~\cite{Adkins:1983ya} (see also ref.~\cite{Jenkins:1993zu}). This is shown in fig.~\ref{fig:baryon_spectrum}, where the mass splittings between baryons of different spin are plotted against each other. 

An interesting comparison of lattice results with large-$N$ predictions for baryons was also presented in ref.~\cite{Jenkins:2009wv}, in which a set of configurations in $N=3$~QCD (with dynamical fermions)~\cite{WalkerLoud:2008bp} was used to test the baryon mass splitting predicted in a $1/N$-expansion~\cite{Jenkins:1995td}. By varying flavor-breaking terms via a change in the quark mass, this work found that the results from lattice simulations of QCD with $N=3$~colors are consistent with $1/N$-flavor scaling laws. Related ideas were also discussed in ref.~\cite{WalkerLoud:2011ab}.

\begin{figure}[tb!]
\begin{center}
\begin{minipage}[t]{12 cm}
\begin{centering}
\hspace{-5mm} \epsfig{file=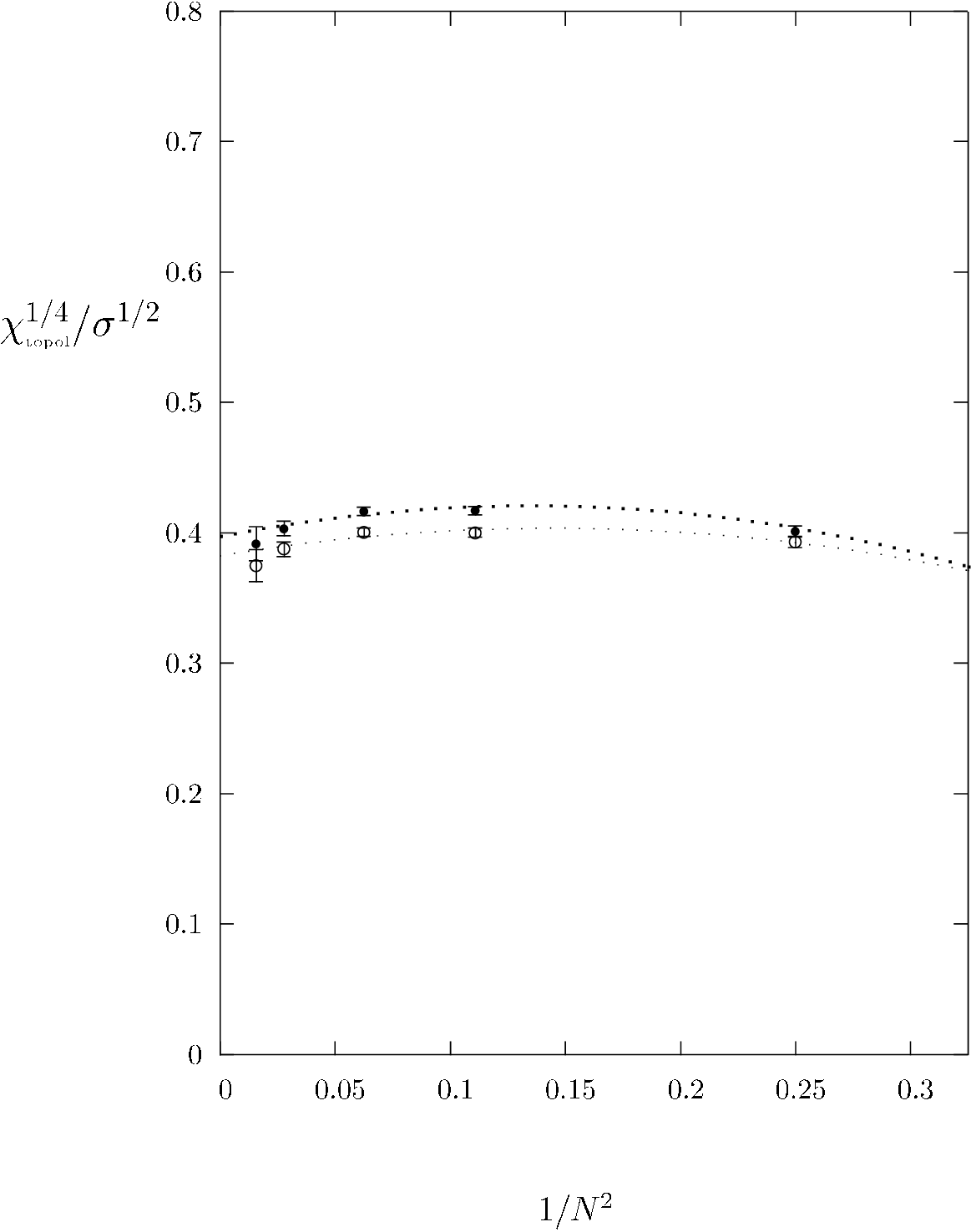}
\end{centering}
\vspace{5mm}\\
\end{minipage}
\begin{minipage}[t]{16.5 cm}
\caption{The results presented in ref.~\cite{Lucini:2004yh} for (the fourth root of) the topological susceptibility $\topsusc$, expressed in units of (the square root of) the string tension $\sigma$, in Yang-Mills theory with a different number of colors, from $2$ to $8$, show a smooth dependence on $N$, and a finite large-$N$ limit, close to the value corresponding to the value in the theory with $N=3$ color charges. The open and closed symbols refer to slightly different definitions of the lattice topological charge operator (see ref.~\cite{Lucini:2004yh} for details). The figure also shows the corresponding fits in $1/N^2$.\label{fig:fig1_heplat0401028}}
\end{minipage}
\end{center}
\end{figure}

The topological properties of large-$N$ QCD at zero temperature have been investigated in various lattice studies~\cite{Lucini:2001ej, Lucini:2001rc, Cundy:2002hv, DelDebbio:2002xa, Lucini:2004yh, DelDebbio:2006df}, and are discussed in the review~\cite{Vicari:2008jw}. The main findings are:
\begin{itemize}
\item The number density of instantons is exponentially suppressed when $N$ becomes large (as predicted by general arguments~\cite{Witten:1978bc}), and, for fixed $N$, the density of instantons of small radius $\rho$ scales like $\rho^{\frac{11}{3}N-5}$~\cite{Lucini:2001ej}.
\item The topological susceptibility tends to a non-vanishing value for $N \to \infty$:
\begin{equation}
\label{topsusc_Biagio_Mike}
\frac{ \topsusc^{1/4} }{ \sigma^{1/2} } = 0.376(20) + \frac{0.43(10)}{N^2}.
\end{equation}
\end{itemize}
Similar values were also reported in ref.~\cite{DelDebbio:2002xa}:
\begin{equation}
\label{topsusc_Luigi_Haris_Ettore}
\frac{ \topsusc^{1/4} }{ \sigma^{1/2} } =  0.386(6) + \frac{0.24(8)}{N^2},
\end{equation}
and in ref.~\cite{Lucini:2004yh}:
\begin{equation}
\label{topsusc_Biagio_Mike_Urs}
\frac{ \topsusc^{1/4} }{ \sigma^{1/2} } = 0.382(7) + \frac{0.30(13)}{N^2} - \frac{1.02(42)}{N^4}.
\end{equation}
The results obtained in ref.~\cite{Lucini:2004yh} are shown in fig.~\ref{fig:fig1_heplat0401028}.

In short, these studies indicate that, in the 't~Hooft limit of QCD, the topological susceptibility has a non-vanishing value about $(170~\mbox{MeV})^4$. This is quite close to the value in the physical case of $N=3$ colors, for which $\topsusc \simeq(180~\mbox{MeV})^4$.

The lattice studies reviewed so far addressed the setup of theories at zero temperature. There exist also a number of works investigating the finite-temperature properties via Monte Carlo simulations~\cite{Lucini:2002ku, Lucini:2003zr, Lucini:2005vg, Bursa:2005yv, Bringoltz:2005rr, Bringoltz:2005xx, deForcrand:2005rg, Panero:2008mg, Panero:2009tv, Datta:2009jn, Datta:2010sq, Mykkanen:2012ri, Lucini:2012wq, Bonati:2013tt}. These works give convincing evidence that all $\SU(N)$ Yang-Mills theories undergo a physical deconfining transition at a critical temperature $T_c$, which remains finite when expressed in some appropriately defined non-perturbative scale of the theory (e.g. the zero-temperature string tension $\sigma$). In particular, the deconfinement transition can be associated with the spontaneous breakdown of the exact global center symmetry of the pure-glue theory, and takes place at temperatures which, if the scale is expressed in physical units, are in the range between $250$ and $300~\mbox{MeV}$, depending on the number of colors. Note that, by contrast, center symmetry is not an exact global symmetry in QCD with physical quarks, since the latter break it \emph{explicitly}. In addition, the finiteness of the light quark masses implies that chiral symmetry is explicitly broken, too. As a consequence, the deconfinement transition in real-world QCD is rather an analytic crossover, taking place at temperatures in the ballpark of $160~\mbox{MeV}$~\cite{Aoki:2006br, Aoki:2009sc, Bazavov:2010sb, Bazavov:2011nk} (see also refs.~\cite{DeTar:2009ef, Petreczky:2012rq, Philipsen:2012nu} for reviews). Nevertheless, the pure-glue setup is an interesting theoretical laboratory, in which the deconfinement transition at finite temperature can be analyzed unambiguously, and captures most of the physically relevant features of the phenomenon. The analysis of large-$N$ gauge theories at finite temperature is particularly interesting, as there exist a number of important implications~\cite{Thorn:1980iv, Gocksch:1982en, Greensite:1982be, Pisarski:1983db, McLerran:1985uh, Toublan:2005rq}.

\begin{figure}[tb!]
\begin{center}
\begin{minipage}[t]{16.5 cm}
\begin{centering}
\hspace{5mm} \epsfig{file=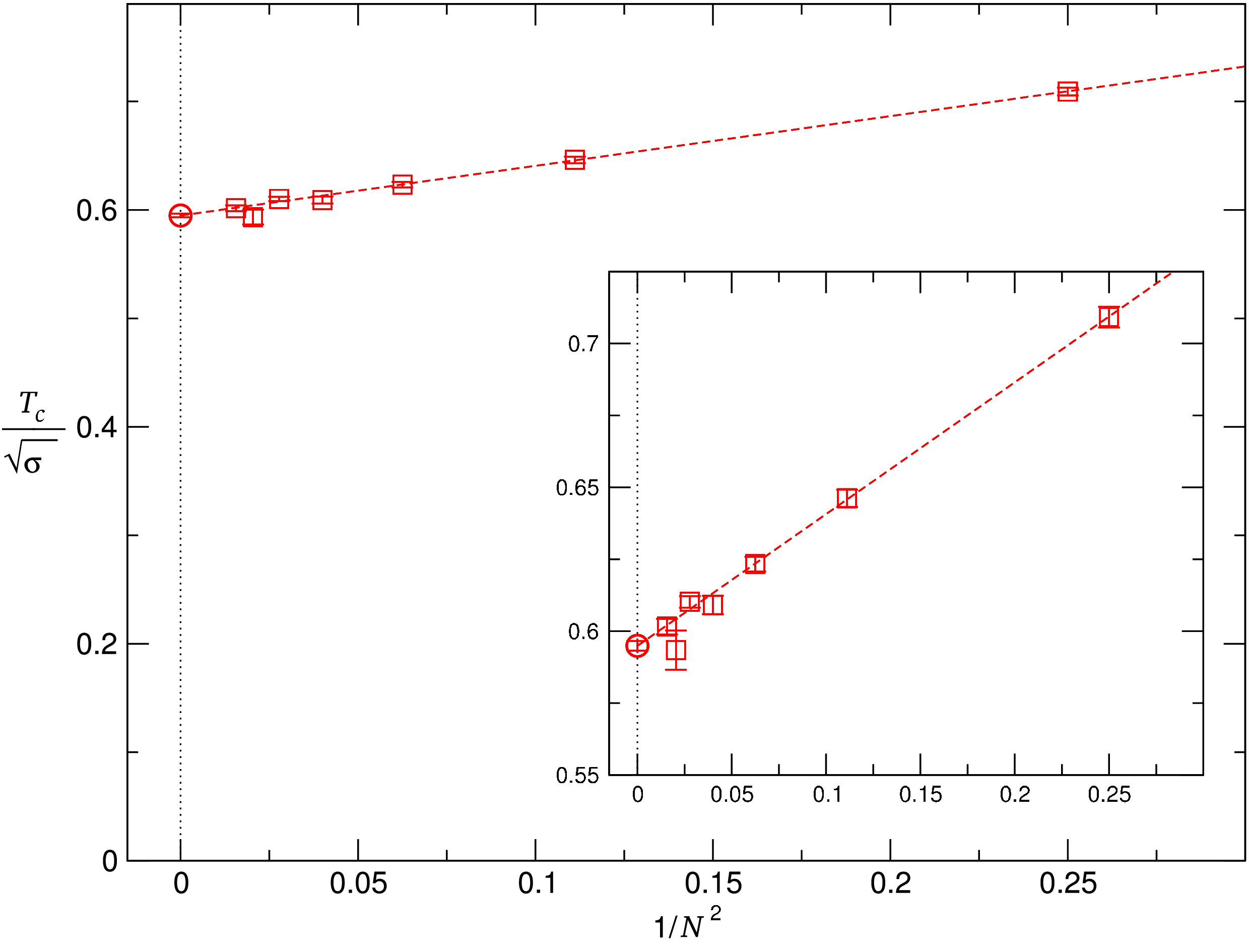,scale=0.55}
\end{centering}
\vspace{5mm}\\
\end{minipage}
\begin{minipage}[t]{16.5 cm}
\caption{The critical temperature associated with deconfinement in $\SU(N)$ Yang-Mills theory is a slowly decreasing function of the number of colors $N$, with a finite large-$N$ limit~\cite{Lucini:2012wq}. The plot shows the numerical results and their interpolation according to eq.~(\protect\ref{large_N_Tc}). A zoomed view of the simulation results is displayed in the inset.\label{fig:Tc_over_root_sigma}}
\end{minipage}
\end{center}
\end{figure}

Lattice results indicate that the finite-temperature deconfinement transition is of second order for $N=2$ colors~\cite{Engels:1990vr, Fingberg:1992ju, Engels:1994xj}. According to a conjecture due to Svetitsky and Yaffe~\cite{Svetitsky:1982gs}, it is then expected that the behavior of the theory at the critical point should be described in terms of a model in the same universality class as those of a spin model, with degrees of freedom taking values in the center of the gauge group, in one dimension less. For $\SU(2)$ Yang-Mills theory the critical exponents are indeed consistent with those of the corresponding spin model, i.e. the Ising model in three spatial dimensions~\cite{condmat0012164}. For larger values of $N$, the deconfinement transition becomes a discontinuous (i.e. first order) one: this is seen in the $\SU(3)$ theory~\cite{Boyd:1996bx, Borsanyi:2012ve} and---even more clearly---for all $\SU(N)$ theories with $N \ge 4$~\cite{Lucini:2002ku, Lucini:2003zr, Lucini:2005vg, Datta:2009jn}. Intuitively, the change to a (more and more strongly) first-order transition as the number of color charges is increased can be interpreted in terms of a more and more ``violent'' transition, which takes place at the temperature where the free energies of a gas of glueballs (whose number is $O(N^0)$) and of gluons (with $O(N^2)$ degrees of freedom) become equal. Correspondingly, it is also found that the critical temperature is a slightly decreasing function of the number of colors~\cite{Lucini:2012wq}:
\begin{equation}
\label{large_N_Tc}
\frac{T_c}{\sqrt{\sigma}} = 0.5949(17) + \frac{0.458(18)}{N^2}, \qquad \mbox{with: $\chi^2$/d.o.f.}=1.18
\end{equation}
(see also fig.~\ref{fig:Tc_over_root_sigma}).

The first-order nature of the deconfinement transition for $N \ge 3$ is associated to the finiteness of the latent heat $L_h$, which scales like $O(N^2)$ in the large-$N$ limit~\cite{Lucini:2005vg, Panero:2009tv}:
\begin{equation}
\label{latent_heat}
\lim_{N \to \infty} \frac{ L_h^{1/4} }{ N^{1/2} T_c} = 0.759(19).
\end{equation}
Similarly, a first-order deconfinement transition also implies a non-vanishing value for the surface tension associated with interfaces between the confining and deconfined phases~\cite{Lucini:2005vg}
\begin{equation}
\label{confined_deconfined_interface_tension}
\frac{\gamma_W^{c \to d}}{N^2 T_c^3} = 0.0138(3) - \frac{0.104(3)}{N^2}, \qquad \mbox{with: $\chi^2$/d.o.f.}=2.7.
\end{equation}
This quantity is related to the surface tension of 't~Hooft loops~\cite{Bhattacharya:1990hk, Enqvist:1990ae, Bhattacharya:1992qb, KorthalsAltes:1993ca, Giovannangeli:2001bh, Giovannangeli:2002uv, Giovannangeli:2004sg}) and might possibly be of phenomenological interest~\cite{Asakawa:2012yv}, although the relevance of center domains in Minkowski spacetime has been debated~\cite{Smilga:1993vb}. On the other hand, an interesting technical aside of the strongly first-order nature of the deconfinement transition for lattice studies is that it implies suppression of tunneling events between different center sectors---but also of finite-volume effects (see refs.~\cite{Elze:1988zs, Gliozzi:2007jh, Panero:2008mg} for a discussion). Other lattice works studying 't~Hooft loops in large-$N$ Yang-Mills theories at finite temperature include refs.~\cite{Bursa:2005yv, deForcrand:2005rg}.

\begin{figure}[tb!]
\begin{center}
\begin{minipage}[t]{16.5 cm}
\begin{centering}
\hspace{1cm} \epsfig{file=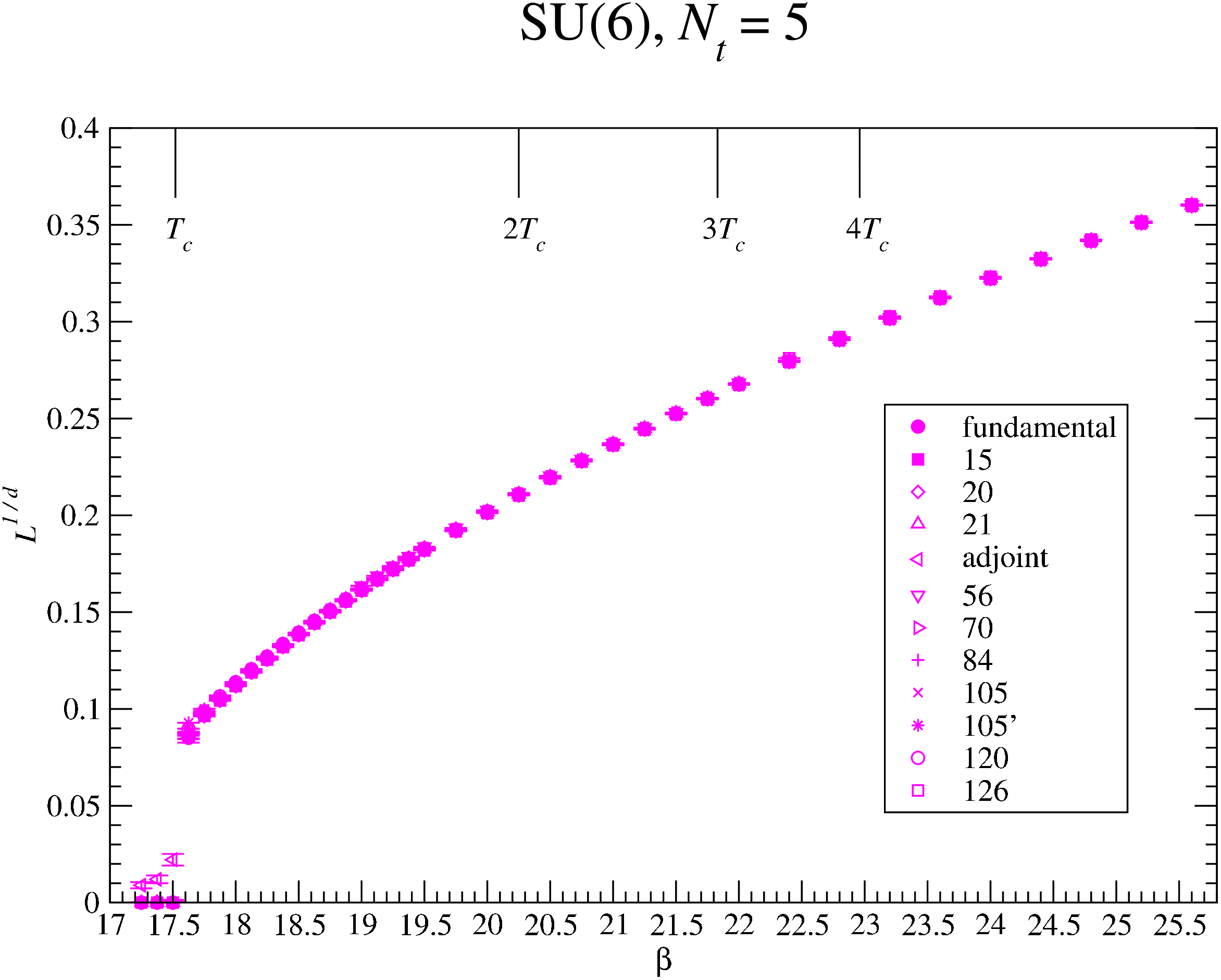,scale=0.55}
\end{centering}
\vspace{5mm}\\
\end{minipage}
\begin{minipage}[t]{16.5 cm}
\caption{Bare Polyakov loops in different irreducible representations in $\SU(N)$ Yang-Mills theory, with their free energies rescaled according to the assumption of perfect Casimir scaling of the corresponding representation, fall on the same curve (indicating consistency with Casimir scaling). The plot shows the results obtained in ref.~\cite{Mykkanen:2012ri} from simulations of the $\SU(6)$ theory, at finite lattice spacing $a=1/(5T)$. The data are plotted as a function of the lattice action parameter $\beta$. For reference, the corresponding values of the temperature are displayed on the upper horizontal axis.\label{fig:bare_Polyakov_loops}}
\end{minipage}
\end{center}
\end{figure}

\begin{figure}[tb!]
\begin{center}
\begin{minipage}[t]{16.5 cm}
\begin{centering}
\hspace{1cm} \epsfig{file=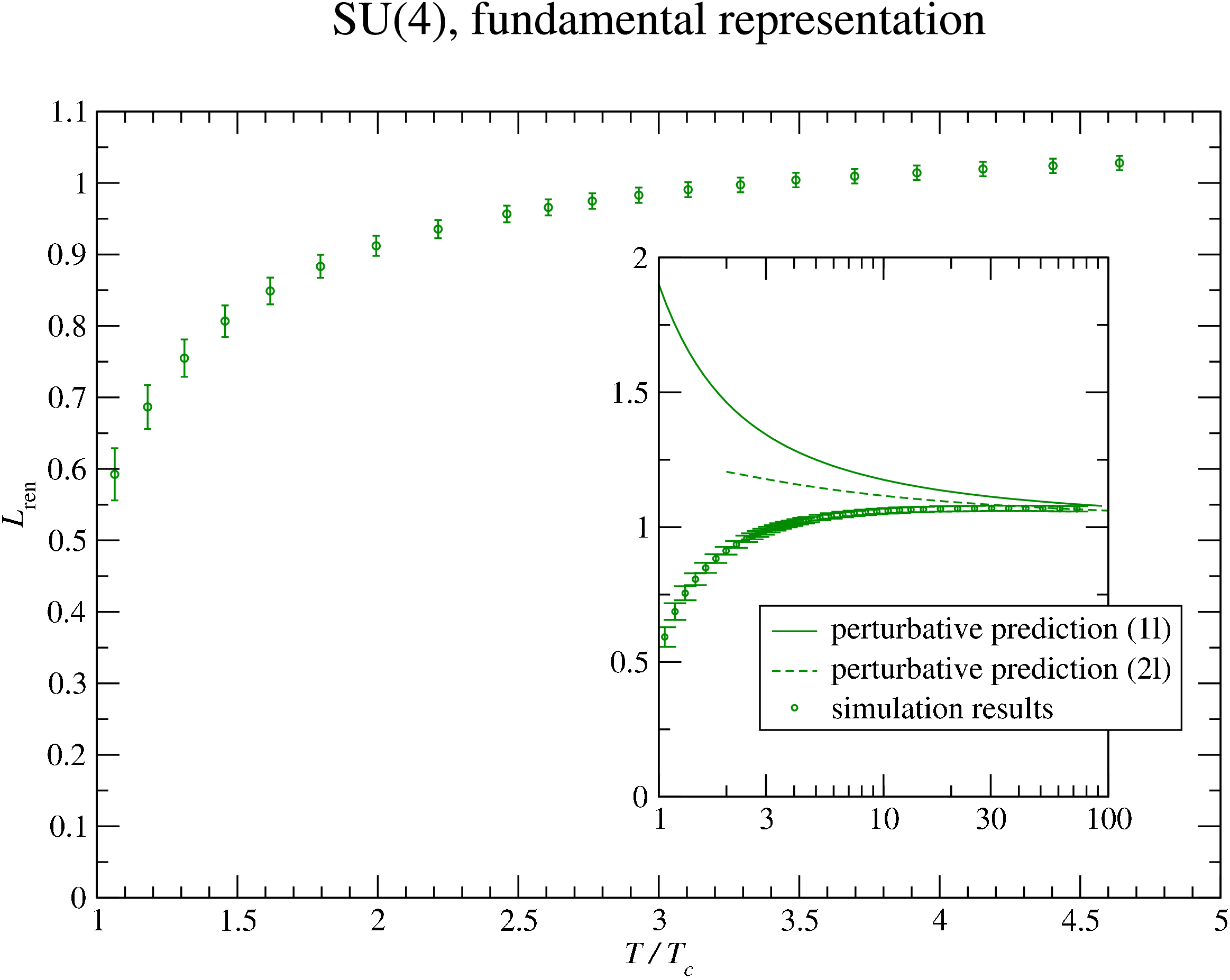,scale=0.55}
\end{centering}
\vspace{5mm}\\
\end{minipage}
\begin{minipage}[t]{16.5 cm}
\caption{The renormalized fundamental Polyakov loop computed in ref.~\cite{Mykkanen:2012ri} for $\SU(N)$ Yang-Mills theories agrees well with the perturbative predictions~\cite{Burnier:2009bk, Brambilla:2010xn} at high temperature, while it exhibits large deviations in the range of temperatures close to $T_c$ (where the physical coupling is larger). The figure shows numerical results for the $\SU(4)$ theory, in comparison with one- (solid line) or two-loop (dashed line) weak-coupling predictions.\label{fig:renormalized_Polyakov_loops}}
\end{minipage}
\end{center}
\end{figure}

A detailed study of the order parameter associated with the finite-temperature deconfinement transition (the Polyakov loop) was presented in ref.~\cite{Mykkanen:2012ri}, where theories with different numbers of colors, from $2$ to $6$, and loops in different irreducible representations of the gauge group, were considered. In particular, this work showed that the free energies of bare Polyakov loops in different representations satisfy Casimir scaling~\cite{Damgaard:1987wh} very accurately, even at temperatures close to $T_c$, for all the gauge groups considered, as shown in fig.~\ref{fig:bare_Polyakov_loops}. In addition, it also showed that the high-temperature behavior of renormalized Polyakov loops is consistent with weak-coupling expansions~\cite{Burnier:2009bk, Brambilla:2010xn}, while large non-perturbative contributions are present at temperatures close to deconfinement---see fig.~\ref{fig:renormalized_Polyakov_loops}. These results are consistent with studies of the $\SU(3)$ theory previously reported in refs.~\cite{Dumitru:2003hp, Gupta:2007ax}.

\begin{figure}[tb!]
\begin{center}
\begin{minipage}[t]{16.5 cm}
\begin{centering}
\hspace{18mm} \epsfig{file=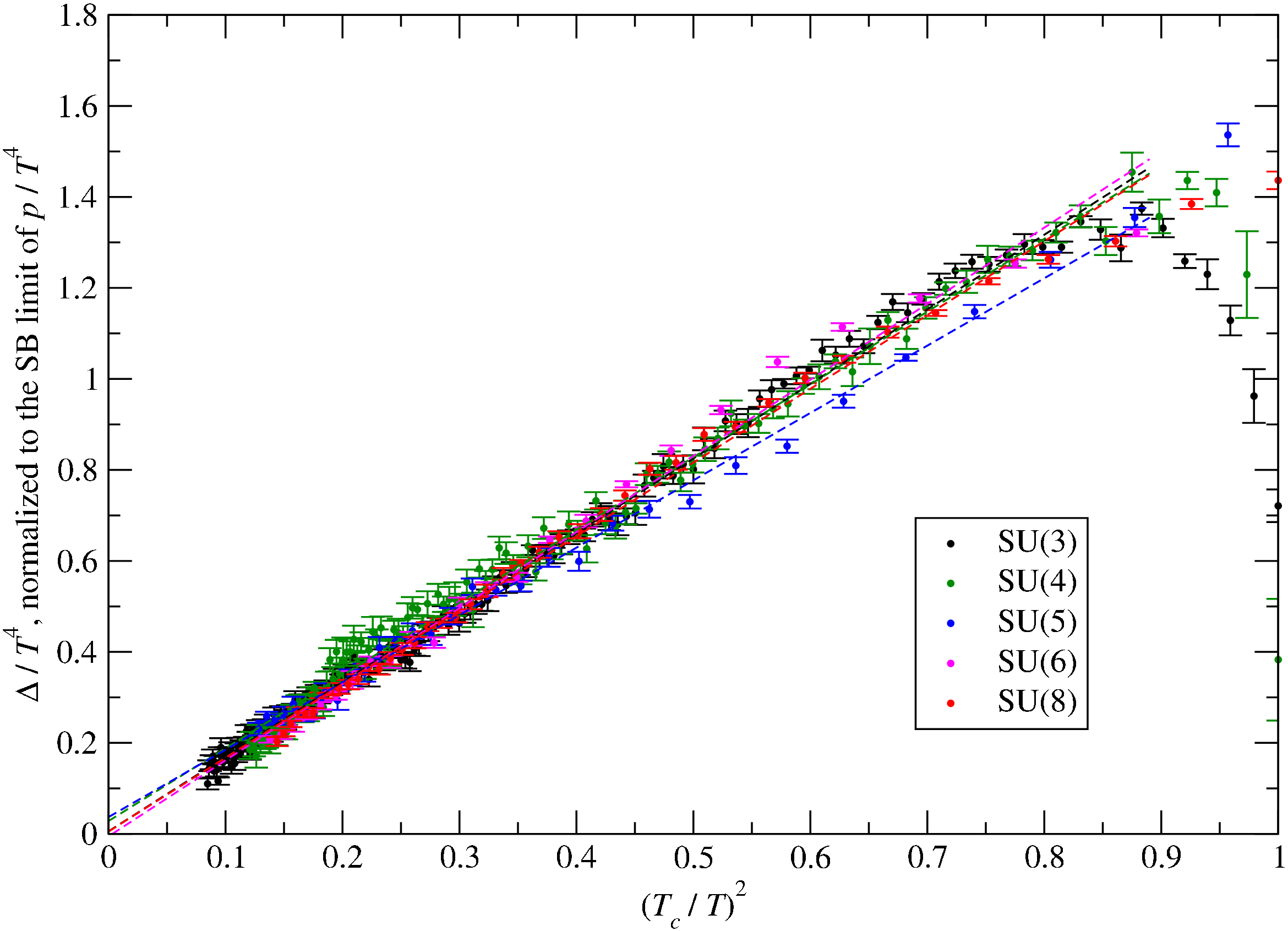,scale=0.5}
\end{centering}
\vspace{5mm}\\
\end{minipage}
\begin{minipage}[t]{16.5 cm}
\caption{For temperatures close to deconfinement (in the deconfined phase), the trace anomaly $\Delta$ is nearly perfectly proportional to $T^2$. The figure, taken from ref.~\cite{Panero:2009tv}, shows the linear dependence of the dimensionless ratio $\Delta/T^4$ per gluon degree of freedom (with an appropriate normalization) on $1/T^2$.\label{fig:Delta_T2}}
\end{minipage}
\end{center}
\end{figure}

The dependence of the free energy of non-Abelian gauge theories on the temperature (or the ``equation of state'') has been studied via lattice simulations in refs.~\cite{Bringoltz:2005rr, Bringoltz:2005xx, Panero:2008mg, Panero:2009tv, Datta:2009jn, Datta:2010sq}: in the deconfined phase at $T > T_c$, all equilibrium thermodynamic quantities (such as the pressure $p$, the trace anomaly $\Delta$, and the energy and entropy densities $\epsilon$ and $s$) are exactly proportional to the number of gluon degrees of freedom, $2(N^2-1)$, with essentially the same dependence on $T/T_c$ in all $\SU(N \ge 3 )$ theories. In fact, the dependence of the equation of state on the number of colors (up to the trivial gluon multiplicity factor) appears to be even milder than for other quantities, so that the equilibrium thermodynamics properties of the ``physical'' theory with $N=3$ colors are basically the same as those of the large-$N$ theory. This result is particularly interesting, and relevant for studies of the QCD plasma based on holographic models~\cite{Son:2007vk, Mateos:2007ay, Shuryak:2008eq, Gubser:2009md, Rangamani:2009xk, CasalderreySolana:2011us}, which implicitly rely on the approximation of an infinite number of colors. In fact, ref.~\cite{Panero:2009tv} reported good agreement between lattice results and holographic computations carried out both in top-down~\cite{Gubser:1996de, Gubser:1998nz} and in bottom-up approaches~\cite{Gursoy:2007cb, Gursoy:2007er, Gursoy:2008bu, Gursoy:2009jd} (see also refs.~\cite{Andreev:2006vy, Kajantie:2006hv, Alanen:2009ej, Alanen:2009xs, Galow:2009kw, Megias:2010ku, Veschgini:2010ws, Jarvinen:2011qe, Alho:2012mh, Mia:2013pca, Arean:2013tja} for related gauge/gravity computations. A similar type of comparison (but considering lattice results for QCD with $N=3$ color charges and including dynamical quarks) was also performed in ref.~\cite{Gubser:2006qh}. Another finding discussed in refs.~\cite{Panero:2009tv, Datta:2010sq} is that, as shown in fig.~\ref{fig:Delta_T2}, at temperatures of the same order of magnitude of $T_c$, the trace anomaly $\Delta$ in the 
deconfined phase seems to be approximately proportional to $T^2$: a behavior possibly due to non-perturbative effects, which has been studied in various phenomenological models~\cite{Meisinger:2001cq, Megias:2005ve, Pisarski:2006yk, Andreev:2007zv, Brau:2009mp, Megias:2009mp, Giacosa:2010vz, Gogokhia:2010dw, Lacroix:2012pt, Dumitru:2012fw}.

Other phenomenologically interesting quantities for the quark-gluon plasma were investigated in the large-$N$ limit in ref.~\cite{Lucini:2005vg}. This work presents results for the Debye mass $m_D$, which describes the phenomenon of color charge screening in the deconfined plasma, and the spatial string tension $\sigma_s$, which characterizes the non-perturbative nature of the physics of ultra-soft modes in the QGP \emph{at all temperatures}. Both quantities turn out to be essentially independent of $N$ over a broad temperature range. Refs.~\cite{Lucini:2004yh, DelDebbio:2004rw, D'Elia:2012vv, Bonati:2013tt}, on the other hand, focused on the interplay between finite temperature and topological properties of $\SU(N)$ Yang-Mills theories; in particular, it was shown that the topological susceptibility is strongly suppressed in the deconfined phase, where it vanishes for $N \to \infty$. Recent works discussing related topics from an analytical point of view include refs.~\cite{Unsal:2012zj, Poppitz:2012nz, Anber:2013sga}.

Although a deconfined state of matter is also expected to exist in QCD at large net quark density~\cite{Itoh:1970uw, Collins:1974ky}, and a number of very interesting expectations have been formulated~\cite{Alford:1998mk, Son:1998uk, Berges:1998rc, Pisarski:1999bf, Pisarski:1999tv, Schafer:1999pb, Rapp:1999qa, Rajagopal:2000wf, Stephanov:2004wx} (including some indicating that the large-$N$ theory could exhibit intriguing novel features~\cite{Cohen:2004cd, Cohen:2004mw, Ohnishi:2006hs, McLerran:2007qj, McLerran:2008ua, Zhitnitsky:2008ha, Glozman:2008fk, Hidaka:2008yy, Kojo:2009ha, Hidaka:2010ph, Torrieri:2010gz, Adam:2010ds, Kojo:2010fe, Torrieri:2011dg, Lottini:2011zp, Kojo:2012hf}), unfortunately there exist no lattice simulation results at large $N$, since lattice QCD at finite density is hampered by the so-called sign problem: in the presence of a finite quark chemical potential $\mu$, the determinant of the Dirac operator is generically complex, and importance sampling in Monte Carlo integration fails~\cite{deForcrand:2010ys, Gupta:2011ma, Philipsen:2011zx, Levkova:2012jd, Aarts:2013bla, Gattringer_Lattice_2013}. Finally, some features of the QCD vacuum and phase diagram can depend on the presence of strong electromagnetic fields~\cite{Mizher:2010zb, Chernodub:2010qx, Agasian:2008tb, Kharzeev:2009fn}, and the large-$N$ limit may have interesting implications for the related phenomena~\cite{Fraga:2012ev, Fraga:2012rr, Anber:2013tra}. However, for the time being, the lattice investigation of QCD under strong electromagnetic fields is limited to QCD with $N=3$ colors~\cite{D'Elia:2010nq, Bali:2011qj, D'Elia:2012zw, Bali:2013esa, Bruckmann:2013oba}.

\subsection{Lattice results for large-$N$ gauge theories in lower spacetime dimensions}
\label{subsec:results_less_than_4D}

Non-Abelian gauge theories in three spacetime dimensions have a gauge coupling with the dimensions of the square root of an energy. They are linearly confining at large distances and become weakly interacting at short distances, with a logarithmic Coulomb potential. They are also renormalizable (in fact, superrenormalizable: the number of divergent diagrams is finite). Hence, they share many qualitative features with non-Abelian gauge theories in four spacetime dimensions, and can serve as useful toy models for QCD.

The lattice formulation of non-Abelian gauge theories in three spacetime dimensions involves the parameter $\beta$, which is here defined as $\beta=2N/(ag^2)=2N^2/(a\lambda)$, with $\lambda$ the dimensionful 't~Hooft coupling. Much like in the four-dimensional case, the phase structure of $\SU(N)$ lattice gauge theories in three spacetime dimensions features a strong-coupling regime (dominated by unphysical lattice artifacts) at small $\beta$, and a weak-coupling regime which is analytically connected to the continuum limit at large $\beta$~\cite{Bursa:2005tk}.

\begin{figure}[tb!]
\begin{center}
\begin{minipage}[t]{16.5 cm}
\begin{centering}
\hspace{15mm} \epsfig{file=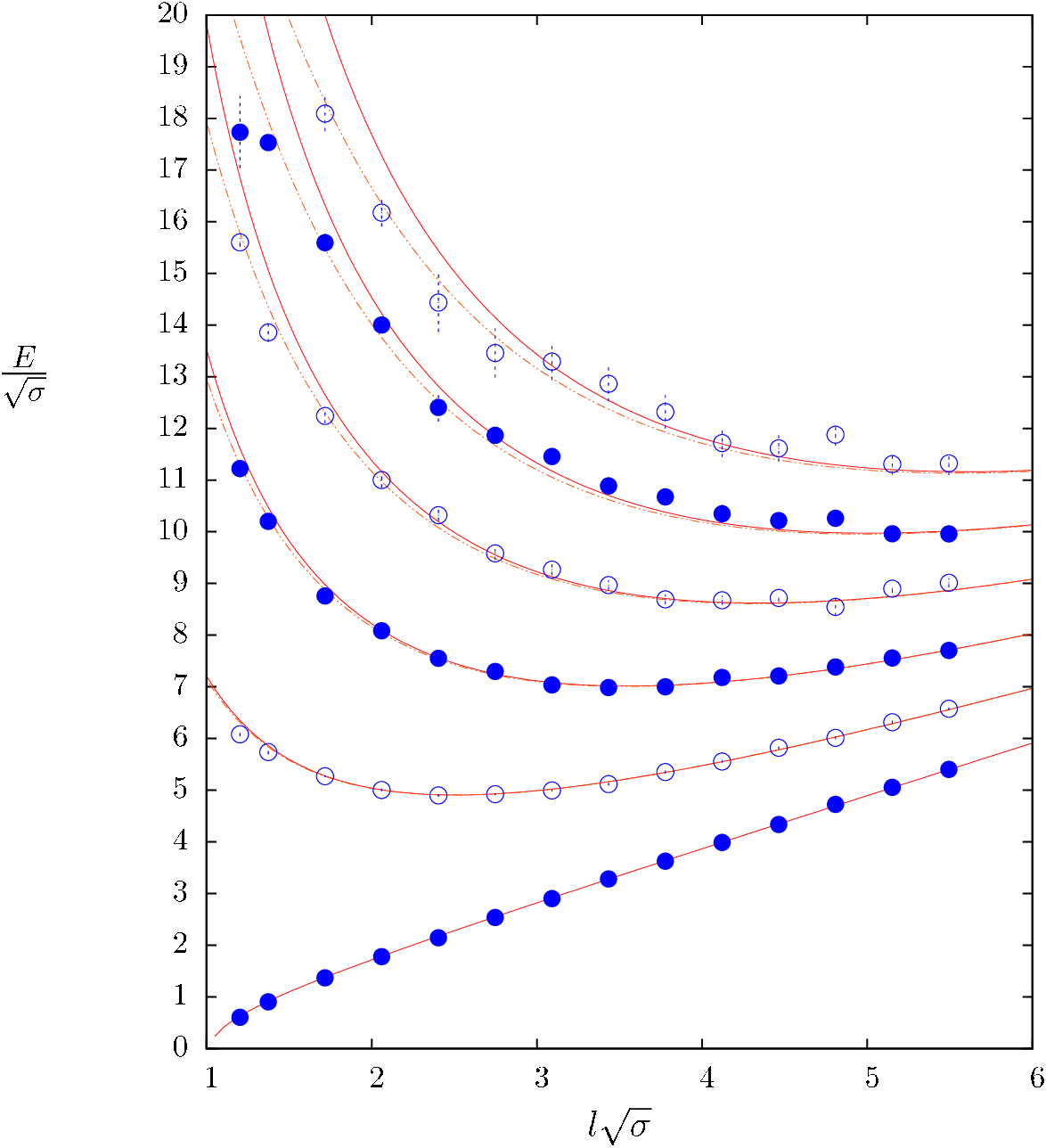}
\end{centering}
\vspace{5mm}\\
\end{minipage}
\begin{minipage}[t]{16.5 cm}
\caption{Spectrum of a closed confining flux tube in $\SU(6)$ Yang-Mills theory, taken from ref.~\cite{Athenodorou:2011rx}. The symbols denote lattice results for different states, as a function of the tube length, while the lines show the predictions from the Nambu-Got{\={o}} bosonic string model. The solid and dashed lines are obtained from two slightly different definitions of the momentum carried by the flux tube (see ref.~\cite{Athenodorou:2011rx} for details).\label{fig:Athenodorou_et_al_11035854_SU6_3d_string_plot_EgsQall_n6f}}
\end{minipage}
\end{center}
\end{figure}

The non-perturbative study of these theories at large $N$ via lattice simulations was initiated during the late 1990's--early 2000's~\cite{Teper:1997tp, Teper:1998te, Lucini:2002wg}, when it was numerically proven that these theories are confining, with a string tension related to the 't~Hooft coupling as~\cite{Lucini:2002wg}
\begin{equation}
\label{sigma_over_lambda_0_3D}
\frac{\sqrt{\sigma}}{\lambda} = 0.19755(34) - \frac{0.1200(29)}{N^2}.
\end{equation}
A more recent determination of $\sigma$ can be found in ref.~\cite{Bringoltz:2006zg}. The lattice results for the confining potential $V(r)$ at intermediate distances show that, in addition to the logarithmic Coulomb term, the potential also includes a $1/r$ contribution, which in three spacetime dimensions is clearly interpreted as a L\"uscher term. In fact, by now there is strong evidence that, also in three spacetime dimensions, confining flux tubes admit a quantitatively very accurate low-energy description in terms of a fluctuating Nambu-Got{\={o}} string~\cite{Teper:1998te, Luscher:2002qv, Athenodorou:2007du, Bringoltz:2008nd, Athenodorou:2008cj, Bialas:2009pt, Athenodorou:2011rx, Caselle:2011vk, Mykkanen:2012dv}, up to subleading corrections which only appear at high orders in an expansion in inverse powers of the string length: see fig.~\ref{fig:Athenodorou_et_al_11035854_SU6_3d_string_plot_EgsQall_n6f} for an example.

In addition to the large-$N$ theories investigated in these studies, other three-dimensional models for which the L\"uscher term has been studied numerically to high precision include $\SU(2)$ Yang-Mills theory~\cite{Majumdar:2002mr, Juge:2004xr, Caselle:2004er, Brandt:2009tc, Brandt:2010bw, Caselle:2011vk} (see also ref.~\cite{Brandt:2013eua} for a very recent summary), as well as various models based on discrete gauge groups~\cite{Caselle:2011vk, Juge:2004xr, Caselle:1996ii, Caselle:2002rm, Caselle:2002ah, Caselle:2004jq, Caselle:2005xy, Caselle:2005vq, Caselle:2006dv, Giudice:2006hw, Caselle:2007yc, Giudice:2007sk, Rajantie:2012zn} and even a random percolation model (with an appropriate definition of the Wilson loop)~\cite{Gliozzi:2005ny, Giudice:2009di}.

The physical spectrum of Yang-Mills theories in three spacetime dimensions consists of glueballs, which are classified by the irreducible representations of the $\SO(2)$ group, and by the quantum numbers of ``mirror'' parity and of charge conjugation (for $N>2$). Lattice results for the spectrum of these states at large $N$ were reported in refs.~\cite{Teper:1997tp, Teper:1998te, Lucini:2002wg, Meyer:2003wx, Meyer:2004jc}: the results show that the mass gap remains finite in the 't~Hooft limit, and that dimensionless ratios of masses in different channels are only mildly dependent on $N$. 

\begin{figure}[tb!]
\begin{center}
\begin{minipage}[t]{16.5 cm}
\begin{centering}
\hspace{15mm} \epsfig{file=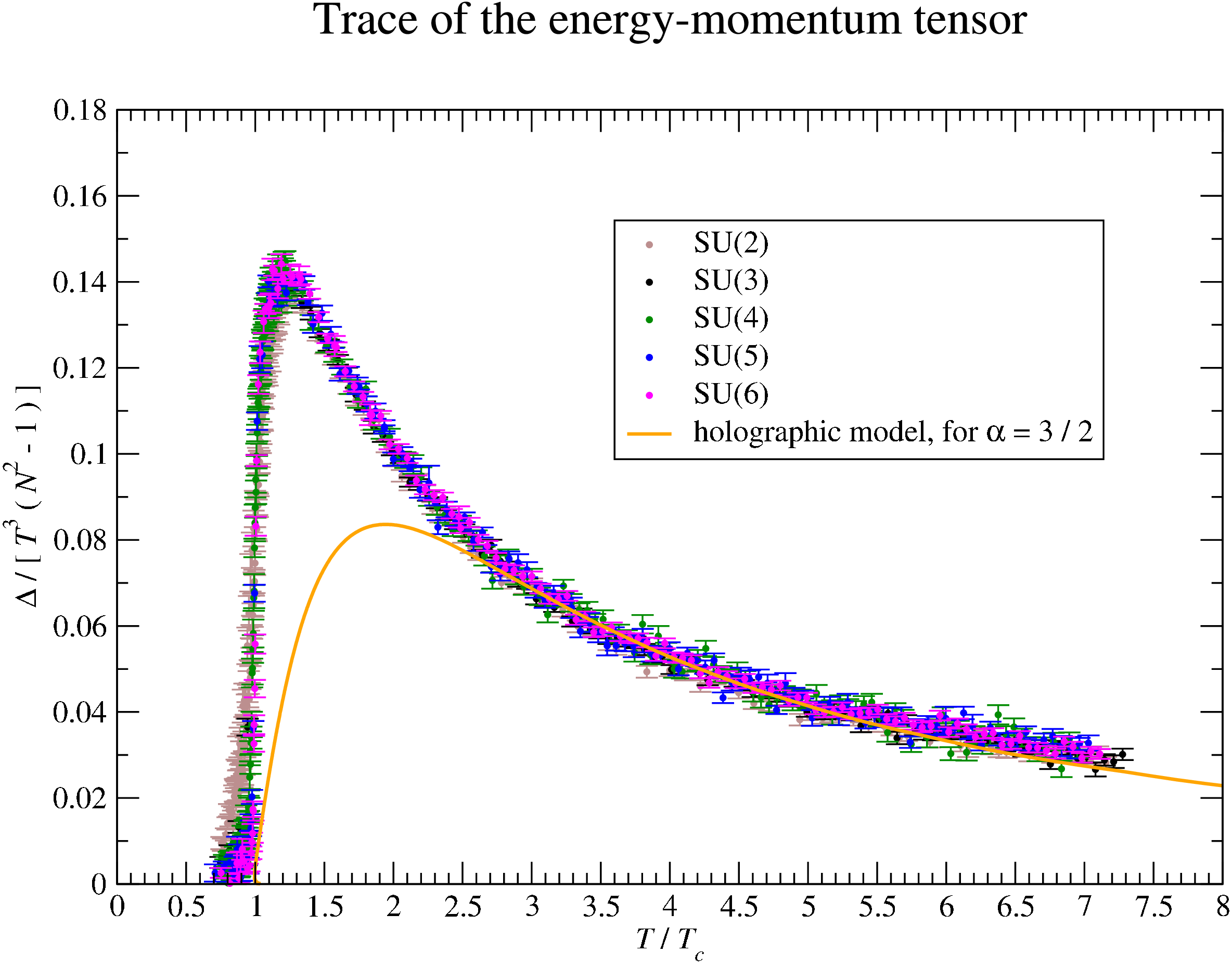,scale=0.5}
\end{centering}
\vspace{5mm}\\
\end{minipage}
\begin{minipage}[t]{16.5 cm}
\caption{Temperature dependence of the trace of the energy-momentum tensor in units of $T^3$ per gluon from simulations in the deconfined phase of 3D $\SU(N)$ Yang-Mills theories, from ref.~\cite{Caselle:2011mn}. The plot shows simulation results for $N=2$ (brown symbols), $3$ (black), $4$ (green), $5$ (blue) and $6$ (magenta), and their comparison with the expectation (yellow curve) obtained from a generalization of the holographic model proposed in refs.~\cite{Gursoy:2008bu, Gursoy:2009jd} to 3D Yang-Mills theory.\label{fig:3D_trace_anomaly_deconfined}}
\end{minipage}
\end{center}
\end{figure}

At finite temperature, three-dimensional $\SU(N)$ Yang-Mills theories undergo a physical phase transition, which separates the confining phase at low temperatures from a deconfined phase at high temperature. The critical temperature of this transition is~\cite{Liddle:2008kk}
\begin{equation}
\label{Tc_root_sigma_3D}
\frac{T_c}{\sqrt{\sigma}} = 0.9026(23) + \frac{0.880(43)}{N^2}.
\end{equation}
The transition is of second order for the $\SU(2)$ and $\SU(3)$ gauge groups~\cite{Christensen:1990vc, Teper:1993gp, Christensen:1991rx, Bialas:2012qz}, while it is probably a first-order one~\cite{Holland:2007ar} for $\SU(4)$ (although some results indicate that it may also be of second order~\cite{deForcrand:2003wa}), and clearly a first-order one for $N \ge 5$~\cite{Liddle:2008kk, Holland:2005nd}. The equation of state of $\SU(N)$ Yang-Mills theories at large $N$ was studied in refs.~\cite{Caselle:2011fy, Caselle:2011mn}: for $T < T_c$, the equation of state can be described by a gas of weakly-interacting, massive glueballs and is essentially independent of $N$ (except for the special case of $\SU(2)$, because, due to the pseudo-real nature of the group, this gauge theory does not admit glueballs that are odd under charge conjugation), while for $T>T_c$ the equilibrium thermodynamic observables are nearly perfectly proportional to the number of gluon degrees of freedom $(N^2-1)$, with a slow approach to the Stefan-Boltzmann limit. An interesting observation reported in ref.~\cite{Caselle:2011mn} (see fig.~\ref{fig:3D_trace_anomaly_deconfined}) is that, like in four spacetime dimensions, also in three spacetime dimension $\Delta$ is proportional to $T^2$ at temperatures above (and of the same order of magnitude as) $T_c$. Analytical studies of $\SU(N)$ models in three spacetime dimensions at finite temperature have a long history~\cite{D'Hoker:1981us}, but continue to attract interest~\cite{Bicudo:2013yza}.

Finally, we mention that there are also studies of large-$N$ QCD or related models in two spacetime dimensions. Several analytical or semi-analytical results have been known since long ago. In particular, the meson spectrum was computed semi-analytically by 't~Hooft~\cite{'tHooft:1974hx}, while the Wilson loop spectral density was studied in ref.~\cite{Durhuus:1980nb}. More recent examples of studies of large-$N$ toy models of QCD in two spacetime dimensions include those reported in refs.~\cite{Rossi:1994xg, Rossi:1996hs, Olesen:2006gt, Narayanan:2007dv, Olesen:2007rf, Blaizot:2008nc, Neuberger:2008mk, Narayanan:2008he, Neuberger:2008ti, Lohmayer:2009aw, Lohmayer:2011nq, Orland:2011rd, Orland:2012sk, Cubero:2012xi, Cubero:2013iga}, as well as a series of articles addressing the issue of large-$N$ baryonic matter at finite density in a world with two spacetime dimensions~\cite{Bringoltz:2008iu, Bringoltz:2009ym, Galvez:2009rq} (see also refs.~\cite{Schon:2000he, Kojo:2011fh} for a discussion).

\section{Conclusions}
\label{sect:conclusions}

In this work, we presented a brief summary of the main ideas underlying the 't~Hooft large-$N$ limit, with a focus on phenomenological implications and on lattice results. For a more extended and more detailed discussion, we recommend the readers to refer to ref.~\cite{Lucini:2012gg} and to the works mentioned therein.

Almost forty years after the seminal work by 't~Hooft~\cite{'tHooft:1973jz}, we personally find it striking---and awe-inspiring---to see how many different, fruitful research directions have stemmed from the generalization of QCD to the large-$N$ limit. In addition to those presented in this work, implications of planarity have been exploited in a large number of contexts, and in very diverse theoretical models for which analytical treatment is possible. In fact, the applications of the large-$N$ limit are not limited to the ones that we briefly mentioned in this work (e.g. in our elementary discussion of the r\^ole of the large-$N$ limit in the gauge/gravity correspondence, or of large-$N$ ``orbifold'' dualities), and the reader should be aware that the references we mentioned for these types of studies represent only a small fraction of the literature.

\vskip1.0cm 
\noindent{\bf Acknowledgments}\\
We are indebted to our collaborators and to many colleagues working in this field, whom we cannot mention one by one, and who deserve most of the credit for the results discussed herein. The responsibility for any error and omission lies solely with us. This work is supported by the Academy of Finland, project 1134018, by the Royal Society (grant UF09003), by the Spanish MINECO's ``Centro de Excelencia Severo Ochoa'' programme under grant SEV-2012-0249, and in part by STFC under grant ST/G000506/1.
\vskip1.0cm

\addcontentsline{toc}{section}{References}
\bibliographystyle{h-elsevier3}
\bibliography{ppnp_bibliography}

\begin{thebibliography}{100}

\bibitem{Creutz:2003qy}
M. Creutz,
\newblock AIP Conf.Proc. 690 (2003) 52, hep-lat/0306024.

\bibitem{Brodsky:1997de}
S.J. Brodsky, H.C. Pauli and S.S. Pinsky,
\newblock Phys.Rept. 301 (1998) 299, hep-ph/9705477.

\bibitem{Lucini:2012gg}
B. Lucini and M. Panero,
\newblock Phys.Rept. 526 (2013) 93, 1210.4997.

\bibitem{Panero:2012qx}
M. Panero,
\newblock PoS Lattice 2012 (2012) 010, 1210.5510.

\bibitem{Brezin_Wadia}
{\'E}. Br{\'e}zin and S.R. Wadia,
\newblock {The Large N Expansion in Quantum Field Theory and Statistical
  Physics: From Spin Systems to 2-Dimensional Gravity} (World Scientific,
  1993).

\bibitem{Witten:1979pi}
E. Witten,
\newblock (1979).

\bibitem{Coleman:1980nk}
S.R. Coleman,
\newblock (1980).

\bibitem{Yaffe:1981vf}
L.G. Yaffe,
\newblock Rev.Mod.Phys. 54 (1982) 407.

\bibitem{Migdal:1984gj}
A.A. Migdal,
\newblock Phys.Rept. 102 (1983) 199.

\bibitem{Das:1984nb}
S.R. Das,
\newblock Rev.Mod.Phys. 59 (1987) 235.

\bibitem{Lebed:2002tj}
R. Lebed,
\newblock {Phenomenology of large N(c) QCD. Proceedings, Tempe, AZ, US, January
  9-11, 2002} (World Scientific, 2002).

\bibitem{Jenkins:1998wy}
E.E. Jenkins,
\newblock Ann.Rev.Nucl.Part.Sci. 48 (1998) 81, hep-ph/9803349.

\bibitem{Lebed:1998st}
R.F. Lebed,
\newblock Czech.J.Phys. 49 (1999) 1273, nucl-th/9810080.

\bibitem{Teper:1998kw}
M.J. Teper,
\newblock (1998), hep-th/9812187.

\bibitem{Manohar:1998xv}
A.V. Manohar,
\newblock (1998) 1091, hep-ph/9802419.

\bibitem{Makeenko:1999hq}
Y. Makeenko,
\newblock (1999), hep-th/0001047.

\bibitem{Makeenko:2004bz}
Y. Makeenko,
\newblock (2004) 14, hep-th/0407028.

\bibitem{Narayanan:2007fb}
R. Narayanan and H. Neuberger,
\newblock PoS LATTICE 2007 (2007) 020, 0710.0098.

\bibitem{Teper:2008yi}
M. Teper,
\newblock PoS LATTICE 2008 (2008) 022, 0812.0085.

\bibitem{Vicari:2008jw}
E. Vicari and H. Panagopoulos,
\newblock Phys.Rept. 470 (2009) 93, 0803.1593.

\bibitem{Teper:2009uf}
M. Teper,
\newblock Acta Phys.Polon. B40 (2009) 3249, 0912.3339.

\bibitem{Bochicchio:2013tfa}
M. Bochicchio and S.P. Muscinelli,
\newblock JHEP 1308 (2013) 064, 1304.6409.

\bibitem{Bochicchio:2013eda}
M. Bochicchio,
\newblock Nucl.Phys. B875 (2013) 621, 1305.0273.

\bibitem{Kaplan:2013dca}
D.B. Kaplan,
\newblock (2013), 1306.5818.

\bibitem{'tHooft:1973jz}
G. 't~Hooft,
\newblock Nucl. Phys. B72 (1974) 461.

\bibitem{Stanley:1968gx}
H. Stanley,
\newblock Phys.Rev. 176 (1968) 718.

\bibitem{Ma:1973zu}
S.K. Ma,
\newblock Rev.Mod.Phys. 45 (1973) 589.

\bibitem{Brezin:1972se}
{\'E}. Br{\'e}zin and D. Wallace,
\newblock (1972).

\bibitem{Brezin:1976qa}
{\'E}. Br{\'e}zin and J. Zinn-Justin,
\newblock Phys.Rev. B14 (1976) 3110.

\bibitem{Bardeen:1976zh}
W.A. Bardeen, B.W. Lee and R.E. Shrock,
\newblock Phys.Rev. D14 (1976) 985.

\bibitem{Okabe:1978nn}
Y. Okabe, M. Oku and R. Abe,
\newblock Prog.Theor.Phys. 59 (1978) 1825.

\bibitem{Veneziano:1976wm}
G. Veneziano,
\newblock Nucl. Phys. B117 (1976) 519.

\bibitem{Maldacena:1997re}
J.M. Maldacena,
\newblock Adv. Theor. Math. Phys. 2 (1998) 231, hep-th/9711200.

\bibitem{Gubser:1998bc}
S. Gubser, I.R. Klebanov and A.M. Polyakov,
\newblock Phys.Lett. B428 (1998) 105, hep-th/9802109.

\bibitem{Witten:1998qj}
E. Witten,
\newblock Adv.Theor.Math.Phys. 2 (1998) 253, hep-th/9802150.

\bibitem{Aharony:1999ti}
O. Aharony et~al.,
\newblock Phys.Rept. 323 (2000) 183, hep-th/9905111.

\bibitem{Petersen:1999zh}
J.L. Petersen,
\newblock Int.J.Mod.Phys. A14 (1999) 3597, hep-th/9902131.

\bibitem{Klebanov:2000me}
I.R. Klebanov,
\newblock (2000) 615, hep-th/0009139.

\bibitem{D'Hoker:2002aw}
E. D'Hoker and D.Z. Freedman,
\newblock (2002) 3, hep-th/0201253.

\bibitem{Maldacena:2003nj}
J.M. Maldacena,
\newblock (2003) 155, hep-th/0309246.

\bibitem{Erdmenger:2007cm}
J. Erdmenger et~al.,
\newblock Eur.Phys.J. A35 (2008) 81, 0711.4467.

\bibitem{Mateos:2007ay}
D. Mateos,
\newblock Class.Quant.Grav. 24 (2007) S713, 0709.1523.

\bibitem{Gubser:2009md}
S.S. Gubser and A. Karch,
\newblock Ann.Rev.Nucl.Part.Sci. 59 (2009) 145, 0901.0935.

\bibitem{CasalderreySolana:2011us}
J. Casalderrey-Solana et~al.,
\newblock (2011), 1101.0618.

\bibitem{'tHooft:1993gx}
G. 't~Hooft,
\newblock (1993), gr-qc/9310026,
\newblock Essay dedicated to Abdus Salam.

\bibitem{Stephens:1993an}
C.R. Stephens, G. 't~Hooft and B.F. Whiting,
\newblock Class.Quant.Grav. 11 (1994) 621, gr-qc/9310006.

\bibitem{Susskind:1994vu}
L. Susskind,
\newblock J.Math.Phys. 36 (1995) 6377, hep-th/9409089.

\bibitem{Susskind:1998dq}
L. Susskind and E. Witten,
\newblock (1998), hep-th/9805114.

\bibitem{Bousso:2002ju}
R. Bousso,
\newblock Rev.Mod.Phys. 74 (2002) 825, hep-th/0203101.

\bibitem{Witten:1998zw}
E. Witten,
\newblock Adv.Theor.Math.Phys. 2 (1998) 505, hep-th/9803131.

\bibitem{Brink:1976bc}
L. Brink, J.H. Schwarz and J. Scherk,
\newblock Nucl.Phys. B121 (1977) 77.

\bibitem{Gliozzi:1976qd}
F. Gliozzi, J. Scherk and D.I. Olive,
\newblock Nucl.Phys. B122 (1977) 253.

\bibitem{Green:1981yb}
M.B. Green and J.H. Schwarz,
\newblock Phys.Lett. B109 (1982) 444.

\bibitem{Schwarz:1983qr}
J.H. Schwarz,
\newblock Nucl.Phys. B226 (1983) 269.

\bibitem{Mandelstam:1982cb}
S. Mandelstam,
\newblock Nucl.Phys. B213 (1983) 149.

\bibitem{Howe:1983sr}
P.S. Howe, K. Stelle and P. Townsend,
\newblock Nucl.Phys. B236 (1984) 125.

\bibitem{Seiberg:1988ur}
N. Seiberg,
\newblock Phys.Lett. B206 (1988) 75.

\bibitem{Polchinski:1995mt}
J. Polchinski,
\newblock Phys.Rev.Lett. 75 (1995) 4724, hep-th/9510017.

\bibitem{Johnson:2000ch}
C.V. Johnson,
\newblock (2000) 129, hep-th/0007170.

\bibitem{Balasubramanian:1998sn}
V. Balasubramanian, P. Kraus and A.E. Lawrence,
\newblock Phys.Rev. D59 (1999) 046003, hep-th/9805171.

\bibitem{Balasubramanian:1998de}
V. Balasubramanian et~al.,
\newblock Phys.Rev. D59 (1999) 104021, hep-th/9808017.

\bibitem{Danielsson:1998wt}
U.H. Danielsson, E. Keski-Vakkuri and M. Kruczenski,
\newblock JHEP 9901 (1999) 002, hep-th/9812007.

\bibitem{Gubser:1996de}
S. Gubser, I.R. Klebanov and A. Peet,
\newblock Phys.Rev. D54 (1996) 3915, hep-th/9602135.

\bibitem{Gubser:1998nz}
S.S. Gubser, I.R. Klebanov and A.A. Tseytlin,
\newblock Nucl.Phys. B534 (1998) 202, hep-th/9805156.

\bibitem{Policastro:2001yc}
G. Policastro, D. Son and A. Starinets,
\newblock Phys.Rev.Lett. 87 (2001) 081601, hep-th/0104066.

\bibitem{novelnum12}
{Novel Numerical Methods for Strongly Coupled Quantum Field Theory and Quantum
  Gravity, Kavli Institute for Theoretical Physics, UCSB, Santa Barbara, CA,
  US, January 17-March 9, 2012},
\newblock \url{http://www.kitp.ucsb.edu/activities/dbdetails?acro=novelnum12}.

\bibitem{Karch:2000gx}
A. Karch and L. Randall,
\newblock JHEP 0106 (2001) 063, hep-th/0105132.

\bibitem{Karch:2002sh}
A. Karch and E. Katz,
\newblock JHEP 0206 (2002) 043, hep-th/0205236.

\bibitem{Polchinski:2000uf}
J. Polchinski and M.J. Strassler,
\newblock (2000), hep-th/0003136.

\bibitem{Maldacena:2000yy}
J.M. Maldacena and C. N{\'u}{\~{n}}ez,
\newblock Phys.Rev.Lett. 86 (2001) 588, hep-th/0008001.

\bibitem{Klebanov:2000hb}
I.R. Klebanov and M.J. Strassler,
\newblock JHEP 0008 (2000) 052, hep-th/0007191.

\bibitem{Aharony:2002up}
O. Aharony,
\newblock (2002) 3, hep-th/0212193.

\bibitem{Kruczenski:2003be}
M. Kruczenski et~al.,
\newblock JHEP 0307 (2003) 049, hep-th/0304032.

\bibitem{Kruczenski:2003uq}
M. Kruczenski et~al.,
\newblock JHEP 0405 (2004) 041, hep-th/0311270.

\bibitem{Sakai:2004cn}
T. Sakai and S. Sugimoto,
\newblock Prog.Theor.Phys. 113 (2005) 843, hep-th/0412141.

\bibitem{Sakai:2005yt}
T. Sakai and S. Sugimoto,
\newblock Prog.Theor.Phys. 114 (2005) 1083, hep-th/0507073.

\bibitem{Nunez:2010sf}
C. N{\'u}{\~n}ez, {\'A}. Paredes and A.V. Ramallo,
\newblock Adv.High Energy Phys. 2010 (2010) 196714, 1002.1088.

\bibitem{Polchinski:2001tt}
J. Polchinski and M.J. Strassler,
\newblock Phys.Rev.Lett. 88 (2002) 031601, hep-th/0109174.

\bibitem{Polchinski:2002jw}
J. Polchinski and M.J. Strassler,
\newblock JHEP 0305 (2003) 012, hep-th/0209211.

\bibitem{Karch:2002xe}
A. Karch, E. Katz and N. Weiner,
\newblock Phys.Rev.Lett. 90 (2003) 091601, hep-th/0211107.

\bibitem{Son:2003et}
D. Son and M. Stephanov,
\newblock Phys.Rev. D69 (2004) 065020, hep-ph/0304182.

\bibitem{Brodsky:2003px}
S.J. Brodsky and G.F. de~T{\'e}ramond,
\newblock Phys.Lett. B582 (2004) 211, hep-th/0310227.

\bibitem{deTeramond:2005su}
G.F. de~T{\'e}ramond and S.J. Brodsky,
\newblock Phys.Rev.Lett. 94 (2005) 201601, hep-th/0501022.

\bibitem{DaRold:2005zs}
L. Da~Rold and A. Pomarol,
\newblock Nucl.Phys. B721 (2005) 79, hep-ph/0501218.

\bibitem{Erlich:2005qh}
J. Erlich et~al.,
\newblock Phys.Rev.Lett. 95 (2005) 261602, hep-ph/0501128.

\bibitem{Hirn:2005nr}
J. Hirn and V. Sanz,
\newblock JHEP 0512 (2005) 030, hep-ph/0507049.

\bibitem{Karch:2006pv}
A. Karch et~al.,
\newblock Phys.Rev. D74 (2006) 015005, hep-ph/0602229.

\bibitem{Csaki:2006ji}
C. Cs{\'a}ki and M. Reece,
\newblock JHEP 0705 (2007) 062, hep-ph/0608266.

\bibitem{Csaki:2008dt}
C. Cs{\'a}ki, M. Reece and J. Terning,
\newblock JHEP 0905 (2009) 067, 0811.3001.

\bibitem{Gursoy:2010fj}
U. G{\"u}rsoy et~al.,
\newblock Lect.Notes Phys. 828 (2011) 79, 1006.5461.

\bibitem{Reece:2011zz}
M. Reece,
\newblock AIP Conf.Proc. 1343 (2011) 117.

\bibitem{Minahan:2002ve}
J. Minahan and K. Zarembo,
\newblock JHEP 0303 (2003) 013, hep-th/0212208.

\bibitem{Bena:2003wd}
I. Bena, J. Polchinski and R. Roiban,
\newblock Phys.Rev. D69 (2004) 046002, hep-th/0305116.

\bibitem{Beisert:2003jj}
N. Beisert,
\newblock Nucl.Phys. B676 (2004) 3, hep-th/0307015.

\bibitem{Beisert:2003tq}
N. Beisert, C. Kristjansen and M. Staudacher,
\newblock Nucl.Phys. B664 (2003) 131, hep-th/0303060.

\bibitem{Beisert:2010jr}
N. Beisert et~al.,
\newblock Lett.Math.Phys. 99 (2012) 3, 1012.3982.

\bibitem{Weinberg:2013cfa}
S. Weinberg,
\newblock Phys. Rev. Lett. 110, 261601 (2013), 1303.0342.

\bibitem{Knecht:2013yqa}
M. Knecht and S. Peris,
\newblock Phys.Rev. D88 (2013) 036016, 1307.1273.

\bibitem{Lebed:2013aka}
R.F. Lebed,
\newblock Phys.Rev. D88 (2013) 057901, 1308.2657.

\bibitem{Okubo:1963fa}
S. Okubo,
\newblock Phys.Lett. 5 (1963) 165.

\bibitem{Zweig:1981pd}
G. Zweig,
\newblock (1964).

\bibitem{Iizuka:1966fk}
J. Iizuka,
\newblock Prog.Theor.Phys.Suppl. 37 (1966) 21.

\bibitem{Geiger:1996re}
P. Geiger and N. Isgur,
\newblock Phys.Rev. D55 (1997) 299, hep-ph/9610445.

\bibitem{Pich:2002xy}
A. Pich,
\newblock (2002) 239, hep-ph/0205030.

\bibitem{Nakamura:2010zzi}
Particle Data Group, K. Nakamura et~al.,
\newblock J.Phys.G G37 (2010) 075021.

\bibitem{Kaiser:2005eu}
R. Kaiser,
\newblock (2005) 144, hep-ph/0502065.

\bibitem{Uehara:2003ax}
M. Uehara,
\newblock (2003), hep-ph/0308241.

\bibitem{Pelaez:2003dy}
J. Pel{\'a}ez,
\newblock Phys.Rev.Lett. 92 (2004) 102001, hep-ph/0309292.

\bibitem{Pelaez:2004xp}
J. Pel{\'a}ez,
\newblock Mod.Phys.Lett. A19 (2004) 2879, hep-ph/0411107.

\bibitem{Pelaez:2006nj}
J. Pel{\'a}ez and G. R{\'{\i}}os,
\newblock Phys.Rev.Lett. 97 (2006) 242002, hep-ph/0610397.

\bibitem{Geng:2008ag}
L. Geng et~al.,
\newblock Eur.Phys.J. A39 (2009) 81, 0811.1941.

\bibitem{Dashen:1993jt}
R.F. Dashen, E.E. Jenkins and A.V. Manohar,
\newblock Phys. Rev. D49 (1994) 4713, hep-ph/9310379.

\bibitem{Jenkins:1993zu}
E.E. Jenkins,
\newblock Phys.Lett. B315 (1993) 441, hep-ph/9307244.

\bibitem{Gervais:1983wq}
J.L. Gervais and B. Sakita,
\newblock Phys.Rev.Lett. 52 (1984) 87.

\bibitem{Carone:1993dz}
C. Carone, H. Georgi and S. Osofsky,
\newblock Phys. Lett. B322 (1994) 227, hep-ph/9310365.

\bibitem{Luty:1993fu}
M.A. Luty and J. March-Russell,
\newblock Nucl. Phys. B426 (1994) 71, hep-ph/9310369.

\bibitem{Jenkins:1996de}
E.E. Jenkins,
\newblock Phys. Rev. D54 (1996) 4515, hep-ph/9603449.

\bibitem{Lutz:2001yb}
M. Lutz and E. Kolomeitsev,
\newblock Nucl.Phys. A700 (2002) 193, nucl-th/0105042.

\bibitem{Haan:1981ks}
O. Haan,
\newblock Phys.Lett. B106 (1981) 207.

\bibitem{Klauder_Skagerstam}
J.R. Klauder and B.S. Skagerstam,
\newblock {Coherent States: Applications in Physics and Mathematical Physics}
  (World Scientific, 1985).

\bibitem{Coleman:1974jh}
S.R. Coleman, R. Jackiw and H.D. Politzer,
\newblock Phys.Rev. D10 (1974) 2491.

\bibitem{Brezin:1977sv}
{\'E}. Br{\'e}zin et~al.,
\newblock Commun.Math.Phys. 59 (1978) 35.

\bibitem{Marchesini:1979yq}
G. Marchesini and E. Onofri,
\newblock J.Math.Phys. 21 (1980) 1103.

\bibitem{Jevicki:1980zq}
A. Jevicki and B. Sakita,
\newblock Phys.Rev. D22 (1980) 467.

\bibitem{Wadia:1980cp}
S.R. Wadia,
\newblock Phys.Lett. B93 (1980) 403.

\bibitem{Greensite:1982mf}
J. Greensite and M. Halpern,
\newblock Nucl.Phys. B211 (1983) 343.

\bibitem{Douglas:1994kw}
M.R. Douglas,
\newblock Phys.Lett. B344 (1995) 117, hep-th/9411025.

\bibitem{Douglas:1994zu}
M.R. Douglas,
\newblock Nucl.Phys.Proc.Suppl. 41 (1995) 66, hep-th/9409098.

\bibitem{Gopakumar:1994iq}
R. Gopakumar and D.J. Gross,
\newblock Nucl.Phys. B451 (1995) 379, hep-th/9411021.

\bibitem{Singer:1994zz}
I.M. Singer,
\newblock (1994).

\bibitem{Accardi:1994gd}
L. Accardi, Y. Lu and I. Volovich,
\newblock (1994), hep-th/9412241.

\bibitem{www.uni-math.gwdg.de/mitch/free.pdf}
P.D. Mitchener,
\newblock {Non-Commutative Probability Theory},
\newblock \url{http:/www.uni-math.gwdg.de/mitch/free.pdf}.

\bibitem{Makeenko:1979pb}
Y. Makeenko and A.A. Migdal,
\newblock Phys.Lett. B88 (1979) 135.

\bibitem{Eguchi:1982nm}
T. Eguchi and H. Kawai,
\newblock Phys.Rev.Lett. 48 (1982) 1063.

\bibitem{Bhanot:1982sh}
G. Bhanot, U.M. Heller and H. Neuberger,
\newblock Phys.Lett. B113 (1982) 47.

\bibitem{Bringoltz:2008av}
B. Bringoltz and S.R. Sharpe,
\newblock Phys.Rev. D78 (2008) 034507, 0805.2146.

\bibitem{GonzalezArroyo:1982hz}
A. Gonz{\'a}lez-Arroyo and M. Okawa,
\newblock Phys.Rev. D27 (1983) 2397.

\bibitem{GonzalezArroyo:1982ub}
A. Gonz{\'a}lez-Arroyo and M. Okawa,
\newblock Phys.Lett. B120 (1983) 174.

\bibitem{GonzalezArroyo:1983ac}
A. Gonz{\'a}lez-Arroyo and C. Korthals~Altes,
\newblock Phys.Lett. B131 (1983) 396.

\bibitem{Aoki:1999vr}
H. Aoki et~al.,
\newblock Nucl.Phys. B565 (2000) 176, hep-th/9908141.

\bibitem{Ambjorn:1999ts}
J. Ambj{\o}rn et~al.,
\newblock JHEP 9911 (1999) 029, hep-th/9911041.

\bibitem{Ambjorn:2000nb}
J. Ambj{\o}rn et~al.,
\newblock Phys.Lett. B480 (2000) 399, hep-th/0002158.

\bibitem{Ambjorn:2000cs}
J. Ambj{\o}rn et~al.,
\newblock JHEP 0005 (2000) 023, hep-th/0004147.

\bibitem{Panero:2006bx}
M. Panero,
\newblock JHEP 0705 (2007) 082, hep-th/0608202.

\bibitem{Panero:2006cs}
M. Panero,
\newblock SIGMA 2 (2006) 081, hep-th/0609205.

\bibitem{Teper:2006sp}
M. Teper and H. Vairinhos,
\newblock Phys.Lett. B652 (2007) 359, hep-th/0612097.

\bibitem{Azeyanagi:2007su}
T. Azeyanagi et~al.,
\newblock JHEP 0801 (2008) 025, 0711.1925.

\bibitem{GonzalezArroyo:2010ss}
A. Gonz{\'a}lez-Arroyo and M. Okawa,
\newblock JHEP 1007 (2010) 043, 1005.1981.

\bibitem{GonzalezArroyo:2012fx}
A. Gonz{\'a}lez-Arroyo and M. Okawa,
\newblock Phys.Lett. B718 (2013) 1524, 1206.0049.

\bibitem{Perez:2013dra}
M.G. P{\'e}rez, A. Gonz{\'a}lez-Arroyo and M. Okawa,
\newblock JHEP 1309 (2013) 003, 1307.5254.

\bibitem{Kovtun:2007py}
P. Kovtun, M. {\"U}nsal and L.G. Yaffe,
\newblock JHEP 0706 (2007) 019, hep-th/0702021.

\bibitem{Hollowood:2009sy}
T.J. Hollowood and J.C. Myers,
\newblock JHEP 0911 (2009) 008, 0907.3665.

\bibitem{Azeyanagi:2010ne}
T. Azeyanagi et~al.,
\newblock Phys.Rev. D82 (2010) 125013, 1006.0717.

\bibitem{Catterall:2010gx}
S. Catterall, R. Galvez and M. {\"U}nsal,
\newblock JHEP 1008 (2010) 010, 1006.2469.

\bibitem{Bringoltz:2009kb}
B. Bringoltz and S.R. Sharpe,
\newblock Phys.Rev. D80 (2009) 065031, 0906.3538.

\bibitem{Bringoltz:2011by}
B. Bringoltz, M. Kore\'n and S.R. Sharpe,
\newblock Phys.Rev. D85 (2012) 094504, 1106.5538.

\bibitem{Cossu:2009sq}
G. Cossu and M. D'Elia,
\newblock JHEP 0907 (2009) 048, 0904.1353.

\bibitem{Hietanen:2009ex}
A. Hietanen and R. Narayanan,
\newblock JHEP 1001 (2010) 079, 0911.2449.

\bibitem{Hietanen:2010fx}
A. Hietanen and R. Narayanan,
\newblock Phys.Lett. B698 (2011) 171, 1011.2150.

\bibitem{Hanada:2013ota}
M. Hanada, J.W. Lee and N. Yamada,
\newblock Phys.Rev. D88 (2013) 025046, 1302.3532.

\bibitem{Lee:2013hk}
J.W. Lee, M. Hanada and N. Yamada,
\newblock PoS Lattice 2012 (2012) 047, 1301.0029.

\bibitem{Okawa_parallel}
A. Gonz{\'a}lez-Arroyo and M. Okawa,
\newblock PoS Lattice 2012 (2012) 046, 1210.7881.

\bibitem{Gonzalez-Arroyo:2013bta}
A. Gonz{\'a}lez-Arroyo and M. Okawa,
\newblock Phys.Rev. D88 (2013) 014514, 1305.6253.

\bibitem{Unsal:2008ch}
M. {\"U}nsal and L.G. Yaffe,
\newblock Phys.Rev. D78 (2008) 065035, 0803.0344.

\bibitem{Myers:2007vc}
J.C. Myers and M.C. Ogilvie,
\newblock Phys.Rev. D77 (2008) 125030, 0707.1869.

\bibitem{Vairinhos:2010ha}
H. Vairinhos,
\newblock (2010), 1010.1253.

\bibitem{Narayanan:2003fc}
R. Narayanan and H. Neuberger,
\newblock Phys.Rev.Lett. 91 (2003) 081601, hep-lat/0303023.

\bibitem{Kiskis:2003rd}
J. Kiskis, R. Narayanan and H. Neuberger,
\newblock Phys.Lett. B574 (2003) 65, hep-lat/0308033.

\bibitem{Kiskis:2009rf}
J. Kiskis and R. Narayanan,
\newblock Phys.Lett. B681 (2009) 372, 0908.1451.

\bibitem{Bershadsky:1998cb}
M. Bershadsky and A. Johansen,
\newblock Nucl.Phys. B536 (1998) 141, hep-th/9803249.

\bibitem{Strassler:2001fs}
M.J. Strassler,
\newblock (2001), hep-th/0104032.

\bibitem{Kovtun:2003hr}
P. Kovtun, M. {\"U}nsal and L.G. Yaffe,
\newblock JHEP 0312 (2003) 034, hep-th/0311098.

\bibitem{Kovtun:2004bz}
P. Kovtun, M. {\"U}nsal and L.G. Yaffe,
\newblock JHEP 0507 (2005) 008, hep-th/0411177.

\bibitem{Kovtun:2005kh}
P. Kovtun, M. {\"U}nsal and L.G. Yaffe,
\newblock Phys.Rev. D72 (2005) 105006, hep-th/0505075.

\bibitem{Unsal:2006pj}
M. {\"U}nsal and L.G. Yaffe,
\newblock Phys.Rev. D74 (2006) 105019, hep-th/0608180.

\bibitem{Armoni:2003gp}
A. Armoni, M. Shifman and G. Veneziano,
\newblock Nucl.Phys. B667 (2003) 170, hep-th/0302163.

\bibitem{Armoni:2003fb}
A. Armoni, M. Shifman and G. Veneziano,
\newblock Phys.Rev.Lett. 91 (2003) 191601, hep-th/0307097.

\bibitem{Lucini:2010kj}
B. Lucini et~al.,
\newblock Phys.Rev. D82 (2010) 114510, 1008.5180.

\bibitem{Catterall:2009it}
S. Catterall, D.B. Kaplan and M. {\"U}nsal,
\newblock Phys.Rept. 484 (2009) 71, 0903.4881.

\bibitem{Unsal:2006qp}
M. {\"U}nsal,
\newblock JHEP 0610 (2006) 089, hep-th/0603046.

\bibitem{Hanada:2007ti}
M. Hanada, J. Nishimura and S. Takeuchi,
\newblock Phys.Rev.Lett. 99 (2007) 161602, 0706.1647.

\bibitem{Anagnostopoulos:2007fw}
K.N. Anagnostopoulos et~al.,
\newblock Phys.Rev.Lett. 100 (2008) 021601, 0707.4454.

\bibitem{Ishii:2008ib}
T. Ishii et~al.,
\newblock Phys.Rev. D78 (2008) 106001, 0807.2352.

\bibitem{Ishiki:2008te}
G. Ishiki et~al.,
\newblock Phys.Rev.Lett. 102 (2009) 111601, 0810.2884.

\bibitem{Gross:1973id}
D. Gross and F. Wilczek,
\newblock Phys.Rev.Lett. 30 (1973) 1343.

\bibitem{Politzer:1973fx}
H. Politzer,
\newblock Phys.Rev.Lett. 30 (1973) 1346.

\bibitem{Wilson:1974sk}
K.G. Wilson,
\newblock Phys.Rev. D10 (1974) 2445.

\bibitem{Kogut:1974ag}
J.B. Kogut and L. Susskind,
\newblock Phys.Rev. D11 (1975) 395.

\bibitem{Osterwalder:1973dx}
K. Osterwalder and R. Schrader,
\newblock Commun.Math.Phys. 31 (1973) 83.

\bibitem{Lang:1982tj}
C. Lang and C. Rebbi,
\newblock Phys.Lett. B115 (1982) 137.

\bibitem{Curci:1983an}
G. Curci, P. Menotti and G. Paffuti,
\newblock Phys.Lett. B130 (1983) 205.

\bibitem{Weisz:1983bn}
P. Weisz and R. Wohlert,
\newblock Nucl.Phys. B236 (1984) 397.

\bibitem{Luscher:1984xn}
M. L{\"u}scher and P. Weisz,
\newblock Commun.Math.Phys. 97 (1985) 59.

\bibitem{Luscher:1985zq}
M. L{\"u}scher and P. Weisz,
\newblock Phys.Lett. B158 (1985) 250.

\bibitem{deForcrand:2009dh}
P. de~Forcrand and M. Fromm,
\newblock Phys.Rev.Lett. 104 (2010) 112005, 0907.1915.

\bibitem{Langelage:2009jb}
J. Langelage and O. Philipsen,
\newblock JHEP 1001 (2010) 089, 0911.2577.

\bibitem{Langelage:2010yn}
J. Langelage and O. Philipsen,
\newblock JHEP 1004 (2010) 055, 1002.1507.

\bibitem{Langelage:2010yr}
J. Langelage, S. Lottini and O. Philipsen,
\newblock JHEP 1102 (2011) 057, 1010.0951.

\bibitem{Fromm:2011qi}
M. Fromm et~al.,
\newblock JHEP 1201 (2012) 042, 1111.4953.

\bibitem{Fromm:2012eb}
M. Fromm et~al.,
\newblock Phys.Rev.Lett. 110 (2013) 122001, 1207.3005.

\bibitem{Wilson:1975id}
K.G. Wilson,
\newblock (1975).

\bibitem{Nielsen:1981xu}
H.B. Nielsen and M. Ninomiya,
\newblock Nucl.Phys. B193 (1981) 173.

\bibitem{Ginsparg:1981bj}
P.H. Ginsparg and K.G. Wilson,
\newblock Phys.Rev. D25 (1982) 2649.

\bibitem{Luscher:1998pqa}
M. L{\"u}scher,
\newblock Phys.Lett. B428 (1998) 342, hep-lat/9802011.

\bibitem{Neuberger:1997fp}
H. Neuberger,
\newblock Phys.Lett. B417 (1998) 141, hep-lat/9707022.

\bibitem{Atiyah:1968mp}
M. Atiyah and I. Singer,
\newblock Annals Math. 87 (1968) 484.

\bibitem{Kaplan:1992bt}
D.B. Kaplan,
\newblock Phys.Lett. B288 (1992) 342, hep-lat/9206013.

\bibitem{Shamir:1993zy}
Y. Shamir,
\newblock Nucl.Phys. B406 (1993) 90, hep-lat/9303005.

\bibitem{Furman:1994ky}
V. Furman and Y. Shamir,
\newblock Nucl.Phys. B439 (1995) 54, hep-lat/9405004.

\bibitem{Teper:1997tq}
M. Teper,
\newblock Phys.Lett. B397 (1997) 223, hep-lat/9701003.

\bibitem{Lucini:2000qp}
B. Lucini and M. Teper,
\newblock Phys.Lett. B501 (2001) 128, hep-lat/0012025.

\bibitem{Lucini:2001nv}
B. Lucini and M. Teper,
\newblock Phys. Rev. D64 (2001) 105019, hep-lat/0107007.

\bibitem{Lucini:2001ej}
B. Lucini and M. Teper,
\newblock JHEP 0106 (2001) 050, hep-lat/0103027.

\bibitem{DelDebbio:2001sj}
L. Del~Debbio et~al.,
\newblock JHEP 0201 (2002) 009, hep-th/0111090.

\bibitem{DelDebbio:2001kz}
L. Del~Debbio et~al.,
\newblock Phys.Rev. D65 (2002) 021501, hep-th/0106185.

\bibitem{DelDebbio:2002yp}
L. Del~Debbio and D. Diakonov,
\newblock Phys.Lett. B544 (2002) 202, hep-lat/0205015.

\bibitem{DelDebbio:2003tk}
L. Del~Debbio, H. Panagopoulos and E. Vicari,
\newblock JHEP 0309 (2003) 034, hep-lat/0308012.

\bibitem{Meyer:2004hv}
H. Meyer and M. Teper,
\newblock JHEP 0412 (2004) 031, hep-lat/0411039.

\bibitem{Lucini:2004my}
B. Lucini, M. Teper and U. Wenger,
\newblock JHEP 0406 (2004) 012, hep-lat/0404008.

\bibitem{Athenodorou:2010cs}
A. Athenodorou, B. Bringoltz and M. Teper,
\newblock JHEP 1102 (2011) 030, 1007.4720.

\bibitem{Lohmayer:2012ue}
R. Lohmayer and H. Neuberger,
\newblock JHEP 1208 (2012) 102, 1206.4015.

\bibitem{Mykkanen:2012dv}
A. Mykk{\"a}nen,
\newblock JHEP 1212 (2012) 069, 1209.2372.

\bibitem{Luscher:1980fr}
M. L{\"u}scher, K. Symanzik and P. Weisz,
\newblock Nucl.Phys. B173 (1980) 365.

\bibitem{Luscher:1980ac}
M. L{\"u}scher,
\newblock Nucl.Phys. B180 (1981) 317.

\bibitem{Nambu:1974zg}
Y. Nambu,
\newblock Phys.Rev. D10 (1974) 4262.

\bibitem{Goto:1971ce}
T. Got{\={o}},
\newblock Prog.Theor.Phys. 46 (1971) 1560.

\bibitem{Polchinski:1992vg}
J. Polchinski,
\newblock (1992), hep-th/9210045.

\bibitem{Luscher:2004ib}
M. L{\"u}scher and P. Weisz,
\newblock JHEP 0407 (2004) 014, hep-th/0406205.

\bibitem{Drummond:2004yp}
J. Drummond,
\newblock (2004), hep-th/0411017.

\bibitem{Billo:2006zg}
M. Bill{\'o}, M. Caselle and L. Ferro,
\newblock JHEP 0602 (2006) 070, hep-th/0601191.

\bibitem{Meyer:2006qx}
H.B. Meyer,
\newblock JHEP 0605 (2006) 066, hep-th/0602281.

\bibitem{Aharony:2009gg}
O. Aharony and E. Karzbrun,
\newblock JHEP 0906 (2009) 012, 0903.1927.

\bibitem{Aharony:2010cx}
O. Aharony and M. Field,
\newblock JHEP 1101 (2011) 065, 1008.2636.

\bibitem{Aharony:2011ga}
O. Aharony, M. Field and N. Klinghoffer,
\newblock JHEP 1204 (2012) 048, 1111.5757.

\bibitem{Aharony:2010db}
O. Aharony and N. Klinghoffer,
\newblock JHEP 1012 (2010) 058, 1008.2648.

\bibitem{Aharony:2011gb}
O. Aharony and M. Dodelson,
\newblock JHEP 1202 (2012) 008, 1111.5758.

\bibitem{Billo:2012da}
M. Bill{\'o} et~al.,
\newblock JHEP 1205 (2012) 130, 1202.1984.

\bibitem{Gomis:2012ki}
J. Gomis, K. Kamimura and J.M. Pons,
\newblock Nucl.Phys. B871 (2013) 420, 1205.1385.

\bibitem{Dubovsky:2012sh}
S. Dubovsky, R. Flauger and V. Gorbenko,
\newblock JHEP 1209 (2012) 044, 1203.1054.

\bibitem{Gliozzi:2012cx}
F. Gliozzi and M. Meineri,
\newblock JHEP 1208 (2012) 056, 1207.2912.

\bibitem{Bali:1994de}
G. Bali, K. Schilling and C. Schlichter,
\newblock Phys.Rev. D51 (1995) 5165, hep-lat/9409005.

\bibitem{Luscher:2002qv}
M. L{\"u}scher and P. Weisz,
\newblock JHEP 0207 (2002) 049, hep-lat/0207003.

\bibitem{Juge:2002br}
K.J. Juge, J. Kuti and C. Morningstar,
\newblock Phys.Rev.Lett. 90 (2003) 161601, hep-lat/0207004.

\bibitem{Bonati:2011nt}
C. Bonati,
\newblock Phys.Lett. B703 (2011) 376, 1106.5920.

\bibitem{Koma:2003gi}
Y. Koma, M. Koma and P. Majumdar,
\newblock Nucl.Phys. B692 (2004) 209, hep-lat/0311016.

\bibitem{Panero:2004zq}
M. Panero,
\newblock Nucl.Phys.Proc.Suppl. 140 (2005) 665, hep-lat/0408002.

\bibitem{Panero:2005iu}
M. Panero,
\newblock JHEP 0505 (2005) 066, hep-lat/0503024.

\bibitem{Amado:2013rja}
A. Amado, N. Cardoso and P. Bicudo,
\newblock (2013), 1309.3859.

\bibitem{Greensite:2006sm}
J. Greensite et~al.,
\newblock Phys.Rev. D75 (2007) 034501, hep-lat/0609050.

\bibitem{Allton:2008ty}
C. Allton, M. Teper and A. Trivini,
\newblock JHEP 0807 (2008) 021, 0803.1092.

\bibitem{Parisi:1980pe}
G. Parisi,
\newblock World Sci.Lect.Notes Phys. 49 (1980) 349.

\bibitem{Lepage:1992xa}
G.P. Lepage and P.B. Mackenzie,
\newblock Phys.Rev. D48 (1993) 2250, hep-lat/9209022.

\bibitem{Luscher:1991wu}
M. L{\"u}scher, P. Weisz and U. Wolff,
\newblock Nucl.Phys. B359 (1991) 221.

\bibitem{Luscher:1992an}
M. L{\"u}scher et~al.,
\newblock Nucl.Phys. B384 (1992) 168, hep-lat/9207009.

\bibitem{Sint:1993un}
S. Sint,
\newblock Nucl.Phys. B421 (1994) 135, hep-lat/9312079.

\bibitem{Lucini:2008vi}
B. Lucini and G. Moraitis,
\newblock Phys. Lett. B668 (2008) 226, 0805.2913.

\bibitem{Luscher:1992zx}
M. L{\"u}scher et~al.,
\newblock Nucl.Phys. B389 (1993) 247, hep-lat/9207010.

\bibitem{Luscher:1993gh}
M. L{\"u}scher et~al.,
\newblock Nucl.Phys. B413 (1994) 481, hep-lat/9309005.

\bibitem{DeGrand:2012qa}
T. DeGrand, Y. Shamir and B. Svetitsky,
\newblock Phys.Rev. D85 (2012) 074506, 1202.2675.

\bibitem{DeGrand:2013uha}
T. DeGrand, Y. Shamir and B. Svetitsky,
\newblock Phys.Rev. D88 (2013) 054505, 1307.2425.

\bibitem{Meyer:2004gx}
H.B. Meyer,
\newblock (2004), hep-lat/0508002.

\bibitem{Meyer:2004jc}
H.B. Meyer and M.J. Teper,
\newblock Phys.Lett. B605 (2005) 344, hep-ph/0409183.

\bibitem{DelDebbio:2007wk}
L. Del~Debbio et~al.,
\newblock JHEP 0803 (2008) 062, 0712.3036.

\bibitem{Bali:2008an}
G.S. Bali and F. Bursa,
\newblock JHEP 0809 (2008) 110, 0806.2278.

\bibitem{Hietanen:2009tu}
A. Hietanen et~al.,
\newblock Phys. Lett. B674 (2009) 80, 0901.3752.

\bibitem{Lucini:2010nv}
B. Lucini, A. Rago and E. Rinaldi,
\newblock JHEP 1008 (2010) 119, 1007.3879.

\bibitem{DeGrand:2012hd}
T. DeGrand,
\newblock Phys.Rev. D86 (2012) 034508, 1205.0235.

\bibitem{Bali:2013kia}
G.S. Bali et~al.,
\newblock JHEP 1306 (2013) 071, 1304.4437.

\bibitem{DeGrand:2013nna}
T. DeGrand,
\newblock Phys.Rev. D89 (2014) 014506, 1308.4114.

\bibitem{Bochicchio:2013sra}
M. Bochicchio,
\newblock (2013), 1308.2925.

\bibitem{Narayanan:2005gh}
R. Narayanan and H. Neuberger,
\newblock Phys.Lett. B616 (2005) 76, hep-lat/0503033.

\bibitem{Adkins:1983ya}
G.S. Adkins, C.R. Nappi and E. Witten,
\newblock Nucl.Phys. B228 (1983) 552.

\bibitem{Jenkins:2009wv}
E.E. Jenkins et~al.,
\newblock Phys.Rev. D81 (2010) 014502, 0907.0529.

\bibitem{WalkerLoud:2008bp}
A. Walker-Loud et~al.,
\newblock Phys.Rev. D79 (2009) 054502, 0806.4549.

\bibitem{Jenkins:1995td}
E.E. Jenkins and R.F. Lebed,
\newblock Phys. Rev. D52 (1995) 282, hep-ph/9502227.

\bibitem{WalkerLoud:2011ab}
A. Walker-Loud,
\newblock Phys.Rev. D86 (2012) 074509, 1112.2658.

\bibitem{Lucini:2001rc}
B. Lucini and M. Teper,
\newblock Nucl.Phys.Proc.Suppl. 106 (2002) 685, hep-lat/0110004.

\bibitem{Cundy:2002hv}
N. Cundy, M. Teper and U. Wenger,
\newblock Phys.Rev. D66 (2002) 094505, hep-lat/0203030.

\bibitem{DelDebbio:2002xa}
L. Del~Debbio, H. Panagopoulos and E. Vicari,
\newblock JHEP 0208 (2002) 044, hep-th/0204125.

\bibitem{Lucini:2004yh}
B. Lucini, M. Teper and U. Wenger,
\newblock Nucl. Phys. B715 (2005) 461, hep-lat/0401028.

\bibitem{DelDebbio:2006df}
L. Del~Debbio et~al.,
\newblock JHEP 0606 (2006) 005, hep-th/0603041.

\bibitem{Witten:1978bc}
E. Witten,
\newblock Nucl.Phys. B149 (1979) 285.

\bibitem{Lucini:2002ku}
B. Lucini, M. Teper and U. Wenger,
\newblock Phys.Lett. B545 (2002) 197, hep-lat/0206029.

\bibitem{Lucini:2003zr}
B. Lucini, M. Teper and U. Wenger,
\newblock JHEP 0401 (2004) 061, hep-lat/0307017.

\bibitem{Lucini:2005vg}
B. Lucini, M. Teper and U. Wenger,
\newblock JHEP 0502 (2005) 033, hep-lat/0502003.

\bibitem{Bursa:2005yv}
F. Bursa and M. Teper,
\newblock JHEP 0508 (2005) 060, hep-lat/0505025.

\bibitem{Bringoltz:2005rr}
B. Bringoltz and M. Teper,
\newblock Phys.Lett. B628 (2005) 113, hep-lat/0506034.

\bibitem{Bringoltz:2005xx}
B. Bringoltz and M. Teper,
\newblock Phys.Rev. D73 (2006) 014517, hep-lat/0508021.

\bibitem{deForcrand:2005rg}
P. de~Forcrand, B. Lucini and D. Noth,
\newblock PoS LAT2005 (2006) 323, hep-lat/0510081.

\bibitem{Panero:2008mg}
M. Panero,
\newblock PoS LATTICE 2008 (2008) 175, 0808.1672.

\bibitem{Panero:2009tv}
M. Panero,
\newblock Phys. Rev. Lett. 103 (2009) 232001, 0907.3719.

\bibitem{Datta:2009jn}
S. Datta and S. Gupta,
\newblock Phys.Rev. D80 (2009) 114504, 0909.5591.

\bibitem{Datta:2010sq}
S. Datta and S. Gupta,
\newblock Phys.Rev. D82 (2010) 114505, 1006.0938.

\bibitem{Mykkanen:2012ri}
A. Mykk{\"a}nen, M. Panero and K. Rummukainen,
\newblock JHEP 1205 (2012) 069, 1202.2762.

\bibitem{Lucini:2012wq}
B. Lucini, A. Rago and E. Rinaldi,
\newblock Phys.Lett. B712 (2012) 279, 1202.6684.

\bibitem{Bonati:2013tt}
C. Bonati et~al.,
\newblock Phys. Rev. Lett. 110 (2013) 252003, 1301.7640.

\bibitem{Aoki:2006br}
Y. Aoki et~al.,
\newblock Phys.Lett. B643 (2006) 46, hep-lat/0609068.

\bibitem{Aoki:2009sc}
Y. Aoki et~al.,
\newblock JHEP 0906 (2009) 088, 0903.4155.

\bibitem{Bazavov:2010sb}
HotQCD collaboration, A. Bazavov and P. Petreczky,
\newblock J.Phys.Conf.Ser. 230 (2010) 012014, 1005.1131.

\bibitem{Bazavov:2011nk}
A. Bazavov et~al.,
\newblock Phys.Rev. D85 (2012) 054503, 1111.1710.

\bibitem{DeTar:2009ef}
C. DeTar and U. Heller,
\newblock Eur.Phys.J. A41 (2009) 405, 0905.2949.

\bibitem{Petreczky:2012rq}
P. Petreczky,
\newblock J.Phys. G39 (2012) 093002, 1203.5320.

\bibitem{Philipsen:2012nu}
O. Philipsen,
\newblock Prog.Part.Nucl.Phys. 70 (2013) 55, 1207.5999.

\bibitem{Thorn:1980iv}
C.B. Thorn,
\newblock Phys.Lett. B99 (1981) 458.

\bibitem{Gocksch:1982en}
A. Gocksch and F. Neri,
\newblock Phys.Rev.Lett. 50 (1983) 1099.

\bibitem{Greensite:1982be}
J. Greensite and M. Halpern,
\newblock Phys.Rev. D27 (1983) 2545.

\bibitem{Pisarski:1983db}
R.D. Pisarski,
\newblock Phys.Rev. D29 (1984) 1222.

\bibitem{McLerran:1985uh}
L.D. McLerran and A. Sen,
\newblock Phys.Rev. D32 (1985) 2794.

\bibitem{Toublan:2005rq}
D. Toublan,
\newblock Phys.Lett. B621 (2005) 145, hep-th/0501069.

\bibitem{Engels:1990vr}
J. Engels et~al.,
\newblock Phys.Lett. B252 (1990) 625.

\bibitem{Fingberg:1992ju}
J. Fingberg, U.M. Heller and F. Karsch,
\newblock Nucl.Phys. B392 (1993) 493, hep-lat/9208012.

\bibitem{Engels:1994xj}
J. Engels, F. Karsch and K. Redlich,
\newblock Nucl.Phys. B435 (1995) 295, hep-lat/9408009.

\bibitem{Svetitsky:1982gs}
B. Svetitsky and L.G. Yaffe,
\newblock Nucl.Phys. B210 (1982) 423.

\bibitem{condmat0012164}
A. Pelissetto and E. Vicari,
\newblock Phys.Rept. 368 (2002) 549, cond-mat/0012164.

\bibitem{Boyd:1996bx}
G. Boyd et~al.,
\newblock Nucl.Phys. B469 (1996) 419, hep-lat/9602007.

\bibitem{Borsanyi:2012ve}
S. Bors{\'a}nyi et~al.,
\newblock JHEP 1207 (2012) 056, 1204.6184.

\bibitem{Bhattacharya:1990hk}
T. Bhattacharya et~al.,
\newblock Phys.Rev.Lett. 66 (1991) 998.

\bibitem{Enqvist:1990ae}
K. Enqvist and K. Kajantie,
\newblock Z.Phys. C47 (1990) 291.

\bibitem{Bhattacharya:1992qb}
T. Bhattacharya et~al.,
\newblock Nucl.Phys. B383 (1992) 497, hep-ph/9205231.

\bibitem{KorthalsAltes:1993ca}
C. Korthals~Altes,
\newblock Nucl.Phys. B420 (1994) 637, hep-th/9310195.

\bibitem{Giovannangeli:2001bh}
P. Giovannangeli and C. Korthals~Altes,
\newblock Nucl.Phys. B608 (2001) 203, hep-ph/0102022.

\bibitem{Giovannangeli:2002uv}
P. Giovannangeli and C. Korthals~Altes,
\newblock Nucl.Phys. B721 (2005) 1, hep-ph/0212298.

\bibitem{Giovannangeli:2004sg}
P. Giovannangeli and C. Korthals~Altes,
\newblock Nucl.Phys. B721 (2005) 25, hep-ph/0412322.

\bibitem{Asakawa:2012yv}
M. Asakawa, S.A. Bass and B. M{\"u}ller,
\newblock Phys.Rev.Lett. 110 (2013) 202301, 1208.2426.

\bibitem{Smilga:1993vb}
A.V. Smilga,
\newblock Annals Phys. 234 (1994) 1.

\bibitem{Elze:1988zs}
H. Elze, K. Kajantie and J.I. Kapusta,
\newblock Nucl.Phys. B304 (1988) 832.

\bibitem{Gliozzi:2007jh}
F. Gliozzi,
\newblock J.Phys. A40 (2007) F375, hep-lat/0701020.

\bibitem{Damgaard:1987wh}
P. Damgaard,
\newblock Phys.Lett. B194 (1987) 107.

\bibitem{Burnier:2009bk}
Y. Burnier, M. Laine and M. Veps{\"a}l{\"a}inen,
\newblock JHEP 1001 (2010) 054, 0911.3480.

\bibitem{Brambilla:2010xn}
N. Brambilla et~al.,
\newblock Phys.Rev. D82 (2010) 074019, 1007.5172.

\bibitem{Dumitru:2003hp}
A. Dumitru et~al.,
\newblock Phys.Rev. D70 (2004) 034511, hep-th/0311223.

\bibitem{Gupta:2007ax}
S. Gupta, K. H{\"u}bner and O. Kaczmarek,
\newblock Phys.Rev. D77 (2008) 034503, 0711.2251.

\bibitem{Son:2007vk}
D.T. Son and A.O. Starinets,
\newblock Ann.Rev.Nucl.Part.Sci. 57 (2007) 95, 0704.0240.

\bibitem{Shuryak:2008eq}
E. Shuryak,
\newblock Prog.Part.Nucl.Phys. 62 (2009) 48, 0807.3033.

\bibitem{Rangamani:2009xk}
M. Rangamani,
\newblock Class.Quant.Grav. 26 (2009) 224003, 0905.4352.

\bibitem{Gursoy:2007cb}
U. G{\"u}rsoy and E. Kiritsis,
\newblock JHEP 0802 (2008) 032, 0707.1324.

\bibitem{Gursoy:2007er}
U. G{\"u}rsoy, E. Kiritsis and F. Nitti,
\newblock JHEP 0802 (2008) 019, 0707.1349.

\bibitem{Gursoy:2008bu}
U. G{\"u}rsoy et~al.,
\newblock Phys.Rev.Lett. 101 (2008) 181601, 0804.0899.

\bibitem{Gursoy:2009jd}
U. G{\"u}rsoy et~al.,
\newblock Nucl.Phys. B820 (2009) 148, 0903.2859.

\bibitem{Andreev:2006vy}
O. Andreev,
\newblock Phys.Rev. D73 (2006) 107901, hep-th/0603170.

\bibitem{Kajantie:2006hv}
K. Kajantie, T. Tahkokallio and J.T. Yee,
\newblock JHEP 0701 (2007) 019, hep-ph/0609254.

\bibitem{Alanen:2009ej}
J. Alanen, K. Kajantie and V. Suur-Uski,
\newblock Phys.Rev. D80 (2009) 075017, 0905.2032.

\bibitem{Alanen:2009xs}
J. Alanen, K. Kajantie and V. Suur-Uski,
\newblock Phys.Rev. D80 (2009) 126008, 0911.2114.

\bibitem{Galow:2009kw}
B. Galow et~al.,
\newblock Nucl.Phys. B834 (2010) 330, 0911.0627.

\bibitem{Megias:2010ku}
E. Meg{\'{\i}}as, H. Pirner and K. Veschgini,
\newblock Phys.Rev. D83 (2011) 056003, 1009.2953.

\bibitem{Veschgini:2010ws}
K. Veschgini, E. Meg{\'{\i}}as and H. Pirner,
\newblock Phys.Lett. B696 (2011) 495, 1009.4639.

\bibitem{Jarvinen:2011qe}
M. J{\"a}rvinen and E. Kiritsis,
\newblock JHEP 1203 (2012) 002, 1112.1261.

\bibitem{Alho:2012mh}
T. Alho et~al.,
\newblock JHEP 1301 (2013) 093, 1210.4516.

\bibitem{Mia:2013pca}
M. Mia,
\newblock (2013), 1307.7732.

\bibitem{Arean:2013tja}
D. Are{\'a}n et~al.,
\newblock JHEP 1311 (2013) 068, 1309.2286.

\bibitem{Gubser:2006qh}
S.S. Gubser,
\newblock Phys.Rev. D76 (2007) 126003, hep-th/0611272.

\bibitem{Meisinger:2001cq}
P.N. Meisinger, T.R. Miller and M.C. Ogilvie,
\newblock Phys.Rev. D65 (2002) 034009, hep-ph/0108009.

\bibitem{Megias:2005ve}
E. Meg{\'i}as, E. Ruiz~Arriola and L. Salcedo,
\newblock JHEP 0601 (2006) 073, hep-ph/0505215.

\bibitem{Pisarski:2006yk}
R.D. Pisarski,
\newblock Prog.Theor.Phys.Suppl. 168 (2007) 276, hep-ph/0612191.

\bibitem{Andreev:2007zv}
O. Andreev,
\newblock Phys.Rev. D76 (2007) 087702, 0706.3120.

\bibitem{Brau:2009mp}
F. Brau and F. Buisseret,
\newblock Phys.Rev. D79 (2009) 114007, 0902.4836.

\bibitem{Megias:2009mp}
E. Meg{\'i}as, E. Ruiz~Arriola and L. Salcedo,
\newblock Phys.Rev. D80 (2009) 056005, 0903.1060.

\bibitem{Giacosa:2010vz}
F. Giacosa,
\newblock Phys.Rev. D83 (2011) 114002, 1009.4588.

\bibitem{Gogokhia:2010dw}
V. Gogokhia and M. Vas{\'u}th,
\newblock (2010), 1007.1573.

\bibitem{Lacroix:2012pt}
G. Lacroix et~al.,
\newblock Phys.Rev. D87 (2013) 054025, 1210.1716.

\bibitem{Dumitru:2012fw}
A. Dumitru et~al.,
\newblock Phys.Rev. D86 (2012) 105017, 1205.0137.

\bibitem{DelDebbio:2004rw}
L. Del~Debbio, H. Panagopoulos and E. Vicari,
\newblock JHEP 0409 (2004) 028, hep-th/0407068.

\bibitem{D'Elia:2012vv}
M. D'Elia and F. Negro,
\newblock Phys.Rev.Lett. 109 (2012) 072001, 1205.0538.

\bibitem{Unsal:2012zj}
M. {\"U}nsal,
\newblock Phys.Rev. D86 (2012) 105012, 1201.6426.

\bibitem{Poppitz:2012nz}
E. Poppitz, T. Sch{\"a}fer and M. {\"U}nsal,
\newblock JHEP 1303 (2013) 087, 1212.1238.

\bibitem{Anber:2013sga}
M.M. Anber,
\newblock Phys.Rev. D88 (2013) 085003, 1302.2641.

\bibitem{Itoh:1970uw}
N. Itoh,
\newblock Prog.Theor.Phys. 44 (1970) 291.

\bibitem{Collins:1974ky}
J.C. Collins and M. Perry,
\newblock Phys.Rev.Lett. 34 (1975) 1353.

\bibitem{Alford:1998mk}
M.G. Alford, K. Rajagopal and F. Wilczek,
\newblock Nucl.Phys. B537 (1999) 443, hep-ph/9804403.

\bibitem{Son:1998uk}
D. Son,
\newblock Phys.Rev. D59 (1999) 094019, hep-ph/9812287.

\bibitem{Berges:1998rc}
J. Berges and K. Rajagopal,
\newblock Nucl.Phys. B538 (1999) 215, hep-ph/9804233.

\bibitem{Pisarski:1999bf}
R.D. Pisarski and D.H. Rischke,
\newblock Phys.Rev. D61 (2000) 051501, nucl-th/9907041.

\bibitem{Pisarski:1999tv}
R.D. Pisarski and D.H. Rischke,
\newblock Phys.Rev. D61 (2000) 074017, nucl-th/9910056.

\bibitem{Schafer:1999pb}
T. Sch{\"a}fer and F. Wilczek,
\newblock Phys.Rev. D60 (1999) 074014, hep-ph/9903503.

\bibitem{Rapp:1999qa}
R. Rapp et~al.,
\newblock Annals Phys. 280 (2000) 35, hep-ph/9904353.

\bibitem{Rajagopal:2000wf}
K. Rajagopal and F. Wilczek,
\newblock (2000), hep-ph/0011333.

\bibitem{Stephanov:2004wx}
M.A. Stephanov,
\newblock Prog.Theor.Phys.Suppl. 153 (2004) 139, hep-ph/0402115.

\bibitem{Cohen:2004cd}
T.D. Cohen,
\newblock Phys.Rev.Lett. 93 (2004) 201601, hep-ph/0407306.

\bibitem{Cohen:2004mw}
T.D. Cohen,
\newblock Phys.Rev. D70 (2004) 116009, hep-ph/0410156.

\bibitem{Ohnishi:2006hs}
K. Ohnishi, M. Oka and S. Yasui,
\newblock Phys.Rev. D76 (2007) 097501, hep-ph/0609060.

\bibitem{McLerran:2007qj}
L. McLerran and R.D. Pisarski,
\newblock Nucl.Phys. A796 (2007) 83, 0706.2191.

\bibitem{McLerran:2008ua}
L. McLerran, K. Redlich and C. Sasaki,
\newblock Nucl.Phys. A824 (2009) 86, 0812.3585.

\bibitem{Zhitnitsky:2008ha}
A.R. Zhitnitsky,
\newblock Nucl.Phys. A813 (2008) 279, 0808.1447.

\bibitem{Glozman:2008fk}
L.Y. Glozman,
\newblock Phys.Rev. D79 (2009) 037504, 0812.1101.

\bibitem{Hidaka:2008yy}
Y. Hidaka, L.D. McLerran and R.D. Pisarski,
\newblock Nucl.Phys. A808 (2008) 117, 0803.0279.

\bibitem{Kojo:2009ha}
T. Kojo et~al.,
\newblock Nucl.Phys. A843 (2010) 37, 0912.3800.

\bibitem{Hidaka:2010ph}
Y. Hidaka et~al.,
\newblock Nucl.Phys. A852 (2011) 155, 1004.2261.

\bibitem{Torrieri:2010gz}
G. Torrieri and I. Mishustin,
\newblock Phys.Rev. C82 (2010) 055202, 1006.2471.

\bibitem{Adam:2010ds}
C. Adam, J. S{\'a}nchez-Guill{\'e}n and A. Wereszczy{\'n}ski,
\newblock Phys.Rev. D82 (2010) 085015, 1007.1567.

\bibitem{Kojo:2010fe}
T. Kojo, R.D. Pisarski and A. Tsvelik,
\newblock Phys.Rev. D82 (2010) 074015, 1007.0248.

\bibitem{Torrieri:2011dg}
G. Torrieri et~al.,
\newblock Acta Phys.Polon.Supp. 5 (2012) 897, 1110.6219.

\bibitem{Lottini:2011zp}
S. Lottini and G. Torrieri,
\newblock Phys.Rev.Lett. 107 (2011) 152301, 1103.4824.

\bibitem{Kojo:2012hf}
T. Kojo,
\newblock Nucl.Phys. A899 (2013) 76, 1208.5661.

\bibitem{deForcrand:2010ys}
P. de~Forcrand,
\newblock PoS LAT2009 (2009) 010, 1005.0539.

\bibitem{Gupta:2011ma}
S. Gupta,
\newblock PoS LATTICE2010 (2010) 007, 1101.0109.

\bibitem{Philipsen:2011zx}
O. Philipsen,
\newblock Acta Phys.Polon.Supp. 5 (2012) 825, 1111.5370.

\bibitem{Levkova:2012jd}
L. Levkova,
\newblock PoS LATTICE2011 (2011) 011, 1201.1516.

\bibitem{Aarts:2013bla}
G. Aarts,
\newblock PoS LATTICE2012 (2012) 017, 1302.3028.

\bibitem{Gattringer_Lattice_2013}
C. Gattringer,
\newblock PoS LATTICE 2013 (2013) 002.

\bibitem{Mizher:2010zb}
A.J. Mizher, M. Chernodub and E.S. Fraga,
\newblock Phys.Rev. D82 (2010) 105016, 1004.2712.

\bibitem{Chernodub:2010qx}
M. Chernodub,
\newblock Phys.Rev. D82 (2010) 085011, 1008.1055.

\bibitem{Agasian:2008tb}
N. Agasian and S. Fedorov,
\newblock Phys.Lett. B663 (2008) 445, 0803.3156.

\bibitem{Kharzeev:2009fn}
D.E. Kharzeev,
\newblock Annals Phys. 325 (2010) 205, 0911.3715.

\bibitem{Fraga:2012ev}
E.S. Fraga, J. Noronha and L.F. Palhares,
\newblock Phys.Rev. D87 (2013) 114014, 1207.7094.

\bibitem{Fraga:2012rr}
E.S. Fraga,
\newblock Lect.Notes Phys. 871 (2013) 121, 1208.0917.

\bibitem{Anber:2013tra}
M.M. Anber and M. {\"U}nsal,
\newblock (2013), 1309.4394.

\bibitem{D'Elia:2010nq}
M. D'Elia, S. Mukherjee and F. Sanfilippo,
\newblock Phys.Rev. D82 (2010) 051501, 1005.5365.

\bibitem{Bali:2011qj}
G. Bali et~al.,
\newblock JHEP 1202 (2012) 044, 1111.4956.

\bibitem{D'Elia:2012zw}
M. D'Elia, M. Mariti and F. Negro,
\newblock Phys.Rev.Lett. 110 (2013) 082002, 1209.0722.

\bibitem{Bali:2013esa}
G. Bali et~al.,
\newblock JHEP 1304 (2013) 130, 1303.1328.

\bibitem{Bruckmann:2013oba}
F. Bruckmann, G. Endr{\H{o}}di and T.G. Kov{\'a}cs,
\newblock JHEP 1304 (2013) 112, 1303.3972.

\bibitem{Bursa:2005tk}
F. Bursa and M. Teper,
\newblock Phys.Rev. D74 (2006) 125010, hep-th/0511081.

\bibitem{Teper:1997tp}
M. Teper,
\newblock Nucl.Phys.Proc.Suppl. 53 (1997) 715, hep-lat/9701004.

\bibitem{Teper:1998te}
M.J. Teper,
\newblock Phys.Rev. D59 (1999) 014512, hep-lat/9804008.

\bibitem{Lucini:2002wg}
B. Lucini and M. Teper,
\newblock Phys.Rev. D66 (2002) 097502, hep-lat/0206027.

\bibitem{Bringoltz:2006zg}
B. Bringoltz and M. Teper,
\newblock Phys.Lett. B645 (2007) 383, hep-th/0611286.

\bibitem{Athenodorou:2007du}
A. Athenodorou, B. Bringoltz and M. Teper,
\newblock Phys. Lett. B656 (2007) 132, 0709.0693.

\bibitem{Bringoltz:2008nd}
B. Bringoltz and M. Teper,
\newblock Phys.Lett. B663 (2008) 429, 0802.1490.

\bibitem{Athenodorou:2008cj}
A. Athenodorou, B. Bringoltz and M. Teper,
\newblock JHEP 0905 (2009) 019, 0812.0334.

\bibitem{Bialas:2009pt}
P. Bialas et~al.,
\newblock Nucl.Phys. B836 (2010) 91, 0912.0206.

\bibitem{Athenodorou:2011rx}
A. Athenodorou, B. Bringoltz and M. Teper,
\newblock JHEP 1105 (2011) 042, 1103.5854.

\bibitem{Caselle:2011vk}
M. Caselle et~al.,
\newblock JHEP 1104 (2011) 020, 1102.0723.

\bibitem{Majumdar:2002mr}
P. Majumdar,
\newblock Nucl.Phys. B664 (2003) 213, hep-lat/0211038.

\bibitem{Juge:2004xr}
K.J. Juge, J. Kuti and C. Morningstar,
\newblock (2004) 233, hep-lat/0401032.

\bibitem{Caselle:2004er}
M. Caselle, M. Pepe and A. Rago,
\newblock JHEP 0410 (2004) 005, hep-lat/0406008.

\bibitem{Brandt:2009tc}
B.B. Brandt and P. Majumdar,
\newblock Phys. Lett. B682 (2009) 253, 0905.4195.

\bibitem{Brandt:2010bw}
B.B. Brandt,
\newblock JHEP 1102 (2011) 040, 1010.3625.

\bibitem{Brandt:2013eua}
B.B. Brandt,
\newblock PoS EPS-HEP 2013 (2013) 540, 1308.4993.

\bibitem{Caselle:1996ii}
M. Caselle et~al.,
\newblock Nucl. Phys. B486 (1997) 245, hep-lat/9609041.

\bibitem{Caselle:2002rm}
M. Caselle, M. Panero and P. Provero,
\newblock JHEP 0206 (2002) 061, hep-lat/0205008.

\bibitem{Caselle:2002ah}
M. Caselle, M. Hasenbusch and M. Panero,
\newblock JHEP 0301 (2003) 057, hep-lat/0211012.

\bibitem{Caselle:2004jq}
M. Caselle, M. Hasenbusch and M. Panero,
\newblock JHEP 0405 (2004) 032, hep-lat/0403004.

\bibitem{Caselle:2005xy}
M. Caselle, M. Hasenbusch and M. Panero,
\newblock JHEP 0503 (2005) 026, hep-lat/0501027.

\bibitem{Caselle:2005vq}
M. Caselle, M. Hasenbusch and M. Panero,
\newblock JHEP 0601 (2006) 076, hep-lat/0510107.

\bibitem{Caselle:2006dv}
M. Caselle, M. Hasenbusch and M. Panero,
\newblock JHEP 0603 (2006) 084, hep-lat/0601023.

\bibitem{Giudice:2006hw}
P. Giudice, F. Gliozzi and S. Lottini,
\newblock JHEP 0701 (2007) 084, hep-th/0612131.

\bibitem{Caselle:2007yc}
M. Caselle, M. Hasenbusch and M. Panero,
\newblock JHEP 0709 (2007) 117, 0707.0055.

\bibitem{Giudice:2007sk}
P. Giudice, F. Gliozzi and S. Lottini,
\newblock JHEP 0705 (2007) 010, hep-th/0703153.

\bibitem{Rajantie:2012zn}
A. Rajantie, K. Rummukainen and D.J. Weir,
\newblock Phys.Rev. D86 (2012) 125040, 1210.1106.

\bibitem{Gliozzi:2005ny}
F. Gliozzi et~al.,
\newblock Nucl.Phys. B719 (2005) 255, cond-mat/0502339.

\bibitem{Giudice:2009di}
P. Giudice, F. Gliozzi and S. Lottini,
\newblock JHEP 0903 (2009) 104, 0901.0748.

\bibitem{Meyer:2003wx}
H.B. Meyer and M.J. Teper,
\newblock Nucl.Phys. B668 (2003) 111, hep-lat/0306019.

\bibitem{Liddle:2008kk}
J. Liddle and M. Teper,
\newblock (2008), 0803.2128.

\bibitem{Christensen:1990vc}
J. Christensen and P. Damgaard,
\newblock Nucl.Phys. B354 (1991) 339.

\bibitem{Teper:1993gp}
M. Teper,
\newblock Phys.Lett. B313 (1993) 417.

\bibitem{Christensen:1991rx}
J. Christensen et~al.,
\newblock Nucl.Phys. B374 (1992) 225.

\bibitem{Bialas:2012qz}
P. Bialas et~al.,
\newblock Nucl.Phys. B871 (2013) 111, 1211.3304.

\bibitem{Holland:2007ar}
K. Holland, M. Pepe and U.J. Wiese,
\newblock JHEP 0802 (2008) 041, 0712.1216.

\bibitem{deForcrand:2003wa}
P. de~Forcrand and O. Jahn,
\newblock Nucl.Phys.Proc.Suppl. 129 (2004) 709, hep-lat/0309153.

\bibitem{Holland:2005nd}
K. Holland,
\newblock JHEP 0601 (2006) 023, hep-lat/0509041.

\bibitem{Caselle:2011fy}
M. Caselle et~al.,
\newblock JHEP 1106 (2011) 142, 1105.0359.

\bibitem{Caselle:2011mn}
M. Caselle et~al.,
\newblock JHEP 1205 (2012) 135, 1111.0580.

\bibitem{D'Hoker:1981us}
E. D'Hoker,
\newblock Nucl.Phys. B201 (1982) 401.

\bibitem{Bicudo:2013yza}
P. Bicudo, R.D. Pisarski and E. Seel,
\newblock Phys.Rev. D88 (2013) 034007, 1306.2943.

\bibitem{'tHooft:1974hx}
G. 't~Hooft,
\newblock Nucl.Phys. B75 (1974) 461.

\bibitem{Durhuus:1980nb}
B. Durhuus and P. Olesen,
\newblock Nucl.Phys. B184 (1981) 461.

\bibitem{Rossi:1994xg}
P. Rossi and E. Vicari,
\newblock Phys.Lett. B349 (1995) 177, hep-lat/9412090.

\bibitem{Rossi:1996hs}
P. Rossi, M. Campostrini and E. Vicari,
\newblock Phys.Rept. 302 (1998) 143, hep-lat/9609003.

\bibitem{Olesen:2006gt}
P. Olesen,
\newblock Nucl.Phys. B752 (2006) 197, hep-th/0606153.

\bibitem{Narayanan:2007dv}
R. Narayanan and H. Neuberger,
\newblock JHEP 0712 (2007) 066, 0711.4551.

\bibitem{Olesen:2007rf}
P. Olesen,
\newblock Phys.Lett. B660 (2008) 597, 0712.0923.

\bibitem{Blaizot:2008nc}
J.P. Blaizot and M.A. Nowak,
\newblock Phys.Rev.Lett. 101 (2008) 102001, 0801.1859.

\bibitem{Neuberger:2008mk}
H. Neuberger,
\newblock Phys.Lett. B666 (2008) 106, 0806.0149.

\bibitem{Narayanan:2008he}
R. Narayanan, H. Neuberger and E. Vicari,
\newblock JHEP 0804 (2008) 094, 0803.3833.

\bibitem{Neuberger:2008ti}
H. Neuberger,
\newblock Phys.Lett. B670 (2008) 235, 0809.1238.

\bibitem{Lohmayer:2009aw}
R. Lohmayer, H. Neuberger and T. Wettig,
\newblock JHEP 0905 (2009) 107, 0904.4116.

\bibitem{Lohmayer:2011nq}
R. Lohmayer and H. Neuberger,
\newblock Phys.Rev.Lett. 108 (2012) 061602, 1109.6683.

\bibitem{Orland:2011rd}
P. Orland,
\newblock Phys.Rev. D84 (2011) 105005, 1108.0058.

\bibitem{Orland:2012sk}
P. Orland,
\newblock Phys.Rev. D86 (2012) 045023, 1205.1763.

\bibitem{Cubero:2012xi}
A. Cort{\'e}s~Cubero,
\newblock Phys.Rev. D86 (2012) 025025, 1205.2069.

\bibitem{Cubero:2013iga}
A. Cort{\'e}s~Cubero and P. Orland,
\newblock Phys.Rev. D88 (2013) 025044, 1306.1930.

\bibitem{Bringoltz:2008iu}
B. Bringoltz,
\newblock Phys.Rev. D79 (2009) 105021, 0811.4141.

\bibitem{Bringoltz:2009ym}
B. Bringoltz,
\newblock Phys.Rev. D79 (2009) 125006, 0901.4035.

\bibitem{Galvez:2009rq}
R. Galvez, A. Hietanen and R. Narayanan,
\newblock Phys.Lett. B672 (2009) 376, 0812.3449.

\bibitem{Schon:2000he}
V. Sch{\"o}n and M. Thies,
\newblock Phys.Rev. D62 (2000) 096002, hep-th/0003195.

\bibitem{Kojo:2011fh}
T. Kojo,
\newblock Nucl.Phys. A877 (2012) 70, 1106.2187.

\end{thebibliography}

\end{document}